\documentclass[a4paper,twocolumn,11pt,accepted=2026-10-03]{quantumarticle}
\pdfoutput=1
\usepackage[utf8]{inputenc}
\usepackage[english]{babel}
\usepackage[T1]{fontenc}
\usepackage{graphicx}
\usepackage{tabularx}
\usepackage{subcaption}
\usepackage{adjustbox}
\usepackage{csquotes}
\usepackage{comment}
\usepackage[linesnumbered,ruled,vlined]{algorithm2e}
\usepackage{tikz}
\usetikzlibrary{quantikz2}
\usepackage{imakeidx}
\usepackage{amsmath,amssymb,amsthm}
\usepackage{multirow}
\usepackage{xcolor}
\usepackage{braket}
\usepackage[margin=2.5cm]{geometry}
\usepackage[toc,acronym,nopostdot,nonumberlist]{glossaries}
\usepackage{tikzpagenodes}
\usepackage{adjustbox}
\usepackage[ddmmyyyy]{datetime}
\usepackage{soul}
\usepackage{cancel}
\DeclareMathOperator*{\argmax}{arg\,max}

\usepackage{setspace}
\usepackage{indentfirst}
\usepackage{lipsum}
\newcommand{\note}[1]{\textcolor{blue}{#1}}
\newcommand{\new}[1]{\textcolor{violet}{#1}}
\usepackage[numbers,sort&compress]{natbib}
\usepackage{url}
\usepackage{hyperref}
\hypersetup{
     colorlinks=true,
     linkcolor=blue,
     filecolor=blue,
     citecolor = magenta,      
     urlcolor=blue,
     }
\setstcolor{blue}
\newcommand{\Cancel}[2][black]{{\color{#1}\cancel{\color{black}#2}}}

\begin{document}

\title{Sampled-Based Guided Quantum Walk: Non-variational quantum algorithm for combinatorial optimization}

\author{Ugo Nzongani}
\affiliation{Aix-Marseille Université, CNRS, LIS, 13288 Marseille, France}
\affiliation{UMA, ENSTA, Institut Polytechnique de Paris, Palaiseau, France}
\email{ugo.nzongani@lis-lab.fr}

\author{Dylan Laplace Mermoud}
\affiliation{Sorbonne Université, CNRS, LIP6, F-75005, Paris, France}

\author{Giuseppe Di Molfetta}
\affiliation{Aix-Marseille Université, CNRS, LIS, 13288 Marseille, France}
\affiliation{Institut Universitaire de France, Paris, France}

\author{Andrea Simonetto}
\affiliation{UMA, ENSTA, Institut Polytechnique de Paris, Palaiseau, France}
\affiliation{CEDRIC, Conservatoire National des Arts et Métiers, 75003 Paris, France}

\maketitle

\begin{abstract}
    We introduce {{\textsc{SamBa--GQW}}}, a novel quantum algorithm for solving binary combinatorial optimization problems of arbitrary degree with no use of any classical optimizer. The algorithm is based on a continuous-time quantum walk on the solution space represented as a graph. The walker explores the solution space to find its way to vertices that minimize the cost function of the optimization problem. The key novelty of our algorithm is an offline classical sampling protocol that gives information about the spectrum of the problem Hamiltonian. Then, the extracted information is used to guide the walker to high quality solutions via a quantum walk with a time-dependent hopping rate. We investigate the performance of \textsc{SamBa--GQW} on several quadratic problems, namely MaxCut, maximum independent set, portfolio optimization, and higher-order polynomial problems such as LABS, MAX-$k$-SAT and a quartic reformulation of the travelling salesperson problem. We empirically demonstrate that \textsc{SamBa--GQW} finds high quality approximate solutions on problems up to a size of $n=30$ qubits by only sampling poly($n$) states among $2^n$ possible decisions. Furthermore, \textsc{SamBa--GQW} compares in par with classically optimized variational approaches, such as the variational guided quantum walk and QAOA (the latter when run on deep circuits). This places \textsc{Samba--GQW} as a promising heuristic to tackle combinatorial problems beyond the NISQ regime.
\end{abstract}

\section{Introduction}

Combinatorial optimization is a natural benchmark for quantum algorithms. On the one hand, its hard classical nature makes its problems very challenging to solve classically at scale, since the solution space grows exponentially in the number of variables. On the other hand, by its very nature, a quantum computer offers new, compact ways of exploring an exponential space with only a linear number of qubits. As a result, new quantum methods are emerging and may offer an advantage over their classical counterparts~\cite{abbas2024challenges}.

On current noisy devices (NISQ), common approaches for solving combinatorial optimization problems on quantum computers leverage variational quantum algorithms (VQAs)\cite{cerezo2021variational,mariella2023quantum,LaplaceMermoud2026variationalquantum}, which alternate between a classical optimizer and a parametric quantum observable evaluation. This class of algorithms is tailored to current machines, since the depth of the circuit is often a parameter of the algorithm and can be chosen to be small, trading-off accuracy~\cite{preskill2018quantum}.

Among those algorithms, the quantum approximate optimization algorithm (QAOA) \cite{farhi2014quantumapproximateoptimizationalgorithm,Zhou2020} is surely the most studied, 
and it comes with many variants \cite{hadfield2019quantum,Egger2021warmstartingquantum,yu2022quantum,herrman2022multi,PhysRevLett.125.260505,Sack2021quantumannealing,11668574}. The performance and potential advantage of QAOA is still under debate, but some (albeit limited) empirical evaluations show that it could have the potential of offering scaling advantage over classical methods for some classically intractable problems~\cite{doi:10.1126/sciadv.adm6761,Boulebnane2024}. However, QAOA, as many (if not all) variational methods, suffers from the presence of barren plateaus \cite{larocca2025barren}, which make the classical optimization phase very difficult, limiting any potential quantum advantage.

In order to overcome the need for a classical optimization step, we turn to quantum walks. Quantum Walks (QWs) are a coherent transportation model on graphs and a fundamental tool for simulating physics \cite{PhysRevA.89.032322,DIMOLFETTA2014157,PhysRevA.88.042301,PhysRevA.106.032408}, and designing quantum algorithms \cite{10.1145/780542.780552,PhysRevA.70.022314,doi:10.1137/S0097539705447311,PhysRevA.104.022216,10.1007/978-3-031-56943-2_6,gonzales2025efficient}. They are a universal model of quantum computation \cite{PhysRevA.81.042330,PhysRevLett.102.180501}, defined in discrete \cite{PhysRevA.48.1687} and continuous-time \cite{PhysRevA.58.915}. Furthermore, one can interpret QAOA on $n$ qubits as continuous-time QW on the $n$-dimensional hypercube where each computational basis state corresponds to a vertex of the hypercube. 

Several QWs schemes have been proposed for solving combinatorial optimization problems. In the discrete-time settings, some works propose a quantum version of Metropolis-Hasting algorithm with coined QWs \cite{Lemieux2020efficientquantum,campos2023quantum}. In the continuous-time setting, QWs were shown to be a good candidate for finding ground states of spin glass problems \cite{Callison_2019}, which is equivalent to finding the optimal solution of an optimization problem. Another example is the quantum walk optimization algorithm (QWOA) \cite{PhysRevResearch.2.023302,10.3389/fphy.2021.730856,Slate2021quantumwalkbased}, which combines the traditional QAOA with a mixer based on a continuous-time QW. The QW evolves on a graph that only connects feasible solutions. This latter approach however requires classically enumerating and encoding all feasible solutions, which can be as hard as the original optimization problem.

Other works introduce the concept of multistage QWs for tackling optimization problems \cite{PRXQuantum.2.010338,Banks2024continuoustime}, which consists of alternating several QWs with different piece-wise constant hopping rates. In this context, the guided quantum walk (GQW) scheme~\cite{PhysRevResearch.6.013312} is an extension of multistage QWs, and is a mixture between QWs and quantum annealing. The authors show that solving a combinatorial optimization problem is equivalent to finding an optimal hopping rate for the quantum walker to guide it towards high quality solutions. This hopping rate is strictly monotonous, positive, and it depends on the spectrum of the Hamiltonian encoding the cost function. They then turn to a classical optimizer to determine optimal control points of a Bézier curve that models the hopping rate. The numerical results are very encouraging and compare very favorably with respect to quantum annealing and standard quantum walks; yet, the classical optimization leads to exponential scalings~\cite{PhysRevResearch.6.013312} of the number of required iterations. Interestingly, other recent methods also use time-dependent scheduling to solve combinatorial optimization problems by using Planck's principle rather than adiabatic principle \cite{PhysRevA.111.032414}, and without any classical optimization \cite{PhysRevResearch.7.L022010}. See also~\cite{yan2022analytical} for further interesting work on the non-adiabatic regime.

In this paper, we offer the following contributions.

First, we overcome the scaling problems of GQW \cite{PhysRevResearch.6.013312} by introducing an offline classical sampling protocol and present Sampled-Based Guided Quantum Walk (\textsc{SamBa--GQW}). This algorithm is non-variational: the continuous-time QW is specified after the sampling protocol is run and we do not need any classical optimizer. This avoids the scaling issues and potential barren plateaus. 

Second, we present a quantum circuit implementation for \textsc{SamBa--GQW} whose depth is polynomial with respect to the number of qubits. This step turns the continuous-time QW into a circuit implementable on gate-based quantum computers. 

Third, we showcase our algorithm on several combinatorial optimization problems, obtaining consistently high quality approximate solutions, if not the optimal one with high probability. The algorithm compares very well with respect to the original GQW, reducing its execution time by at least one order of magnitude, and in par with QAOA in terms of approximation ratios, when run on deep circuits. 

The worst-case time complexity of \textsc{SamBa--GQW} in the continuous-time regime is inversely proportional to the spectral gap of the problem Hamiltonian, which is characteristic from adiabatic evolutions. As such, the found circuit depth also depends on the spectral gap. However, in our simulations, we observe average-case short evolution times, with respect to the number of qubits, for almost all problems. Furthermore, we observe that one does not have to perform the total evolution for the state vector to be well localized. Thus, for long time evolution, one could only perform a part of the quantum walk evolution, measure prematurely several times to obtain an approximation of the distribution, which is well localized with respect to the initial distribution, and classically recover the best solutions efficiently.

{\bf Organization. } After a short background section (Section~\ref{sec:background}), we  present our contribution: \textsc{SamBa--GQW}, in Section~\ref{sec:samba}. We formally introduce the set of combinatorial problems on which we test our algorithm in Section~\ref{sec:problems}, and we showcase our numerical results in Section~\ref{sec:results}. Lastly, we compare our method with the state of the art: GQW and QAOA in Section~\ref{sec:sota}.

\section{Background}\label{sec:background}

We present here the necessary background. Throughout the paper, we will be interested in solving binary optimization problems in the form,
\begin{equation}\label{prob.combo}
x^* \in \arg\min_{x\in\{0,1\}^n} \, C(x)\quad \textrm{subject to } x \in \mathcal{F},
\end{equation}
where $C:\{0,1\}^n \rightarrow \mathbb{R}$ is the cost, and $\mathcal{F}$ the constraint set. We call $x$ a decision while we label $x^*$ the optimal solution. We also call any $x \in \{0,1\}^n \cap \mathcal{F}$ a feasible decision (i.e., any binary decision that respects the constraints). If $x$ is a feasible decision, then an $\epsilon$-approximate solution ($\epsilon>0$) is a vector, such that, 
$$
C(x) - C(x^*) \leq \epsilon. 
$$
In the following, $C(x)$ will be represented as a polynomial in $x$, while $\mathcal{F}$ will be either the whole space (in such case, we call the problem unconstrained), or a linear constraint.

\subsection{Adiabatic evolution for optimization}

The solution of a combinatorial optimization problem as~\eqref{prob.combo} of $n$ binary variables can be expressed as a bitstring $x=\{0,1\}^n$. When the problem is unconstrained, the space of solutions scales exponentially with the size of the problem as there are $N=2^n$ candidates. 

Naturally, one can use $n$ qubits to represent this candidate space as a $N$-dimensional Hilbert space $\mathcal{H}^{\otimes n}$. One can use then the cost Hamiltonian to apply a bias on each computational basis state as,
\begin{equation}\label{eq:cost_hamiltonian}
    H_C=\sum_{j=0}^{N-1}C(j)\ket{j}\bra{j}.
\end{equation}

Variational quantum algorithms take inspiration from quantum annealing for solving such problems. The principle consists of interpolating between two non-commuting Hamiltonians with a specific annealing schedule. Such adiabatic evolutions are of the form \cite{farhi2000quantum,RevModPhys.90.015002}:
\begin{equation}\label{eq:adiabatic}
    H(t)=\left(1-\frac{t}{T}\right)H_0 + \frac{t}{T}H_1.
\end{equation}
The adiabatic theorem of quantum mechanics states that if the system starts in the ground state of $H_0$ and evolves slowly enough, it will remain in the ground state of $H(t)$ during the evolution. Eventually, it will end up in the ground state of $H_1$ when $t=T$. Therefore, if the Hamiltonians are chosen such that the ground state of $H_0$ is easy to prepare and that of $H_1$ encodes the solution of a combinatorial optimization problem, an adiabatic evolution would lead to the optimal solution \cite{farhi2000quantum,RevModPhys.90.015002}. A perfect candidate for the role of the starting Hamiltonian in the unconstrained case is the transverse magnetic field:
\begin{equation}\label{eq:mixer}
    H_M=-\sum_{j=0}^{n-1}\sigma_j^x,
\end{equation}
where $\sigma^x_j$ is the Pauli-$X$ matrix applied on qubit $j$. The ground state of $H_M$ is the uniform superposition over the computational basis states, which is trivial to prepare with a Hadamard transform on the $n$ input qubits $\ket{0}^{\otimes n}$:
\begin{equation}
    \ket{+}^{\otimes n}=\frac{1}{\sqrt{N}}\sum_{j=0}^{N-1}\ket{j}.
\end{equation}
Thus, setting $H_0=H_M$ and $H_1=H_C$ in Eq. \eqref{eq:adiabatic} would induce a continuous-time evolution for solving combinatorial optimization problems. Moreover, since $H_M$ corresponds to the adjacency matrix of an $n$-dimensional hypercube, this evolution can be interpreted as a continuous-time quantum walk on an hypercube with a vertex-dependent potential field (induced by $H_C$ as it applies a quality-dependent weight on each vertex). Hence, the Hilbert space is spanned by the vertices of the $n$-dimensional hypercube, i.e. each solution $x$ is associated to vertex $\ket{x}$.

\subsection{Guided Quantum Walk}


The authors of Ref.~\cite{PhysRevResearch.6.013312} revisit the continuous evolution~\eqref{eq:adiabatic} by proposing to use a more general time-dependent hopping rate (later indicated with $\Gamma(t)$) for the annealing schedule, as:
\begin{equation}\label{eq:hamiltonian}
    H(t)=\Gamma(t)\,H_M + H_C.
\end{equation}

To determine an optimal $\Gamma(t)$, they leverage the theory of local amplitude transfer and the energy levels of the Hamiltonians. The intuition is that if the energy levels of $\Gamma(t)\,H_M$ and $H_C$ are balanced, then the transfer between the two Hamiltonians is the greatest and it is more probable that the evolution favors high quality approximate solutions. 

Let $\Gamma$ be a function of the energy $E$: $\Gamma = \Gamma(E)$. Then, one can compute the local energy gaps between vertices $j$ and $k$ for $H_M$ and $H_C$ to be $|\Delta^M_{jk}|=2$ and $\Delta^C_{jk}=|\bra{j}H_C\ket{j}-\bra{k}H_C\ket{k}|$, respectively. Moreover, the transfer of amplitude from the highest to the lowest local energy state is maximized when the relative strength between $H_M$ and $H_C$ is balanced, i.e. $|\Gamma(E)\Delta^M_{jk}|\approx\Delta^C_{jk}$. By using such balancing condition, one can derive the hopping rate as \cite{PhysRevResearch.6.013312}:
\begin{equation}\label{eq:hopping}
    \Gamma(E) = \frac{\langle\Delta^C\rangle(E)}{2},
\end{equation}
where $\langle\Delta^C\rangle(E)$ is the average of largest energy gaps at energy $E$.

Now, we can naturally map $\Gamma(E)$ to a function of time: the hopping rate starts at high values and monotonically decreases such that $\Gamma(t=0\equiv E_{\max})$ and $\Gamma(t=T\equiv E_{\min})$. This time-dependent hopping rate will sequentially activate different regions of the solution graph ($n$-dimensional hypercube) and gradually confine the walker to low-energy regions, i.e., optimal solutions. Since Eq.~\eqref{eq:hopping} requires knowledge of the spectrum of $H_C$, its computation is exponential in $n$. To overcome this issue, the authors of Ref.~\cite{PhysRevResearch.6.013312} use an ansatz parametrized with six parameters to approximate the shape of $\Gamma(t)$. However, their approach leads to an exponential scaling of the number of iterations of the variational approach with respect to the problem size, for the classical optimization.

\section{Sampled-Based Guided Quantum Walk}\label{sec:samba}

We are now ready to propose a one-shot algorithm for solving combinatorial optimization on quantum computers without using any classical optimizer. 

The key idea is that one does not need to have the exact spectrum of $H_C$ nor optimize a complex non-linear curve to obtain satisfactory optimization results. A sufficiently good {\emph{estimation}} of the spectrum of $H_C$ is enough. This estimation can be obtain classically by sampling only a polynomial number of points, instead of $2^n$ for the exact spectrum. In fact, we show numerically that $n^2$ points are sufficient in most cases. 

Our algorithm then consists of two parts. The first classical offline part is an efficient sampling protocol for approximating the optimal hopping rate of Eq. \eqref{eq:hopping}, while the second quantum online part consists of running the continuous-time evolution of Eq. \eqref{eq:hamiltonian}. We also propose a quantum circuit implementation of this evolution for it to be run on a gate-based quantum computer. We present the complete algorithm in Algorithm~\ref{alg:full_algorithm}.

\begin{algorithm}[t]
\DontPrintSemicolon
\footnotesize
\caption{Sampled-Based Guided Quantum Walk (\textsc{SamBa--GQW})}\label{alg:full_algorithm}
\KwIn{Optimization problem of $n$ variables}
\KwOut{An approximate solution}
\BlankLine
\emph{Classical offline part}\;
Perform the sampling protocol (\textsc{Sampler}, Alg.~\ref{alg:sampling}) for $q=\text{poly}(n)$ samples\;
Build the annealing schedule $\Gamma(t)$ via \textsc{Builder}, Alg.~\ref{alg:builder}\;
\emph{Quantum online part}\;
Run the continuous-time quantum walk evolution of Eq.~\eqref{eq:rescaled_hamiltonian} (or its discrete version Eq.~\eqref{eq:qaoa_like})\;
\emph{Measurements}\;
Perform $\text{poly}(n)$ measurements in the computational basis and keep the best solution\;
\end{algorithm}

\subsection{Classical offline part: sampling}

Instead of using a variational method, i.e. classical optimization to fine-tune the parameters, we propose to approximate Eq.~\eqref{eq:hopping} with a sampling protocol. We compute $\langle\Delta^C\rangle$ for $q$ candidate points $x\in\{0,1\}^n$. If $q=2^n$ the computation of Eq. \eqref{eq:hopping} is exact, but exponentially costly; while if $q = \textrm{poly}(n)$, the computation of Eq.~\eqref{eq:hopping} is approximated but computationally advantageous. We will see later than $q = n^2$ will be sufficient\footnote{This is not extremely surprising, since we know that smooth curves can be well approximated with poly log inputs~\cite{JMLR:v21:19-1015}. }. At each iteration we proceed as follows, first we randomly select $x\in \mathcal{S}\coloneqq\{0,1\}^n$. Then, we compute the energy gap with $x$ and its neighbors states and we store the largest gap with its associated energy. The set of neighbors of $x$ depends on the connectivity of the mixer. In the case of the $X$-mixer of Eq. \eqref{eq:mixer}, it encodes a $n$-dimensional hypercube, therefore, each state has $n$ neighbors\footnote{In that case, a state $\ket{x}$ is connected to another state $\ket{y}$ if their Hamming distance is 1, i.e. if they only differ by 1 bit.}. Lastly, we remove $x$ from $\mathcal{S}$. Once this protocol is performed for the $q$ solutions to sample, we compute the mean of the stored gaps for their corresponding energy value and we obtain an approximation of Eq. \eqref{eq:hopping}. Lastly, we interpolate between discrete energy gaps the obtain a continuous and smooth hopping rate as a function of the energy. The formal description of the sampling protocol is presented in Algorithm~ \ref{alg:sampling}, labelled \textsc{Sampler}. We note here the similarity of this protocol to a local neighborhood search in combinatorial optimization.   

The complexity of the sampling depends on three elements: the number $q$ of sampled candidates, here $q = \textrm{poly}(n)$; the maximum number of neighbors per state (which depends on the mixer connectivity), $N_{\mathrm{neigh}}$, and the complexity of classically querying the cost function: $\mathsf{C}_{\mathrm{cost}}$. Therefore, the complexity of the sampling protocol is:
\begin{equation}\label{eq:sampling}
    \mathsf{C}_{\mathrm{sampling}} \leq q\,  N_{\mathrm{neigh}} \, \mathsf{C}_{\mathrm{cost}}.
\end{equation}

Since we are mainly concerned with polynomial problems, we also know that the complexity of querying the cost function of a polynomial problem of degree $d$ is $\mathsf{C}_{\mathrm{cost}}=\mathcal{O}(n^d)$. 

If we put all this together, the complexity of the sampling is $\text{poly}(n) N_{\mathrm{neigh}}$. For the $X$-mixer of Eq. \eqref{eq:mixer}, we have that $N_{\mathrm{neigh}}=n$, and the whole complexity is still polynomial. 

\begin{algorithm}[t]
\DontPrintSemicolon
\footnotesize
\caption{\textsc{Sampler}}\label{alg:sampling}
\KwIn{$1 \leq q \leq N=2^n$}
\KwOut{Set of energies $\mathcal{E}$}
\BlankLine
$\mathcal{S}_0 \gets$ set of candidates with $|\mathcal{S}_0|\leq 2^n$\;
$\mathcal{S}_1 \gets \emptyset$\;
$E \gets$ empty dictionary\;
$\mathrm{B}(x) \gets$ returns the neighbors of state $x$\;
\For{$i \in [1, q]$}{
  Randomly generate $x \in \mathcal{S}_0 \setminus \mathcal{S}_1$\;
  $\Delta \gets \{\Delta^C_{x,y} / \lvert\Delta^M_{x,y}\rvert \mid C(x) > C(y), \forall y \in \mathrm{B}(x)\}$\footnote{Notice how we only consider positive energy gaps instead of taking their absolute values, we discuss this choice in Appendix \ref{app:sampling_absolute}.}\;
  $E[C(x)] \gets \text{mean}(E[C(x)] + \max \Delta)$\;
  Add $x$ to $\mathcal{S}_1$\;
}
$\mathcal{E} \gets$ values of $E$\;
Add $0$ to $\mathcal{E}$\;
Remove all duplicates from $\mathcal{E}$\;
Sort $\mathcal{E}$ in decreasing order\;
\end{algorithm}

\begin{algorithm}
\DontPrintSemicolon
\footnotesize
\caption{\textsc{Builder}}\label{alg:builder}
\KwIn{List of sampled energy gaps $\mathcal{E}$ sorted in decreasing order}
\KwOut{The hopping schedule $\Gamma(t)$}
\BlankLine
$dt\_list \gets [0]$\;
$t \gets 0$\;
\For{$\gamma \in \mathcal{E}$}{
  $t_\gamma \gets \frac{\pi}{2\sqrt{2}}\gamma^{-1}$ 
  Add $t_\gamma$ to $dt\_list$\;
    $t \gets t + t_\gamma$\;
      Remove $\gamma$ from $\mathcal{E}$\;
}
$\Gamma(t) \gets$ interpolation of $\mathcal{E}$ where the distance between successive sampled points $\mathcal{E}[i]$ and $\mathcal{E}[j]$ ($i<j$) is $dt\_list[i+1]$\;
\end{algorithm}

\subsection{Hopping rate determination}

Having sampled the energy gaps in the Hamiltonians, we are now ready to set the hopping rate. We leverage once more~\cite{PhysRevResearch.6.013312}. If we start in a local superposition of states $\ket{j}$ and $\ket{k}$, the probability of measuring the lower energy state $\ket{k}$ in this \(2\)-state system is given by:
\begin{multline}\label{eq:guided}
    \mathbb{P}_{jk}(t, \Gamma(E))=\frac{1}{2}+\\ \frac{\Gamma(E)\delta_{jk}}{\Gamma(E)^2 + \delta_{jk}^2}\sin\left(t\sqrt{\Gamma(E)^2 + \delta_{jk}^2}\right)^2,
\end{multline}
with the short-hand notation $\delta_{jk} = \Delta^C_{jk}/2$. Moreover, as amplitude transfer from $\ket{j}$ to $\ket{k}$ is maximized when the relative strength between $H_M$ and $H_C$ is balanced, the matching hopping rate is $\Gamma^*_{jk} = \delta_{jk}$. Thus, we can rewrite Eq.~\eqref{eq:guided} as: 
\begin{equation}
    \mathbb{P}_{jk}(t,\Gamma^*) = \frac{1}{2} + \frac{1}{2}\sin\left(t\Gamma^*_{jk} \sqrt{2}\right)^2.
\end{equation}
To obtain a high probability of measuring a good low energy solution, we need to find the evolution time which maximizes $\mathbb{P}_{jk}(t,\Gamma^*)$. This corresponds to solving $\mathbb{P}_{jk}(t,\Gamma^*) = 1$:
\begin{equation}\label{eq:time}
    t_{jk}^* = \frac{\pi}{2\sqrt{2}} \left( \Gamma^*_{jk} \right)^{-1},
\end{equation}
in the following, we refer to this time as the \emph{transfer time}.

In order to maximize the probability of measuring {\em all} the lowest-energy states, i.e. optimal solutions, one should run the quantum walk evolution of Eq.~\eqref{eq:hamiltonian} for a total duration:
\begin{equation}\label{eq:time_opt}
    T^* = \frac{\pi}{2\sqrt{2}} \sum_{(jk) \in E_M} (\Gamma^*_{jk})^{-1},
\end{equation}
with $E_M$ the set of edges of the mixer underlying graph. In this way, the walker has time to hop through all the energy gaps. Each term of the sum in Eq.~\eqref{eq:time_opt} is upper bounded by $\Delta^{-1}$, where $\Delta$ the spectral gap of $H_C$, i.e. the modulus difference between its ground and first excited states. Therefore, the scaling of the exact time evolution is $T^*=\mathcal{O}(\Delta^{-1})$. In addition, in the case of the $n$-dimensional hypercube, \(\lvert E_M\rvert=n2^{n-1}\), so that  $T^*=\mathcal{O}(n2^{n-1}\Delta^{-1})$. 

In practice, one does not have access to the perfect hopping rate $\Gamma^*$, but rather to that of Eq.~\eqref{eq:hopping}, which corresponds to the average of the largest energy gaps. Additionally, we approximate $\Gamma$ by sampling the cost function. Therefore, we employ the approximation of $T^*$ as,
\begin{equation}\label{eq:t_approx}
    T = \frac{\pi}{2\sqrt{2}}\sum_{(jk) \in E_S}\Gamma^{-1}_{jk},
\end{equation}
with \(E_S\) being the set of sampled edges. As we approximate \(\Gamma\) using a number of samples that is polynomial in \(n\), we obtain $T=\mathcal{O}(\Delta^{-1}\mathrm{poly}(n))$. 

Finally, as seen in Eq.~\eqref{eq:time}, the time input of $\Gamma$ depends on the energy. With the rescaling $t_{ij} = \Gamma^{-1}_{jk}$, we can write the evolution as:
\begin{equation}\label{eq:rescaled_hamiltonian}
    H(t_{jk})=\Gamma \Big( \hspace{1pt} \frac{\pi}{2\sqrt{2}} t_{jk} \Big) H_M + H_C.
\end{equation}

\begin{figure}
    \centering

    \begin{subfigure}[b]{0.49\linewidth}
        \centering
        \includegraphics[width=\linewidth]{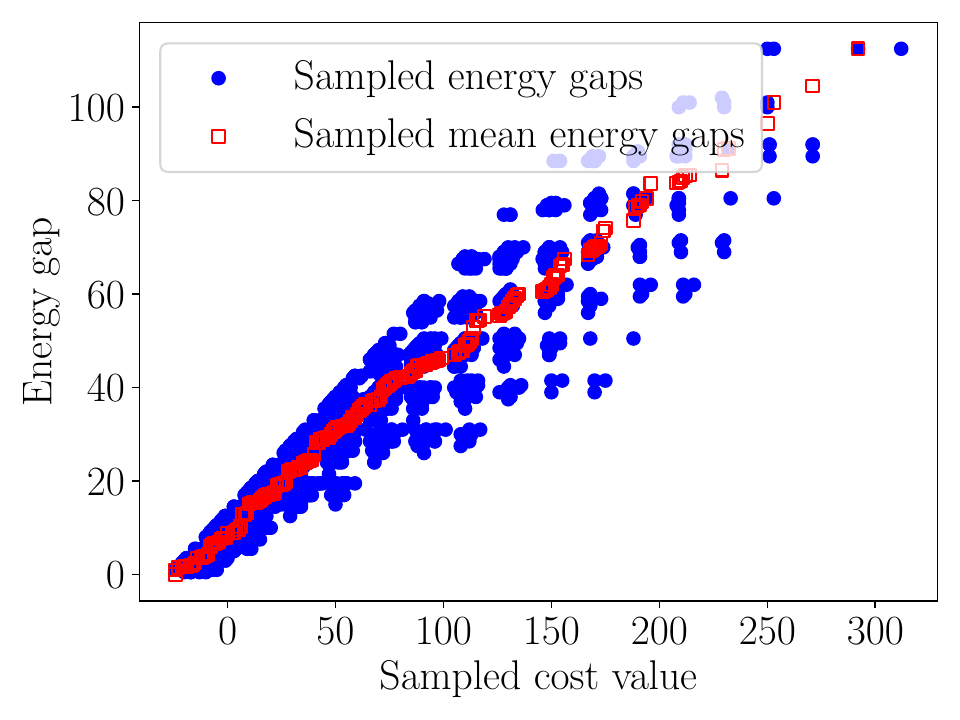}
        \caption{}
        \label{fig:sample_exact}
    \end{subfigure}
    \begin{subfigure}[b]{0.49\linewidth}
        \centering
        \includegraphics[width=\linewidth]{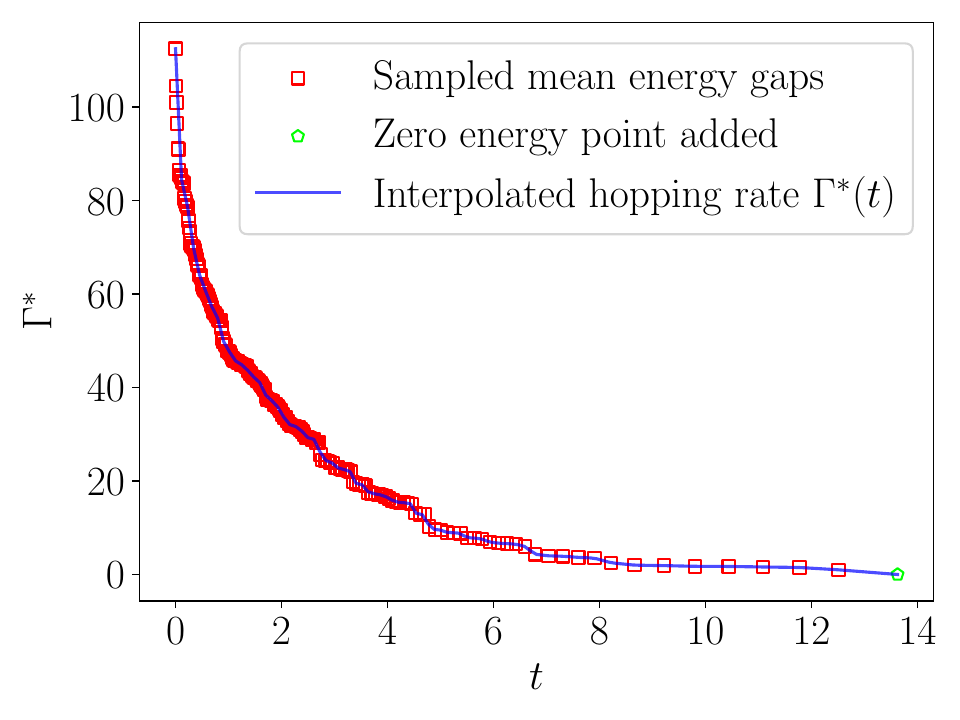}
        \caption{}
        \label{fig:interpolation_exact}
    \end{subfigure}

    \begin{subfigure}[b]{0.49\linewidth}
        \centering
        \includegraphics[width=\linewidth]{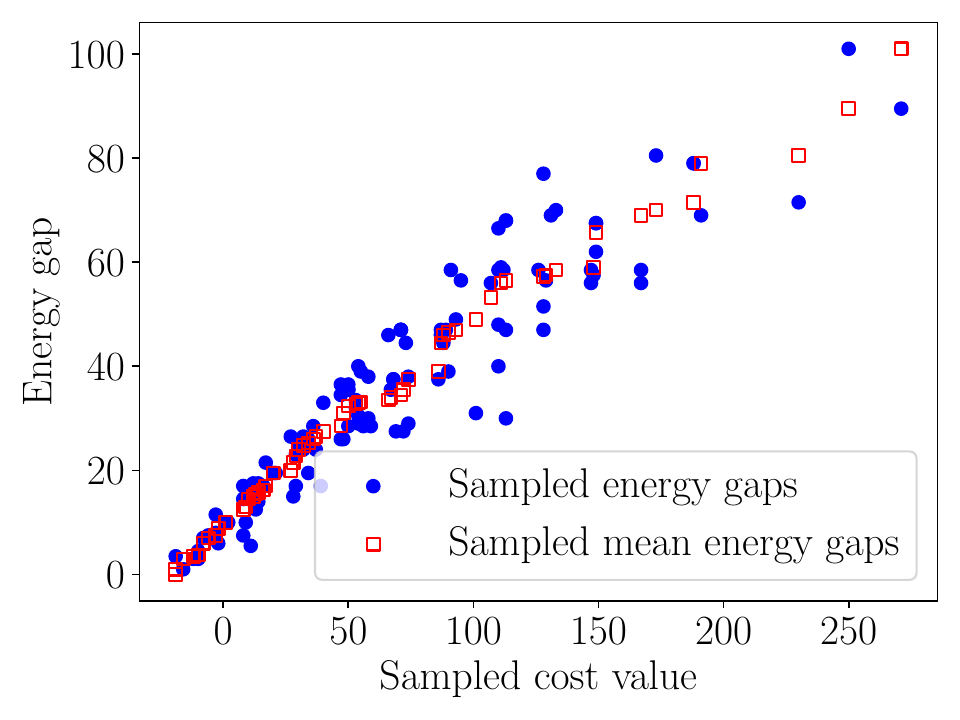}
        \caption{}
        \label{fig:sample_approx}
    \end{subfigure}
    \begin{subfigure}[b]{0.49\linewidth}
        \centering
        \includegraphics[width=\linewidth]{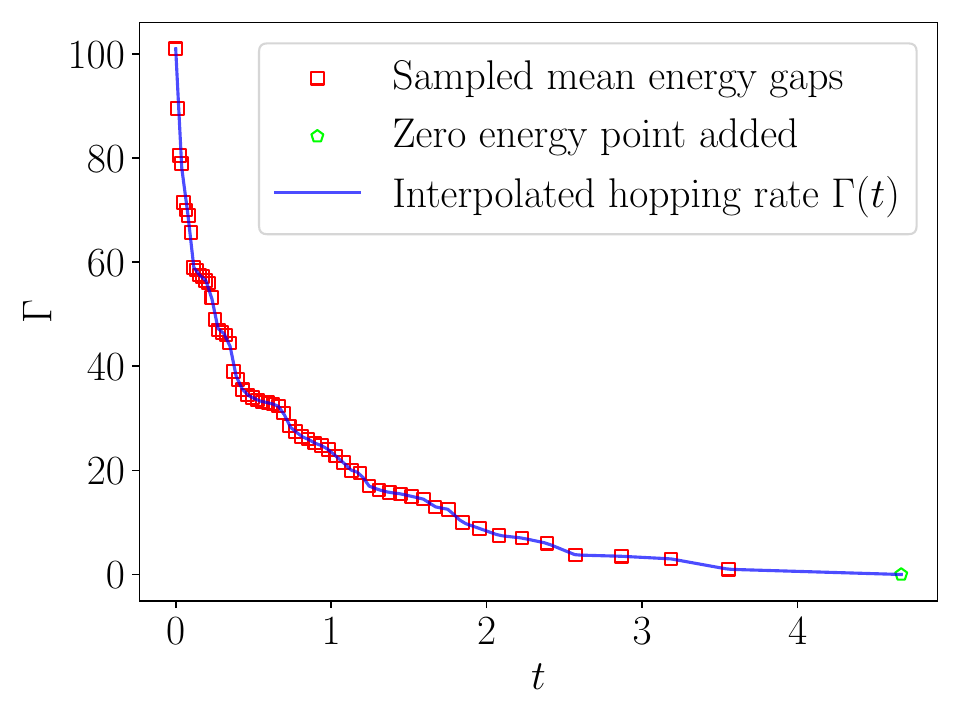}
        \caption{}
        \label{fig:interpolation_approx}
    \end{subfigure}

    \caption{Exact (top) and approximated (bottom) construction of the hopping rate $\Gamma$ for $n=11$ qubits (for UD-MIS instance). (a) Exact sampling with $q=2^n$ samples. (b) Linear interpolation of the hopping rate $\Gamma$. The optimal evolution time is given by Eq. \eqref{eq:time_opt}. The last interpolated point is the zero energy added in Alg. \ref{alg:sampling} to ensure $\Gamma(t=T)=0$. (c) Approximated sampling with $q=n^2$ samples. (d) Linear interpolation of the hopping rate $\Gamma$. The approximated evolution time is given by Eq. \eqref{eq:t_approx}. }
    \label{fig:samba_curves}
\end{figure}

We illustrate the exact and approximated sampling, and resulting interpolation of the hopping rate for $n=11$ qubits on Fig. \ref{fig:samba_curves}, and the complete algorithm in Algorithm~\ref{alg:builder}, labelled \textsc{Builder}. We note that the approximate and the exact $\Gamma(E)$ are close as a function of energy, but not necessarily as a function of time, since $\Gamma(t)$ is rescaled with maximal time dependent on the sum of all energies (Eq.~\ref{eq:t_approx}). The exact $\Gamma(t)$ decays therefore slower than the approximate one, which is a side advantage for us. Moreover, we present the hopping rate as a function of energy, instead of time, in Appendix \ref{app:hopping_rate}.

Once the hopping schedule is computed, then we can run the quantum walk and perform measurements. In the continuous-time regime, this implies running~\eqref{eq:rescaled_hamiltonian} for a time $T$. In a gate-based quantum computer, this would mean discretize in time the evolution and implement a circuit. 

\subsection{Circuit implementation}\label{subsec:circuit}

Having described our main continuous-time quantum walk, we now turn to the question of how to implement it on a gate-based quantum computer.

\begin{figure}
    \centering
    \includegraphics[width=1\linewidth]{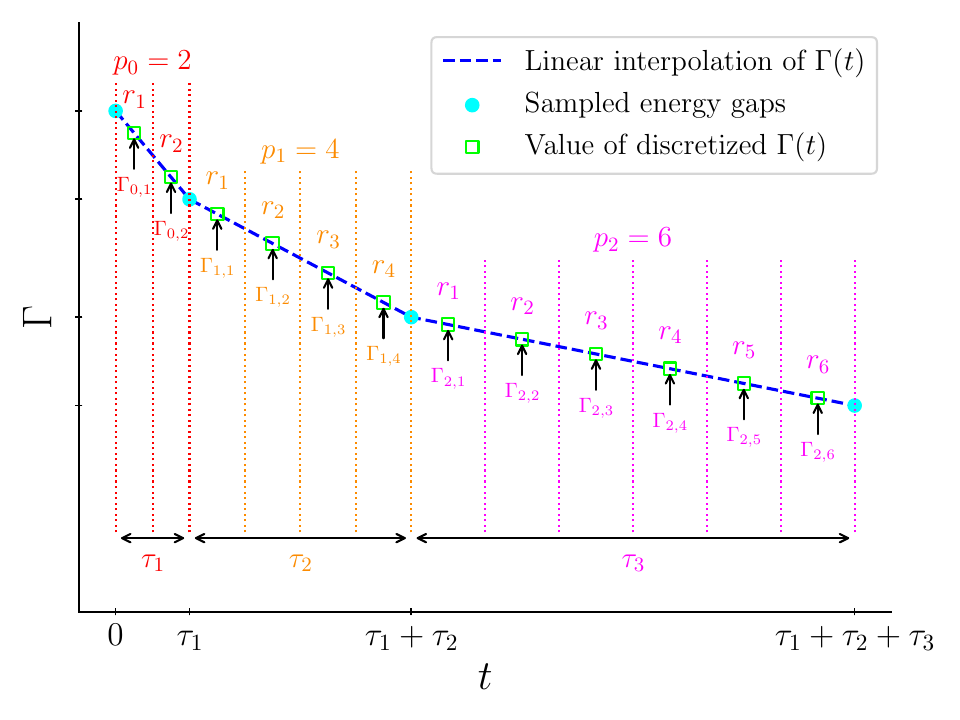}
    \caption{Discretization of the hopping rate $\Gamma$ with $p=(p_0=2,p_1=4,p_2=6)$. Each interval $T_l$ is cut into $p_l$ subintervals indexed $r_i$ with $i\in\{1,\dots,p_l\}$. The value of the discretized hopping rate $\Gamma_{l,r}$ corresponds to the average in each subinterval $r$. The higher the values in $p$, the more accurate the approximation of $\Gamma$. Each subinterval $r_i$ corresponds to the application of a single layer of cost and mixer Hamiltonians, thus the number of layers in the circuit is $\bar{p}=\sum_lp_l$.}
    \label{fig:hopping_discretization}
\end{figure}

Because our Hamiltonian is not a simple convex combination of simpler Hamiltonians due to the hopping rate, the discretization of the dynamics via trotterization is not trivial. The key to our approach (to maintain the circuit depth manageable and the accuracy reasonable) is to have two nested discretizations of the time interval. First we divide the time interval into periods $\tau_\ell$'s that represents the time in which the approximate hopping rate is linear. Then, we subdivide these periods into smaller time intervals $p_\ell$'s, on which we apply the trotterization. We can see a depiction of our method in Figure~\ref{fig:hopping_discretization}. This way of proceeding yields better accuracy than a global trotterization on the Hamiltonian (with a constant sampling period, often employed in QAOA) and it is consistent with the interpolation of the hopping rate.

With this in mind, the evolution we aim for is described by the time-dependent Schrödinger equation:
\begin{equation}\label{eq: schrodinger}
    i \hbar \hspace{0.7pt} \partial_t \ket{\psi(t)} = H(t) \ket{\psi(t)},
\end{equation}
with $\ket{\psi(t)}=U(t)\ket{\psi(0)}$ and we work in units in which $\hbar=1$. 

Since the Hamiltonian is time-dependent, the corresponding propagator \(U(t)\) is the time-ordered exponential 
\[
U(t) = \mathcal{T} \exp \left( -i \int_0^t H(s) \mathrm{d} s \right). 
\]
Here, \(\mathcal{T} \) denotes the \emph{time-ordering operator}, which acts on a product of time-dependent operators by reordering them chronologically, with later times place to the left. Formally, for two time-dependent operators \(A\) and \(B\), 
\[
\mathcal{T} \left[ A(t) B(s) \right] = \begin{cases}
A(t) B(s) & \text{if } t \geq s, \\
B(s) A(t) & \text{if } s > t. 
\end{cases}
\]
The time-ordered exponential is then understood as the limit of the ordered product
\[ \begin{aligned}
& \mathcal{T} \exp \left( -i \int_0^t H(s) \mathrm{d} s \right) \\ & \qquad\qquad\qquad = \lim_{p \to \infty} \prod_{\ell = p}^1 \exp \left( -i \frac{t}{p} H(t_\ell) \right), 
\end{aligned} \]
where \(t_\ell = \ell t/p\) and the product is ordered with \(\ell = p\) on the left, reflecting the fact that later-time factors act last on the state. When \(H(t)\) and \(H(s)\) commute for all \(t\) and \(s\), the time-ordering is useless and the expression reduces to the ordinary exponential \(\exp \left( -i \int_0^t H(s) \mathrm{d}s \right)\). 

To approximate this on a quantum circuit, we follow a two-level discretization strategy. Let \(\mathcal{E}\) be the set of all energy levels corresponding to the values that the hopping rate can take, ordered in a decreasing order. From this set, we define \(\mathcal{D}\) as the set of all durations \(\tau_\ell\) required for the optimal evolution when the hopping rate takes the value \(e_l \in \mathcal{E}\). Then, the elements of \(\mathcal{D}\) dissect the interval \(\left[0, t \right]\) with \(t = \sum_{\tau \in \mathcal{D}} \tau\) into subintervals
\begin{equation*}
	T_\ell = \left[\sum_{j=0}^{\ell-1}\tau_j, \tau_\ell + \sum_{j=0}^{\ell-1}\tau_j\right],
\end{equation*}
where \(\tau_0 = 0\). The two sets \(\mathcal{E}\) and \(\mathcal{D}\) have cardinality \(q\), the number of samples\footnote{In reality, as duplicates of sampled energies are removed, these sets can contain fewer than $q$ elements. Thus, we have $\lvert\mathcal{E}\rvert=\lvert\mathcal{T}\rvert\leq q$.}. At the coarse level, the intervals \(T_\ell\) are chosen so that \(\Gamma\) is well-approximated by a linear interpolation on each \(T_\ell\). At the fine level, each \(T_\ell\) is subdivided into \(p_\ell\) slices of duration \(\tau_\ell / p_\ell\), on each of which \(\Gamma\) is approximately constant. On such a slice, the Hamiltonian is therefore approximately time-independent, and the first-order Trotter-Suzuki formula~\cite{suzuki1985decomposition, PhysRevX.11.011020} yields a controlled approximation. 

For \(r \in [p_\ell]\), we denote by \(U^{(p,r)}(T_\ell)\) the unitary approximating
\[
U^{(p,r)}(T_\ell) \! \simeq U \! \left( \! \left[ \tau_{< \ell} + r \frac{\tau_\ell}{p_\ell}, \ \tau_{< \ell} + (r+1) \frac{\tau_\ell}{p_\ell} \right] \! \right)\!, 
\]
with \(\tau_{< \ell} \coloneqq \sum_{j < \ell} \tau_j\). On the subinterval
\[
\displaystyle\left[ \tau_{< \ell} + r \frac{\tau_\ell}{p_\ell}, \tau_{< \ell} + (r+1) \frac{\tau_\ell}{p_\ell} \right],
\]
the average value of the hopping rate is given by
\[ \textstyle \begin{aligned} 
\Gamma_{\ell, r} = \Gamma_\ell - \left( r + \frac{1}{2} \right) \left( \frac{\Gamma_\ell - \Gamma_{\ell+1}}{p_\ell} \right). 
\end{aligned} \]
Therefore, we choose the unitary \(U^{(p,r)}(T_\ell)\) to be
\[
U^{(p,r)}(T_\ell) = \mathrm{e}^{- i \frac{\tau_\ell}{p_\ell} \Gamma_{\ell\!, r} H_M } \mathrm{e}^{- i \frac{\tau_\ell}{p_\ell} H_C}.
\]
In this way, the larger all the \(p_\ell\) are, the more the evolution follows the linear interpolation of the values of the hopping rate we sampled earlier. We display the discretization process of the hopping rate in Fig.~\ref{fig:hopping_discretization}. The whole circuit can be expressed in the following form
\begin{equation}\label{eq:qaoa_like}
U(t) \simeq \prod_{\ell = 1}^q \left( \prod_{r = 1}^{p_\ell} U^{(p,r)}(T_\ell) \right). 
\end{equation}
To decompose this circuit in terms of primitive gates, we split each 
\[
U^{(p, r)}(T_\ell) = U^{(p, r)}_M(T_\ell) U_C^{(p)}(T_\ell)
\]
with
\[\left\{\begin{aligned}
U_M^{(p, r)}(T_\ell) & = \exp \left( -i \frac{\tau_\ell}{p_\ell} \Gamma_{\ell, r} H_M \right),\\
U_C^{(p)}(T_\ell) & = \exp \left( - i \frac{\tau_\ell}{p_\ell} H_C \right).
\end{aligned}\right.
\]

The quantum circuit implementation of \(U_C^{(p)}(T_\ell)\) is quite standard, and consists of CNOT and \(R_Z\) for QUBO~\cite{Lucas_2014} and HUBO~\cite{10313783} problems. As for the mixer \(U_M^{(p, r)}(T_\ell)\), its implementation is straightforward, even if it depends on the hopping rate and on the connectivity that we select. For a hypercube, it can be computed as:
\begin{equation}\label{eq:mixer_angle}
\begin{aligned}
U_M^{(p, r)}(T_\ell) = \bigotimes_{b = 0}^{n-1} R^{(b)}_X \!\left( -2 \frac{\tau_\ell}{p_\ell} \Gamma_{\ell\!,r} \right). 
\end{aligned}
\end{equation}
Note that for maximization problems, the minus sign in the arguments of the rotation gate must be discarded. We depict an example of the quantum circuit implementation in Figure~\ref{fig:circuit}.

One may notice that the unitary of Eq. \eqref{eq:qaoa_like} induces a QAOA-like evolution. However, unlike QAOA-like evolutions, we do not have to use a classical optimizer to fine-tune the parameters as we guide the walker to the optimal solutions \cite{PhysRevResearch.6.013312}.

The depth of the circuit is:
\begin{equation} \label{eq:depth} \begin{aligned} 
d(n, p) & = d_{SP}(n) + \sum_{l = 1}^q p_l \Big( d_M(n) + d_C(n) \Big) \\ 
& = d_{SP}(n) + \overline{p} \big( d_M(n) + d_C(n) \big), 
\end{aligned} \end{equation}
where \(\overline{p} = \sum_{\ell = 1}^q p_\ell\), \(d_{SP}(n)\) and \(d_M(n)\) are the depth of the implementation of the state preparation and the mixer unitary, and \(d_C(n)\) is the depth of a single layer application of the cost Hamiltonian, which can be reduced by adding ancillary qubits~\cite{10.1145/3718348}. In the case of \(X\)-mixer, we have \(d_{SP}(n) = d_M(n) = 1\). 

In practice, we take all the \(p_\ell\) equal, i.e., \(p_\ell = p_\star\) for all \(\ell \in [q]\) (see Appendix \ref{app:constant_discretisation}). Then, we have that \(\overline{p} = q p_\star\), hence
\begin{equation}\label{eq: depth-2}
d(n, p) = d_{SP}(n) + q p_\star \left( d_M(n) + d_C(n) \right). 
\end{equation}
Therefore, the depth of the circuit depends on the number of samples \(q\), the number of steps in the discretization, and the depth of the cost Hamiltonian, that is problem-specific. 

We present the general algorithm to construct the quantum circuit in Alg.~\ref{alg:circuit}. In Section~\ref{sec:problems}, we give the depth of the circuit according to the problem under study.


\begin{figure}
\hspace{-2pt}\begin{adjustbox}{width=\columnwidth}
    \begin{quantikz}
        & \gate{H}\gategroup[4,steps=1,style={dashed,rounded corners, inner xsep=2pt},background,label style={label
position=below,anchor=north,yshift=-0.2cm}]{{\sc State Preparation}} & \gate[4]{U_C\left( T_1 \right)}\gategroup[4,steps=1,style={dashed,rounded corners, inner xsep=2pt},background,label style={label
position=above,anchor=north,yshift=0.4cm}]{{\sc Cost}} & \gate{R_X(\theta_{1, 1})}\gategroup[4,steps=1,style={dashed,rounded corners, inner xsep=2pt},background,label style={label
position=below,anchor=north,yshift=-0.2cm}]{{\sc Mixer}} & \qw & \cdots\wireoverride{n} & \wireoverride{n} & \gate[4]{U_C\left( T_q \right)}\gategroup[4,steps=1,style={dashed,rounded corners, inner xsep=2pt},background,label style={label
position=above,anchor=north,yshift=0.4cm}]{{\sc Cost}} & \gate{R_X \left( \theta_{q, p_q} \right) }\gategroup[4,steps=1,style={dashed,rounded corners, inner xsep=2pt},background,label style={label
position=below,anchor=north,yshift=-0.2cm}]{{\sc Mixer}} & \meter{}\gategroup[4,steps=1,style={dashed,rounded corners, inner xsep=2pt},background,label style={label
position=above,anchor=north,yshift=0.4cm}]{{\sc Measurements}} \\ 
        & \gate{H} & & \gate{R_X(\theta_{1, 1})} & \qw & \cdots\wireoverride{n} & \wireoverride{n} & \qw & \gate{R_X \left( \theta_{q, p_q} \right)} & \meter{} \\ 
        \wireoverride{n} & \wireoverride{n}\vdots & \wireoverride{n} & \wireoverride{n}\vdots & \wireoverride{n} & \wireoverride{n} & \wireoverride{n} & \wireoverride{n} & \wireoverride{n}\vdots & \vdots\wireoverride{n} \\
        & \gate{H} & & \gate{R_X(\theta_{1, 1})} & \qw & \cdots\wireoverride{n} & \wireoverride{n} & \qw & \gate{R_X \left(\theta_{q, p_q} \right)} & \meter{} 
    \end{quantikz}
    \end{adjustbox}
    \caption{Quantum circuit implementation of the algorithm with $X$-mixer. The circuit is QAOA-like since it contains layers of cost $U_C$ and mixer $U_M$ unitaries. The angle of the mixer unitaries depends both on \(l\) and \(r\), while the cost unitaries depend only on \(l\). The value of the angle $\theta_{l,r}$ is given in Eq.~\eqref{eq:mixer_angle}. The circuit contains $\bar{p}=\sum_l p_l$ cost and mixer Hamiltonians layers.}
    \label{fig:circuit}
\end{figure}

\begin{algorithm}\footnotesize
\DontPrintSemicolon
\caption{Quantum circuit}\label{alg:circuit}
\KwIn{List of sampled energy gaps $\mathcal{E}$ sorted in decreasing order, list $p$ of layer numbers}
\KwOut{A quantum circuit implementation of the continuous-time quantum walk}
\BlankLine
Add state preparation of initial state (depends on $H_M$)\;
\For{$l \in \{0, \ldots, \lvert E_{\mathcal S} \rvert - 1\}$}{
  \For{$r \in \{1, \ldots, p_\ell \}$}{
    $\Gamma_\ell \gets E_{\mathcal S}[\ell]$\;
    $\tau_\ell \gets \Gamma_\ell^{-1} \cdot \frac{\pi}{2\sqrt{2}}$\;
    $\Gamma_{\ell, r} \gets \Gamma_\ell - \left(r + \frac{1}{2} \right) \cdot \frac{\Gamma_\ell - \Gamma_{\ell + 1}}{p_\ell}$\;
    Add $\exp(-i \Delta_t^{(\ell)} H_C)$ for $\Delta_t^{(\ell)} = \tau_\ell / p_\ell$\;
    Add $\exp(-i \Delta_t^{(\ell, r)} H_M)$ for $\Delta_t^{(\ell, r)} = \Gamma_{\ell, r} \cdot \tau_\ell / p_\ell$\;
  }
}
Add measurements in the computational basis\;
\end{algorithm}

\section{Set of problems}\label{sec:problems}

In this section, we give a brief description of the combinatorial optimization problems on which we test our algorithm. While some problems are well studied, some technicalities for the less studied ones need to be discussed before we present the simulations.

\subsection{Quadratic problems}

The first set of problems we study are quadratic unconstrained binary optimization problems, or QUBOs. These are combinatorial optimization problems for which the general problem form~\eqref{prob.combo} has quadratic cost,
$$
C(x) = x^{\top} Qx = \sum_iQ_{ii}x_i + \sum_{i<j}Q_{ij}x_ix_j+c,
$$
for a $Q\in \mathbb{R}^{n\times n}$, and $\mathcal{F}$ is the whole space. For one of the selected problems, we will also add a linear constraint.

It is well-known that finding the optimal solution of a QUBO is equivalent to finding the ground state of its Ising formulation, which can be in turn well implemented into a quantum computer.

As instances of QUBO problems we study the following ones.

\subsubsection{MaxCut}

The weighted Maximum Cut (MaxCut) problem consists of partitioning the vertices of a graph into two disjoint sets so that the number (or weight) of edges between the sets is maximized. This implies defining a graph among the $n$ variables and selecting the cost function,
\begin{equation}\label{eq:maxcut}
    C(x)=-\sum_{i<j}w_{ij}(x_i+x_j-2x_ix_j),
\end{equation}
where $w_{ij}$ is the weight of edge $(i,j)$ connecting the variables $i$ and $j$.



\subsubsection{Maximum Independent Set}

For a graph $\mathcal{G} = (\mathcal{V}, \mathcal{E})$, the Maximum Independent Set (MIS) problem consists of finding the independent subset of vertices that maximizes the weight. A subset $S\subseteq \mathcal{V}$ is said to be independent if no two vertices of $S$ are adjacent. Each vertex $i$ is associated to a binary variable $x_i\in \{0,1\}$, thus we have $n=|\mathcal{V}|$. Moreover, we set $x_i=1$ if $i\in S$. Therefore, cost function of MIS is:
\begin{equation}
    C(x)=-\sum_{i\in \mathcal{V}} w_ix_i+\lambda\sum_{(i,j)\in \mathcal{E}}x_ix_j,
\end{equation}
where $w_i$ is the weight of vertex $i$ and $\lambda>0$ is the penalty coefficient for including adjacent vertices in $S$. To ensure that the final set is independent, we define the penalty coefficient such that:
\begin{equation}
    \lambda = \max_{(i,j)\in \mathcal{E}}(w_i+w_j)+1.
\end{equation}

The study of the MIS problem on quantum computers is of growing interest and particularly on Unit-Disk (UD) graphs \cite{CLARK1990165}, as they are native to neutral-atom technologies that exploit the Rydberg blockade mechanism \cite{doi:10.1126/science.abo6587,dalyac2024graph,dalyac2021qualifying,PhysRevA.111.032611}. In this context, the problem is often referred to as UD-MIS, moreover, its hardness was first assessed \cite{PhysRevResearch.5.043277} and then shown to strongly depend on the graph density, tree-width and thickness \cite{8gjv-ll1f}. Interestingly, recent work introduces an imaginary-time evolution scheme for solving UD-MIS \cite{PhysRevA.111.042403}. 

\begin{figure}
	\centering
	\includegraphics[width=0.6\linewidth]{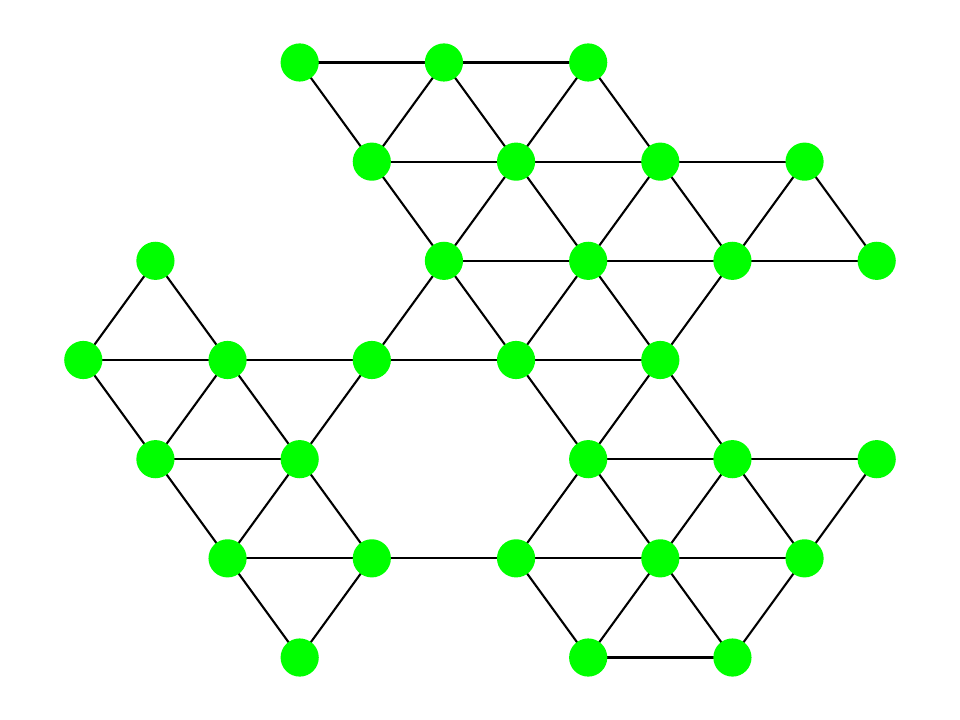}
	\caption{Unit-Disk (UD) graph instance of $n=30$ vertices.}
	\label{fig:ud_graph}
\end{figure}

\subsubsection{Portfolio optimization: constraints}

Portfolio optimization consists of selecting at most $k$ assets among $n$ to obtain the optimal portfolio. Each asset has an expected return and one aims at maximizing profit while minimizing financial risk. Each asset $i$ is associated to a binary variable $x_i\in\{0,1\}$ and $x_i=1$ if asset $i$ is in the portfolio. The limitation in the number of assets corresponds to bounding the Hamming weight of the solutions with a fixed value $k\in[0,n]$, i.e. choosing a constraint set as,
$$
\mathcal{F} = \left\{x \in \{0,1\}^n \hspace{2pt} \middle| \hspace{2pt} \sum_{i=1}^n x_i=k \right\}.
$$

The cost function reads:
\begin{equation}\label{eq:portfolio}
    C(x)=\lambda\sum_{i<j}\sigma_{ij}x_ix_j-\sum_{i}\mu_ix_i,
\end{equation}
with $\sigma\in \mathbb{R}^{n\times n}$ the covariance matrix between assets, $\mu\in \mathbb{R}^n$ is the expected return of the assets and $\lambda>0$ the risk appetite.

One way of encoding constraints is to use specific mixer Hamiltonian that can allow transitions only between feasible solutions. The Hamming weight constraint can be encoded with $XY$-mixers \cite{hadfield2019quantum,PhysRevA.101.012320}, which only induce transitions between states with same Hamming weight. Moreover, the connectivity of $XY$-mixers, i.e. connectivity between states of same Hamming weight, is given by an adjacency matrix $A_{XY}$. Therefore, such mixers Hamiltonian read:
\begin{equation}
    H_{XY} = -\frac{1}{2}\sum_{(i,j)\in A_{XY}}\sigma^x_i\sigma^x_j+\sigma^y_i\sigma^y_j.
\end{equation}
Moreover, an optimal quantum circuit implementation of such permutation mixers is proposed in Ref. \cite{LaplaceMermoud2026variationalquantum}.

Given two qubits $a$ and $b$ such that $\ket{a}\otimes\ket{b}\in \{\ket{i},\ket{j},\ket{k},\ket{l}\}$, $XY$-mixers act locally as:
\begin{equation}
    \begin{split}
        H_{XY}^{(i,j,k,l)} &= -\frac{1}{2}\left((\sigma^x_a\otimes\sigma^x_b)+(\sigma^y_a\otimes\sigma^y_b)\right) \\
        &= 
    \begin{pmatrix}
    0 & 0 & 0 & 0 \\
    0 & 0 & -1 & 0 \\
    0 & -1 & 0 & 0 \\
    0 & 0 & 0 & 0
    \end{pmatrix},
    \end{split}\label{eq:xy_mixer_local}
\end{equation}
we see that it acts as a negative Pauli-$X$ gate on basis states $\ket{j}$ and $\ket{k}$. In this context, the local cost Hamiltonian is:
\begin{equation}
    H_C^{(i,j,k,l)} = \begin{pmatrix}
        0 & 0 & 0 & 0 \\
        0 & \bra{j}H_C\ket{j} & 0 & 0 \\
        0 & 0 & \bra{k}H_C\ket{k} & 0 \\
        0 & 0 & 0 & 0 \\
    \end{pmatrix}.
\end{equation}
Since transfer between higher energy state $j$ to lower energy state $k$ is maximized when $|\Gamma^*\Delta^M_{jk}|\approx \Delta^C_ {jk}$, the general form of the hopping rate reads:
\begin{equation}
    \Gamma^*(E_{jk})=\frac{\Delta^C_{jk}}{|\Delta^M_{jk}|}.
\end{equation}
Looking at Eq. \eqref{eq:xy_mixer_local}, we can see that the local energy gap of Hamiltonian $H_{XY}^{(j,k)}$ is $\lvert\Delta^{XY}_{jk}\rvert=2$, coincidentally the same as for the $X$-mixer. Therefore, the hopping rate for $XY$-mixer is the same as the one in Eq. \eqref{eq:hopping}.

Furthermore, the initial state cannot be the uniform superposition over the computational basis states as some of them correspond to infeasible solutions. The correct initial state is the uniform superposition over the basis states of Hamming weight smaller or equal to the bound $h$. Previously in the sampling protocol, after selecting a random state, we computed its energy gap with its $n$ neighbors as the underlying graph of the $X$-mixer is the $n$-dimensional hypercube. In the case of $XY$-mixers, the number of neighbors depends on the connectivity of the mixer, which is defined by $A_{XY}$. Thus, the complexity of the sampling protocol depends on the connectivity of the mixer.

In our experiments we use $XY$-mixers with ring connectivity. The underlying $XY$-mixer graph is a disconnected graph that contains $n+1$ subgraphs, each composed of ${n \choose k}$ vertices of same Hamming weight $k$. Ring connectivity implies that two states are connected if they only differ by a single permutation (swap) of two adjacent\footnote{Note that we have boundary conditions such that the last bit is adjacent to the first one since it is a ring connectivity.} bits with different values. Therefore, the maximum number of neighbors is $n$, and it is achieved for bitstrings in which every pair of adjacent bits differs. Thus, the asymptotic complexity of the sampling protocol for $XY$-mixers with ring connectivity is the same as for $X$-mixer, i.e.
$$
\mathsf{C}_{\text{sampling}}^{XY_{\text{ring}}}=\mathsf{C}_{\text{sampling}}^{X}=\mathcal{O}(\text{poly}(n)).
$$

\begin{figure}
    \centering
    \includegraphics[width=0.9\linewidth]{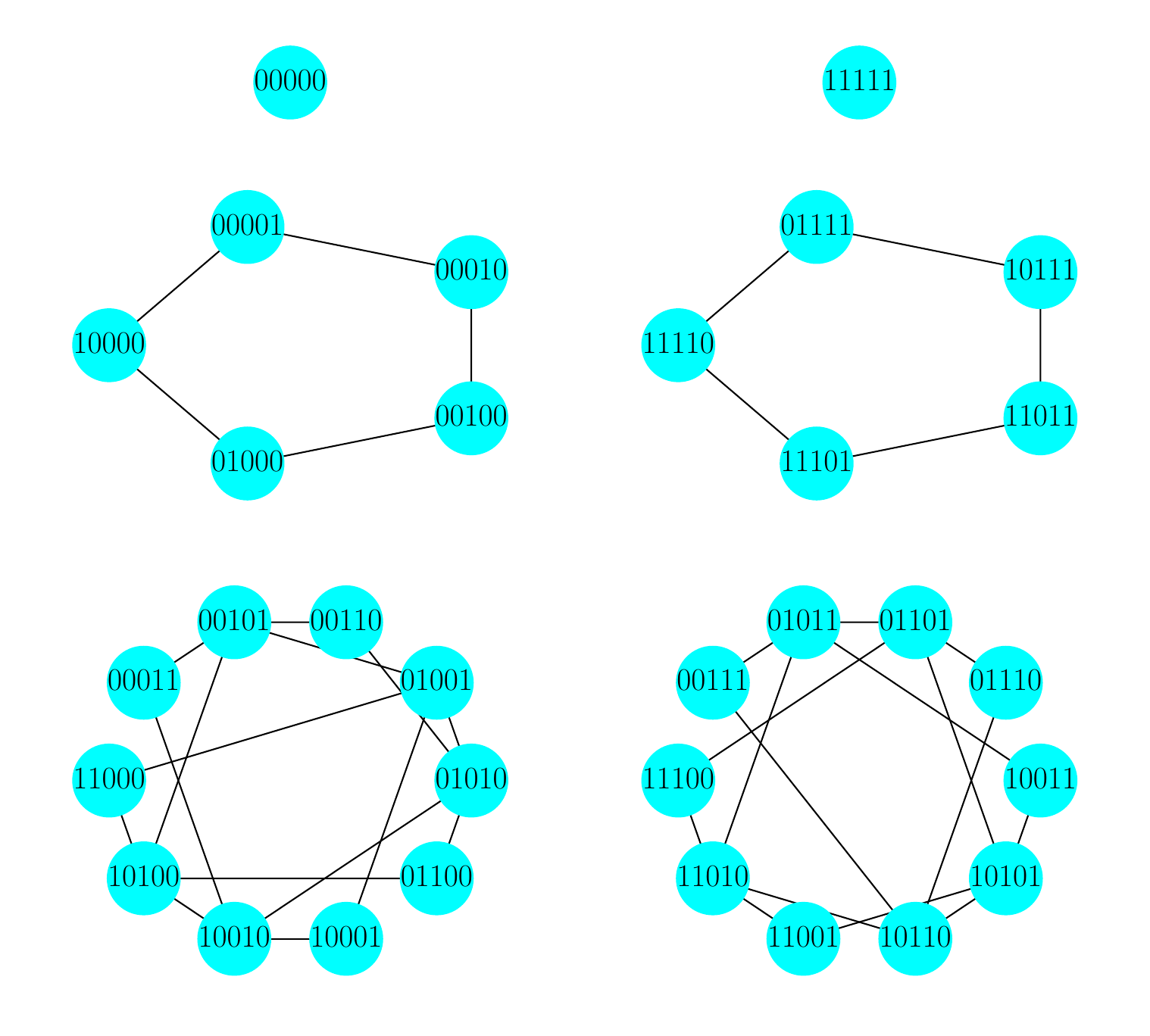}
    \caption{Underlying $XY$-mixer graph with ring connectivity on $n=5$ qubits. The graph is disconnected and contains $n+1=6$ subgraphs of different Hamming weights $k\in \{0,1,2,3,4,5\}$. Subgraph of Hamming weight $k$ contains ${n \choose k}$ vertices. Two vertices $x$ and $y$ are connected if they only differ by a swap of two adjacent different bits, i.e. if $y=\text{swap}_{i,(i+1)\mod n}(x)$ and $x_i\neq x_{(i+1)\mod n}$.}
    \label{fig:xy_mixer_ring}
\end{figure}

As an example, we show the underlying graph of $XY$-mixer with ring connectivity on $n=5$ qubits in Fig. \ref{fig:xy_mixer_ring}. The set of neighbors of bitstring $x$ can be formally expressed as $\mathrm{B}(x)=\{y\in \{0,1\}^n \mid y_i\neq y_{(i+1)\mod n},y=\text{swap}_{i,(i+1)\mod n}(x)\}$, where $\text{swap}_{i,j}(x)$ is the bitstring obtained by exchanging the values of bits $i$ and $j$ in $x$. Additionally, the initial state $\ket{\psi_k}$ is the $k$-Dicke state \cite{10.1007/978-3-030-25027-0_9}, i.e. the uniform superposition over basis states with Hamming weight equal to $k$:
\begin{equation}
    \ket{\psi_k} = \frac{1}{\sqrt{\lvert F_k\rvert}}\sum_{x\in F_k}\ket{x},
\end{equation}
where $F_k=\{x\in\{0,1\}^n \mid \sum_{i=1}^nx_i= k\}$ is the set of bitstring satisfying the $k$-Hamming weight constraint, i.e. the set of feasible solutions.

\subsection{Polynomial problems}

Higher-order unconstrained binary optimization (HUBO) problems\footnote{Also referred to as polynomial unconstrained binary optimization (PUBO) or higher-order unconstrained binary optimization (HUBO) problems.} are a generalization of QUBO as they describe arbitrary polynomial problems. They are special cases of Problem~\eqref{prob.combo}, when the cost is selected as a polynomial,
\begin{equation}\label{eq:HUBO}
    C(x)=\sum_{m \in \pi} \left(\alpha_m \prod_{i\in m} x_i\right),
\end{equation}
where $\pi$ is the polynomial set of monomials of at most $n$ variables and $\alpha_m\in \mathbb{R}$ is the coefficient of monomial $m$. Here, the set $\mathcal{F}$ is selected to be the whole space.

\subsubsection{LABS}

The low autocorrelation binary sequences (LABS) is a quartic combinatorial problem \cite{Packebusch_2016}. This problem is known to be intractable classically as its difficulty increases exponentially with the size $n$ of the sequence. Moreover, it has been shown that QAOA has the potential to outperform classical approaches to solve this problem \cite{doi:10.1126/sciadv.adm6761}. Given a sequence $s=(s_1,\cdots, s_n)$ with $s_i=\pm 1$, one aims at minimizing its energy:
\begin{equation}
    E(s)=\sum_{k=1}^{n-1}\left(\sum_{i=1}^{n-k}s_is_{i+k}\right)^2.
\end{equation}

Equivalently, the problem can be reduced to maximizing the merit factor $F(s)=n^2/2E(s)$. A sequence $s$ of Ising variables can be mapped to Boolean variables $x=(x_1,\cdots, x_n)$ with the relation $s_i=1-2x_i$. Thus, the cost function reads:
\begin{equation}
    C(x)=\sum_{k=1}^{n-1}\left(\sum_{i=1}^{n-k}4x_ix_{i+k}-2(x_i+x_{i+k})+1\right)^2.
\end{equation}

\subsubsection{MAX-$k$-SAT}

The boolean satisfiability problem (SAT) is a famous problem in theoretical computer science as it was the first problem proven to be NP-complete \cite{10.1145/800157.805047}. Several works try to solve this problem with variants of QAOA \cite{hadfield2019quantum,10313612,Boulebnane2024,PhysRevResearch.5.023147}. Given $m$ clauses $\varphi_{i}$ of $k$ literals each, one has to decide if there exists an assignment of $n$ variables that satisfies all the clauses. The combinatorial problem of finding such an assignment is called MAX-$k$-SAT. Formally, a boolean formula in conjunctive normal form of $m$ clauses reads:
\begin{equation}
    \phi=\varphi_1\land \varphi_2\land \cdots \land \varphi_m.
\end{equation}
Each clause $\varphi_{j}$ is a disjunction of $k$ literals $l_{jk}$:
\begin{equation}
    \varphi_j = (l_{j1} \lor l_{j2} \lor \cdots \lor l_{jk}),
\end{equation}
with $l_{ji} \in \{x_1,\bar{x}_1, x_2,\bar{x}_2, \cdots, x_n,\bar{x}_n\}$ where $\bar{x}_j=1-x_j$. The MAX-$k$-SAT consists of finding an assignment $x=x_1x_2\cdots x_n$ that maximizes the number of satisfied clauses. A clause is said to be satisfied if it contains at least one literal equal to 1. This criterion can be expressed as:
\begin{equation}
    S_j(x)=1-\prod_{i=1}^k(1-l_{ji}).
\end{equation}
Therefore, a clause $\varphi_j$ is satisfied with assignment $x$ if $S_j(x)=1$, and unsatisfied if $S_j(x)=0$. One aims at maximizing the number of satisfied clause, thus, the cost function to minimize is:
\begin{equation}
    C(x)=-\sum_{j=1}^m S_j(x).
\end{equation}
This problem is a polynomial of degree $k$.

\subsubsection{Travelling salesperson problem}

\begin{figure}
    \centering
    \includegraphics[width=0.6\linewidth]{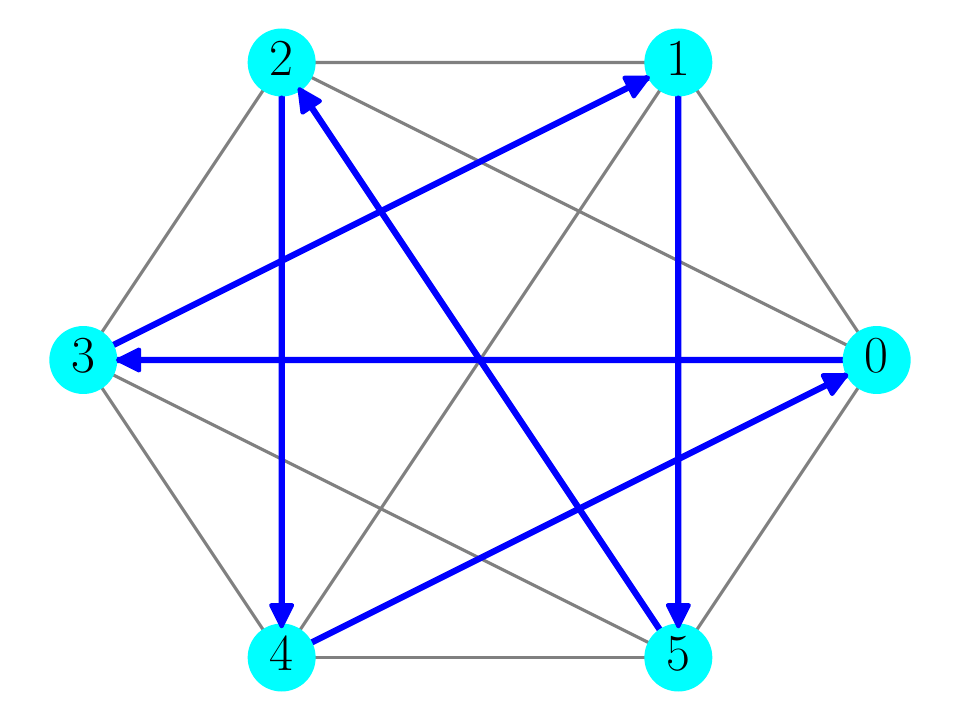}
    \caption{HUBO encoding of TSP with $n=6$ cities labeled with integers from 0 to $n-1$ on a complete graph. The encoding uses $n\lceil \log_2(n)\rceil=18$ qubits, and an assignment is of the form $x=x_0x_1x_2x_3x_4x_5$. The blue route describes the visit of the cities in order: 0, 3, 1, 5, 2, 4. Its corresponding encoding is $x=000011001101010100$.}
    \label{fig:tsp_encoding}
\end{figure}

The travelling salesperson problem (TSP) is an NP-complete decision problem with many applications \cite{Matai10}. Given a list of $n$ cities with specific distance between each of them, is there a route with total distance $k$ that visits all of cities exactly once and returns to the starting point. The related optimization problem consists of finding the shortest route that visits every city exactly once. Formally, the problem is represented with a graph whose vertices and edges respectively correspond to the cities and path between them. The distance between each pair of cities is given by the weight of their connecting edge. Its original QUBO encoding uses $n^2$ qubits \cite{Lucas_2014}. However, one can obtain a more efficient encoding with $n\lceil\log_2(n)\rceil$ qubits at the cost of expressing TSP as a HUBO \cite{glos2022space}. In this higher-order encoding, each city is encoded in binary with $\lceil\log_2(n)\rceil$ bits. Thus, one needs $n$ of these sequences to list all cities. Therefore, each assignment $x$ has two labels, $x_i$ is the $i$-th bitstring sequence and $x_{i,j}$ is the $j$-th bit of sequence $x_i$. We provide an example of this encoding in Fig. \ref{fig:tsp_encoding}. The cost function of the HUBO encoding of TSP reads \cite{glos2022space}:
\begin{multline}\label{eq:tsp}
        C(x)=\lambda\sum_{i=0}^{n-1}C_{1}(x_i)+\gamma\sum_{i=0}^{n-1}\sum_{j=i+1}^{n-1} C_{2}(x_i,x_j) \\ +\mu\sum_{i,j=0, i\neq j}^{n-1}w_{ij}\sum_{k=0}^{n-1}C_{2}(x_k,i)C_{2}(x_{k+1},j),
\end{multline}
where $w_{ij}$ is the weight of edge $(i,j)$ and $\lambda,\gamma>\mu \max_{(i,j)}w_{ij}$. The subcost $C_1$ puts a penalty if a sequence encodes an invalid city, i.e. if its input if greater than $n-1$. Let $n-1=\tilde{x}_{\lceil\log_2(n)\rceil-1}\cdots \tilde{x}_0=\tilde{x}$ and $S$ be the set of indices of $\tilde{x}$ equal to zero, i.e. $S=\{j \mid\tilde{x}_j=0\}$. Hence, the first penalty reads \cite{glos2022space}:
\begin{equation}
    C_1(x_i)=\sum_{j\in S}x_{i,j}\prod_{k=j+1}^{\lceil\log_2(n)\rceil-1}\left(1-(x_{i,k}-\tilde{x}_k)^2\right).
\end{equation}
The second subcost $C_2$ is used to penalize if two sequences $x$ and $y$ encode the same city \cite{glos2022space}:
\begin{equation}
    C_2(x,y)=\prod_{k=0}^{\lceil\log_2(n)\rceil-1}\left(1-(x_k-y_k)^2\right).
\end{equation}

\section{Numerical results for \textsc{SamBa--GQW}}\label{sec:results}

\begin{figure*}
    \centering
	\includegraphics[scale=0.5]{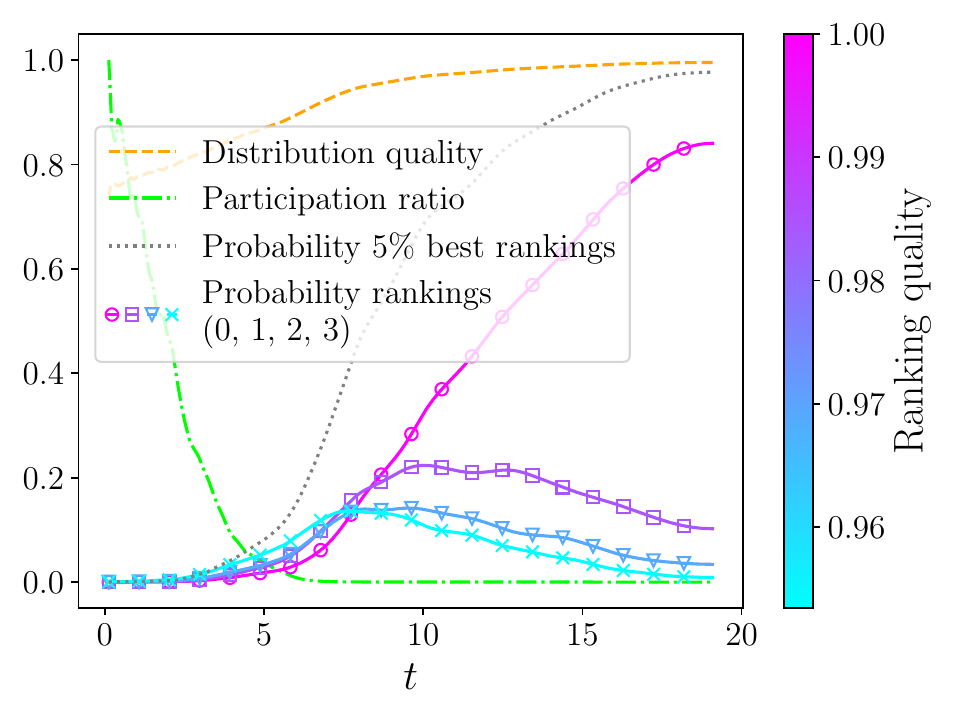}
	\includegraphics[scale=0.5]{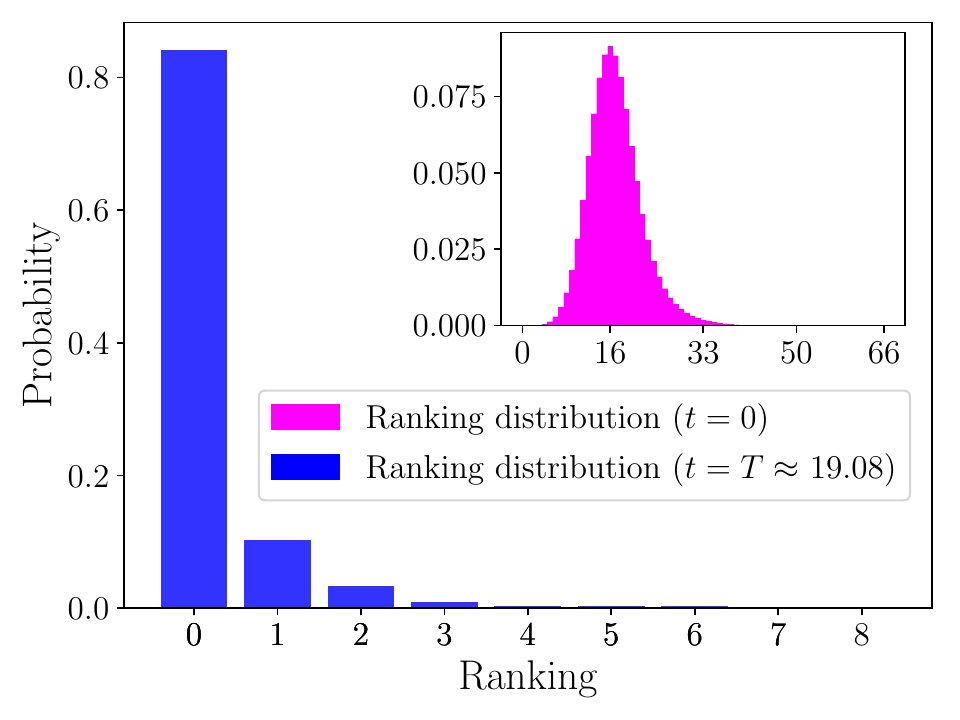}
    \caption{\textsc{SamBa--GQW} Hamiltonian simulations for unweighted MaxCut on random Erdős–Rényi graphs for $n=20$ qubits with a sampling of $q=n^2$ states.}
    \label{fig:unweighted_maxcut}
\end{figure*}

\subsection{Performance metrics}

Throughout, we use several metrics for the evaluation of performance. 

First, the quality of a produced quantum state $\ket{\psi}$ is evaluated with the expectation of its probability distribution as,
\begin{equation}\label{eq:quality}
    \mathbb{E}[\psi] = \sum_{x\in\{0,1\}^n} \mathbf{p}(x)\cdot \mathbf{q}(x),
\end{equation}
where $\mathbf{p}(x)= \lvert \braket{x|\psi} \rvert^2$ is the probability of measuring basis state $\ket{x}$. The function $\mathbf{q}(x)\in [0,1]$\footnote{Since the optimal solution is that minimizing the cost function, we define the quality function as the approximation ratio $\mathbf{q}(x)=\frac{C_{\max}-C(x)}{C_{\max}-C_{\min}}$.} outputs the quality of a decision $x$ such that the optimal and worst decision respectively lead to $\mathbf{q}(x^*)=1$ and $\mathbf{q}(\argmax(C))=0$. In this context, Eq. \eqref{eq:quality} gives information about the distance between $\ket{\psi}$ and $\sum_{x^*}\alpha_{*}\ket{x^*}$ with $\sum_{x^*}\lvert \alpha_{*}\rvert^2=1$. Note that, if constraints are present, we put $\mathbf{q}(x)=0$ for all unfeasible decisions ($\forall x\notin \mathcal{F}$). 

We use then the normalized participation ratio (or PR for short) to quantify the localization of a state vector:
    \begin{equation}
        \text{PR}(\psi)=\left(2^n\sum_{x\in\{0,1\}^n}\mathbf{p}(x)^2\right)^{-1}.
    \end{equation}
    The uniform superposition leads to $\text{PR}=1$ and the fully localized state gives $\text{PR}=1/2^n$. This metric gives information on the sparsity of $\ket{\psi}$.
    
Finally, we label the quality level of a decision $x$ with its ``ranking'' $\mathbf{r}(x)\in\mathbb{N}$. Optimal solutions have a ranking of 0, second-best decisions have a ranking of 1, and so on. Let $S_r$ be the set of decisions with ranking $r$, i.e. $S_r=\{x\in \{0,1\}^n | \mathbf{r}(x)=r\}$. The total number of rankings is equal to the number of distinct cost values, which corresponds to the dimension (cardinality) of the image of the cost function $C : \{0,1\}^n\rightarrow \mathbb{R}$. The probability of measuring a decision of ranking $r$ reads:
    \begin{equation}
        \mathbb{P}_r[\psi]=\sum_{x\in S_r} \lvert \braket{x|\psi} \rvert^2.
    \end{equation}

\subsection{Results}\label{subsec:results}

Hamiltonian simulations are performed for all considered problems, and quantum circuit simulations are additionally carried out for MaxCut, MIS, and LABS. Moreover, for all problems whose instance generation is parameterized (all except LABS) we generate $10$ different instances each time, and we present the average of the results obtained. Across all Hamiltonian simulations, we construct the hopping rate $\Gamma$ by sampling only $q=n^2$ states among $2^n$.

In general, all the simulations consistently support \textsc{Samba--GQW} as a method that in a short evolution time can shift an initial state distribution into a very localized distribution around the optimal solution. The most probable ranking is often the optimal solution, or ranking $\ll 10$. Even then, the optimal solution is within the $5\%$ best rankings and a classical post-processing can easily extract it. In this context, \textsc{Samba--GQW} requires little tuning, is consistent, and scales reasonably well with the number of decisions.

\subsubsection{MaxCut}

\begin{figure*}
    \centering
	\includegraphics[scale=0.5]{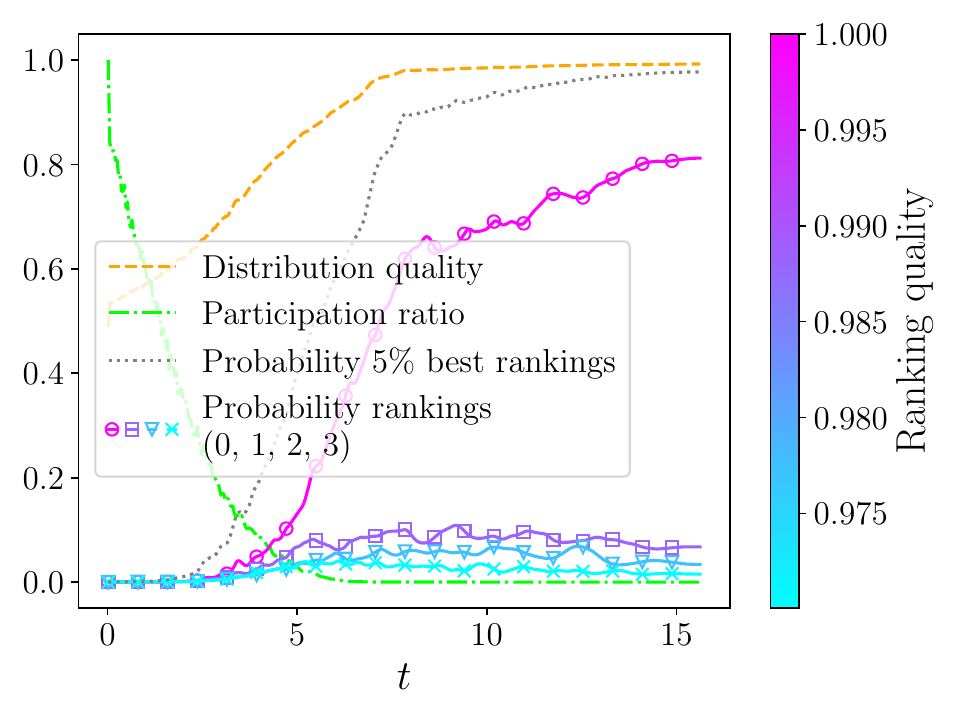}
	\includegraphics[scale=0.5]{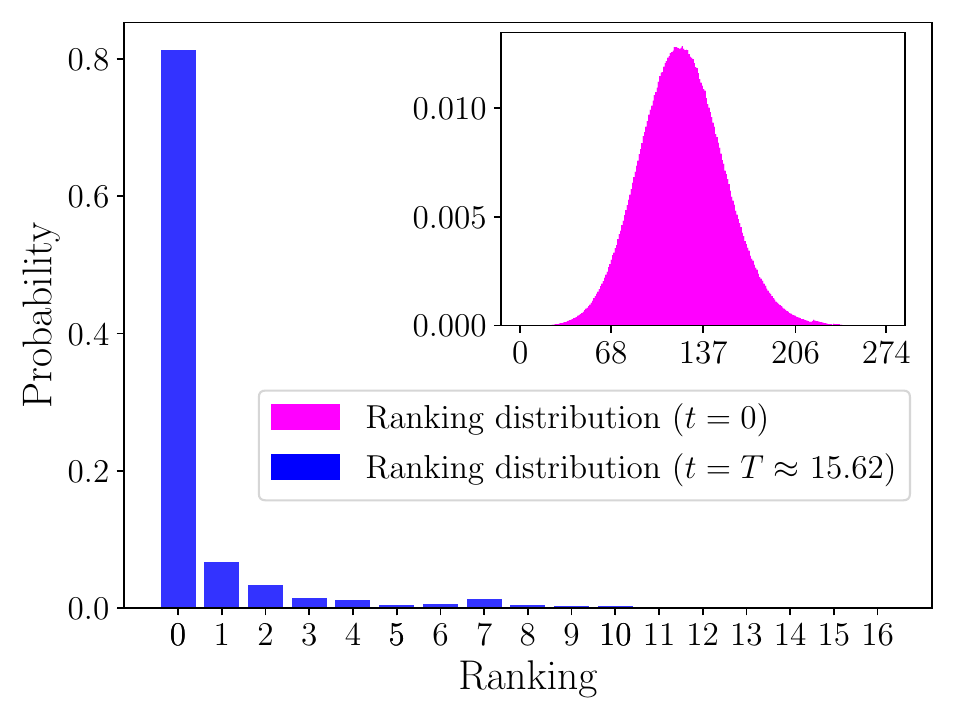}
    \caption{\textsc{SamBa--GQW} Hamiltonian simulations for weighted MaxCut on random Erdős–Rényi graphs for $n=20$ qubits with a sampling of $q=n^2$ states.}
    \label{fig:weighted_maxcut}
\end{figure*}

We generate instances on Erdős–Rényi graphs~\cite{Erdos1960} of $n=20$ vertices, both weighted ($w_{ij}\in [-10,10]$) and unweighted ($w_{ij}=1$), with an edge appearance probability of~\(\frac{1}{2}\). 

\paragraph{Hamiltonian simulations.}
We present the distribution quality, participation ratio, measurement probabilities per ranking over time and the probability distributions over rankings at $t=0$ and $t=T$ on Fig.~ \ref{fig:unweighted_maxcut} (resp. \ref{fig:weighted_maxcut}) for unweighted (resp. weighted) MaxCut. At $t=0$ we show the complete ranking distribution, however,  at $t=T$ we only display rankings with measurement probability higher than $10^{-3}$.

We obtain a high final success probability of measuring the optimal solution, i.e. ranking 0, since it is of order $\mathbb{P}[\psi_T]\approx 0.8$. In both unweighted and weighted scenario, the probability of measuring decisions with ranking among the 5\% best is almost 1. Moreover, at $t=0$ the quantum state is a uniform superposition among the $2^n$ computational basis states resulting in a participation ratio of $\text{PR}(\psi_0)=1$. We observe that this ratio decreases rapidly to converge to $\text{PR}(\psi_T)\approx1/2^n$, resulting in a more localized state. Lastly, we notice that the initial distribution quality is $\mathbb{E}[\psi_0]\approx 0.75$ for unweighted MaxCut and $\mathbb{E}[\psi_0]\approx 0.5$ for weighted MaxCut. This difference means that the unweighted case contains more high quality solutions than weighted MaxCut. We notice that the time evolution is fairly short as it reaches $T\approx 20$ (resp. $T\approx 16$) for unweighted (resp. weighted) MaxCut, this time evolution difference is caused by lower energy gaps in weighted instances.  In both cases, we observe an initial Gaussian ranking distribution, although it is smoother for weighted MaxCut since this problem contains $275$ different rankings compared with $67$ for unweighted MaxCut. Furthermore, as indicated by the distribution quality, the initial distribution for the unweighted case is shifted towards higher quality decisions, while the weighted Gaussian is centered around the middle rankings. After the evolution induced by \textsc{SamBa--GQW}, the distribution is almost fully localized on the optimal solutions for both cases.

\begin{figure*}
    \centering
        \includegraphics[width=0.8\linewidth]{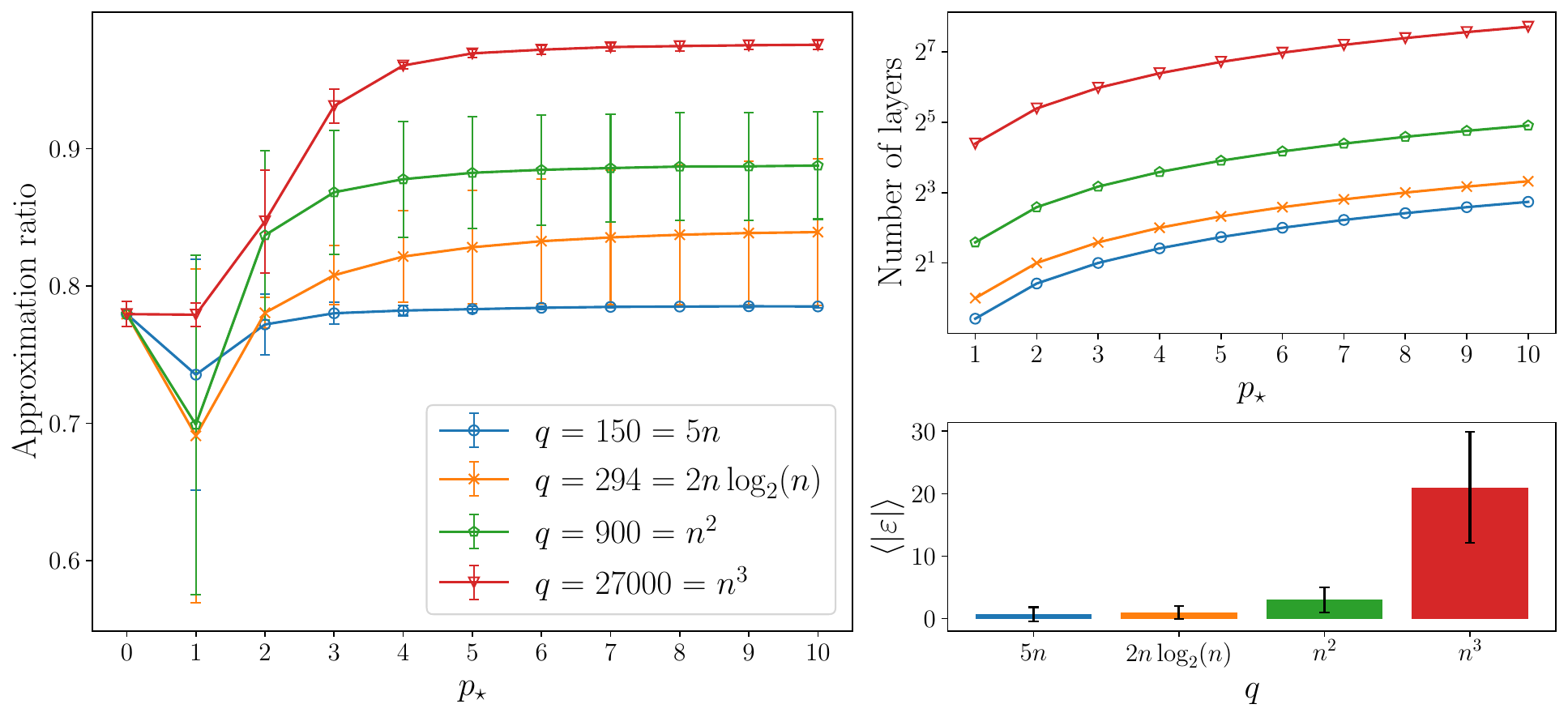}
        \caption{\textsc{SamBa--GQW} quantum circuit simulations for unweighted MaxCut for $n=30$ qubits with different sampling values $q\in\{5n,2n\log_2(n),n^2,n^3\}$. The approximation ratio is shown as a function of the discretization parameter $p_\star\in[0,10]$ ($p_\star=0$ corresponds to the circuit just composed of a single Hadamard layer), indicating the size of the subintervals in which we split the subintervals uniformly, i.e. $p_l=p_\star$, see Fig. \ref{fig:hopping_discretization} (left). The number of layers in the circuit, in which one unit corresponds to the black box application of cost and mixer Hamiltonians (top right). The amount of different sampled energies $\lvert\mathcal{E}\rvert$ as a function of the number of samples $q$ (bottom right).}
        \label{fig:samba_circuit_maxcut_unweighted_30}
\end{figure*}

\begin{figure}[h!]
    \centering
        \includegraphics[width=0.85\linewidth]{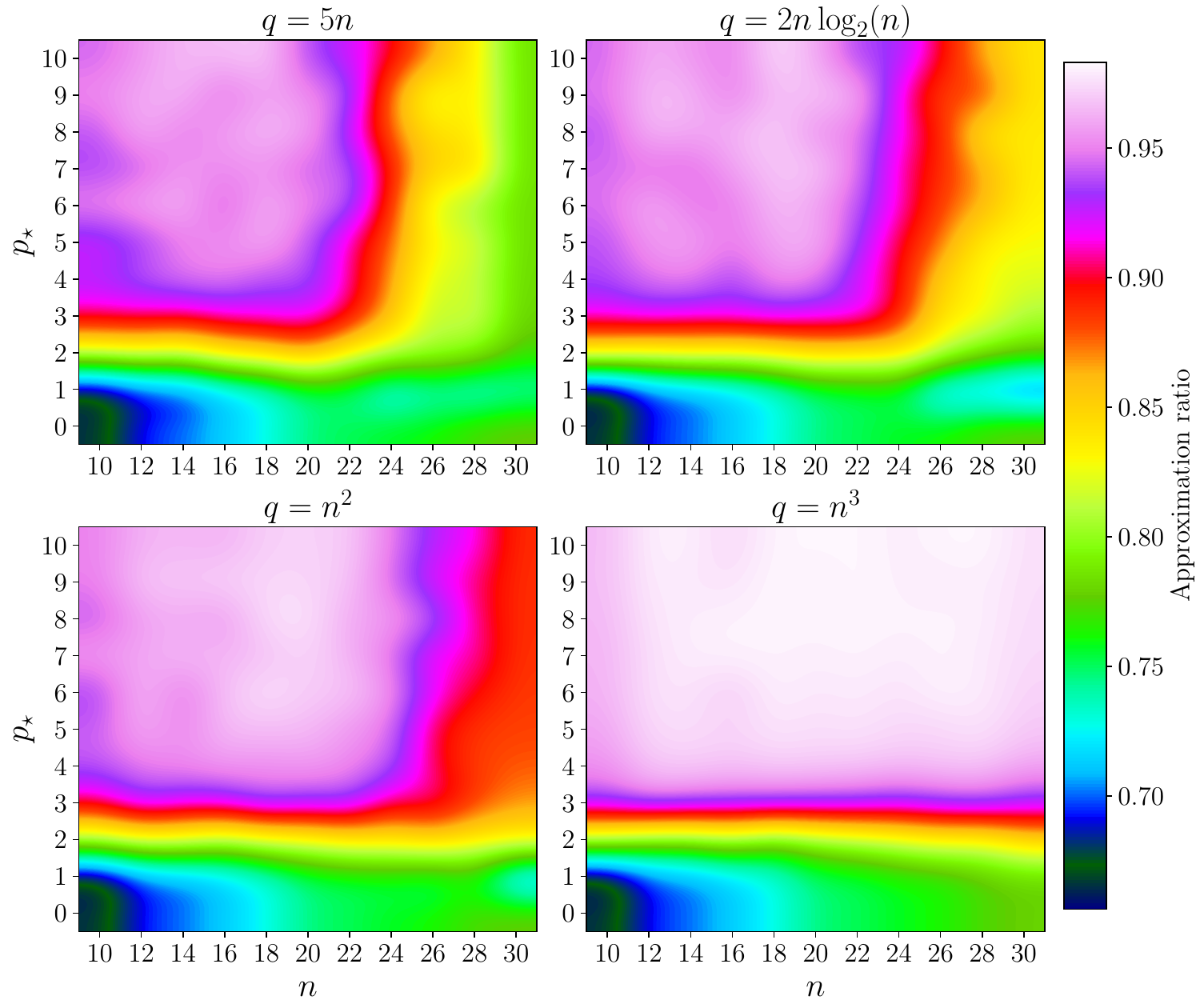}
        \includegraphics[width=0.85\linewidth]{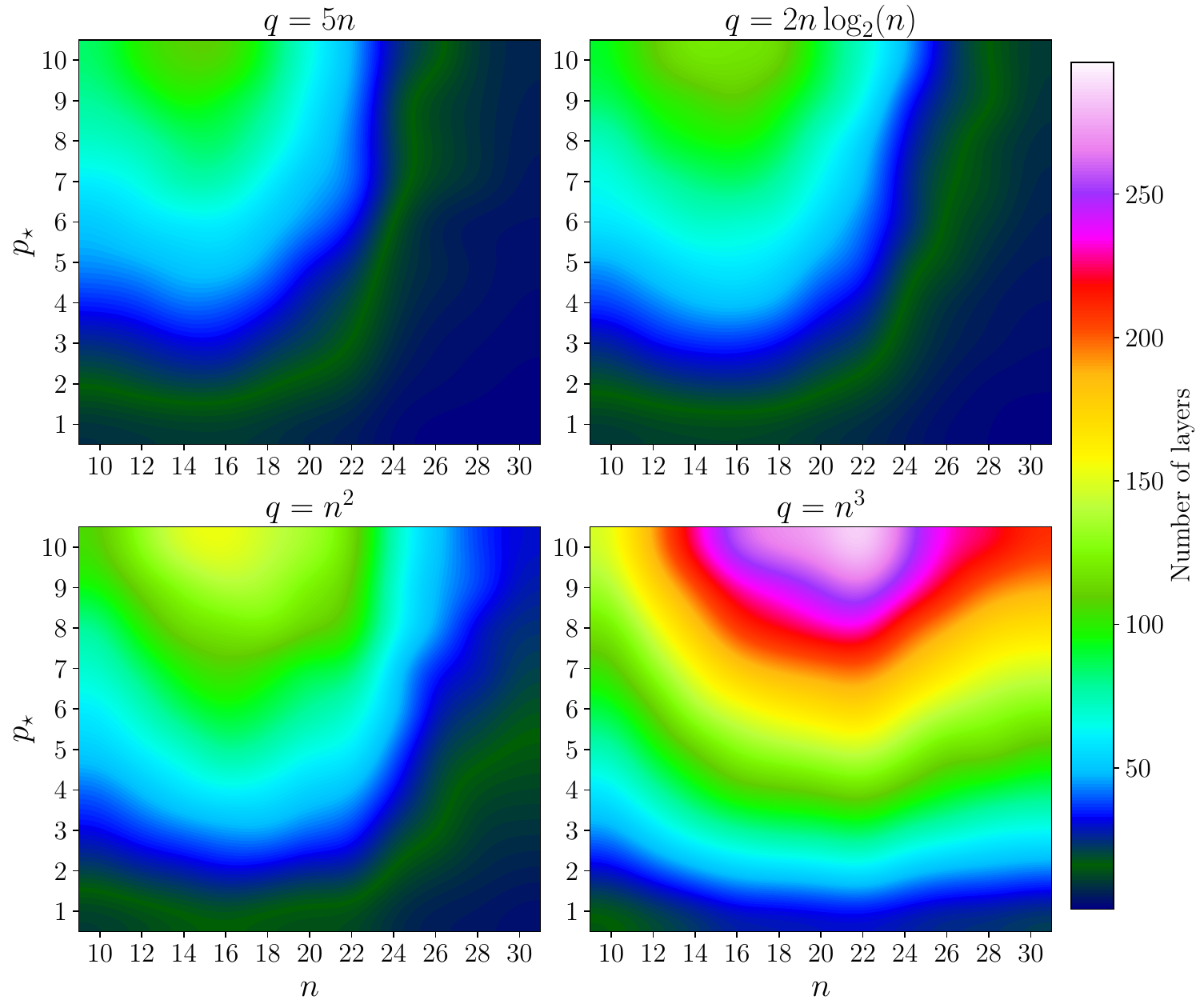}
        \caption{Approximation ratio (top) and number of layers in the circuit (bottom) as a function of the number of qubits $n\in[10,30]$ and the discretization parameter $p_\star\in[1,10]$ for unweighted MaxCut with different sampling values $q\in\{5n,2n\log_2(n),n^2,n^3\}$. We only include $p_\star=0$ for the approximation ratio to show the initial approximation ratio when the circuit only contains a Hadamard layer.}
        \label{fig:samba_circuit_maxcut_unweighted_n}
\end{figure}

\paragraph{Quantum circuit simulations.}
We run quantum circuit simulations for even system sizes $n \in [10, 30]$ and for different sampling values $q \in \{5n, 2n\log_2(n), n^2, n^3\}$. For each sampling value, the hopping-rate schedule is discretized using various values of $p_\star$, where $p_\star$ denotes the number of cuts uniformly applied to each linear subinterval of the hopping rate. Therefore, the resulting quantum circuit has a total of $\bar{p} = |\mathcal{E}| \cdot p_\star$ cost and mixer Hamiltonian layers in the circuit, where $|\mathcal{E}| \leq q$ represents the number of distinct sampled energies, i.e., the number of interpolation points used to construct the hopping rate (excluding the additional zero-energy point).

Detailed results on unweighted MaxCut for $n = 30$ qubits are presented in Fig.~\ref{fig:samba_circuit_maxcut_unweighted_30}. For sampling values $q < n^3$, the number of distinct sampled energies remains small, with $|\mathcal{E}| < 5$, whereas for $q = n^3$, it increases to a range between 10 and 30. This directly impacts performance: the highest approximation ratios, close to 1, are achieved for $q = n^3$, making it the best-performing sampling value among those considered. We also observe a clear trade-off between solution quality and number of circuit layers, as the choice $q = n^3$ leads to the largest circuits. We show the approximation ratio and number of layers in the circuit as a function of the number of qubits $n$ and the discretization parameter $p_\star$ in Fig. \ref{fig:samba_circuit_maxcut_unweighted_n}. The approximation ratio exhibits a clear threshold behavior as a function of the number of circuit layers $p_\star$. For all sampling choices, the performance is relatively poor at very small depth ($p\leq 2$), with the lowest values occurring for small system sizes $n$, and then increases sharply around $p_\star\approx 3$, beyond this point, the ratio remains high over a broad range of $n$ and improves further as $q$ is increased. In particular, for $q=5n$ and $q=2n\log_2 n$, a pronounced degradation appears again at large $n$, whereas for $q=n^2$ this deterioration is pushed to the largest system sizes, and for $q=n^3$ it is essentially absent, with the approximation ratio staying close to its optimal plateau almost uniformly in $n$ once $p_\star\geq 3$. The layer counts show a complementary pattern. They are small at low depth, then grow rapidly with $p_\star$, with the growth becoming much stronger as $q$ increases. For $q=5n$ and $q=2n\log_2 n$, the required number of layers remains moderate and even decreases for the largest $n$, while for $q=n^2$ and especially $q=n^3$ the high-performance region is associated with a substantial overhead, reaching the largest layer counts at large $p_\star$. Overall, increasing $q$ stabilizes high approximation ratios across system sizes, but at the cost of deeper circuits. 

\begin{figure*}
    \centering
        \includegraphics[width=0.8\linewidth]{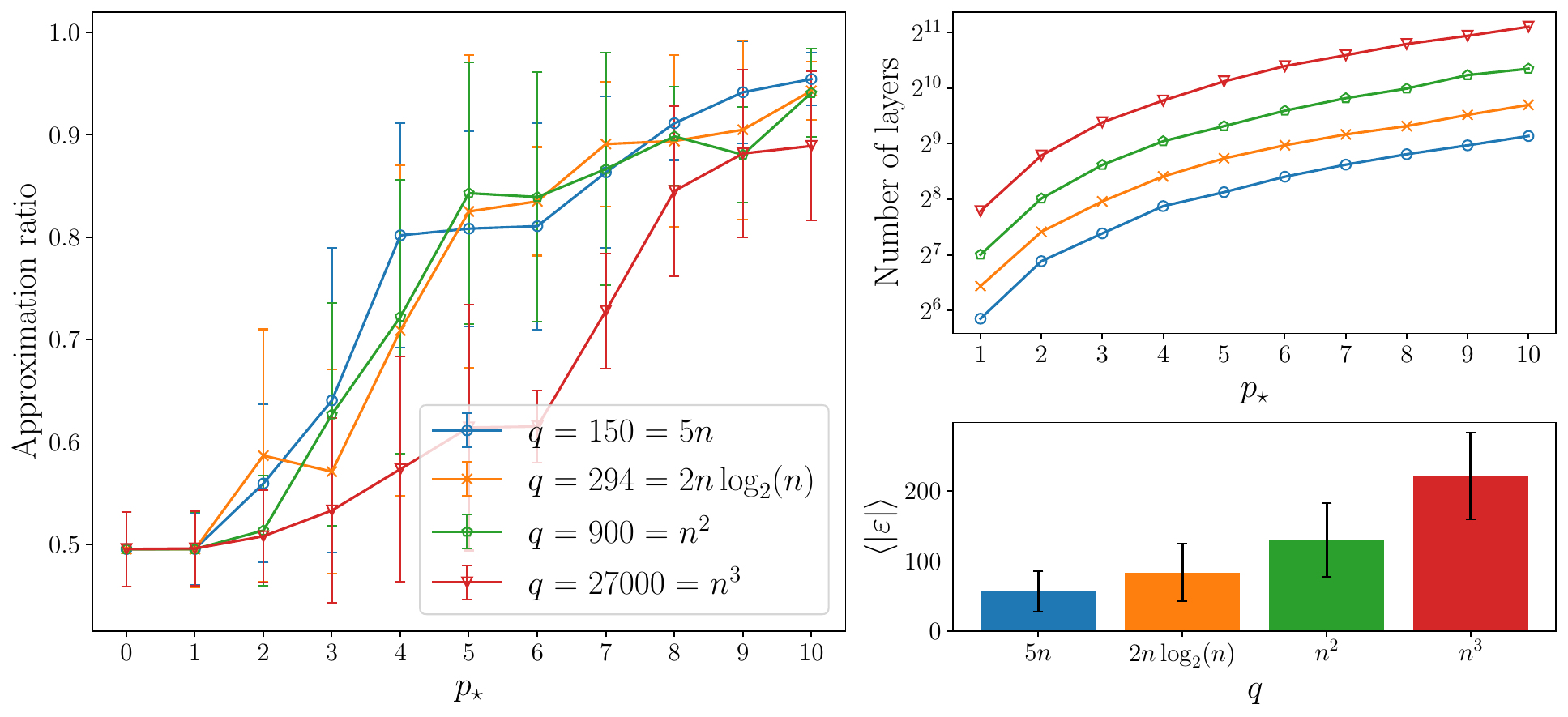}
        \caption{\textsc{SamBa--GQW} quantum circuit simulations for weighted MaxCut for $n=30$ qubits with different sampling values $q\in\{5n,2n\log_2(n),n^2,n^3\}$.}
        \label{fig:samba_circuit_maxcut_weighted_30}
\end{figure*}

\begin{figure}[h!]
    \centering
        \includegraphics[width=.85\linewidth]{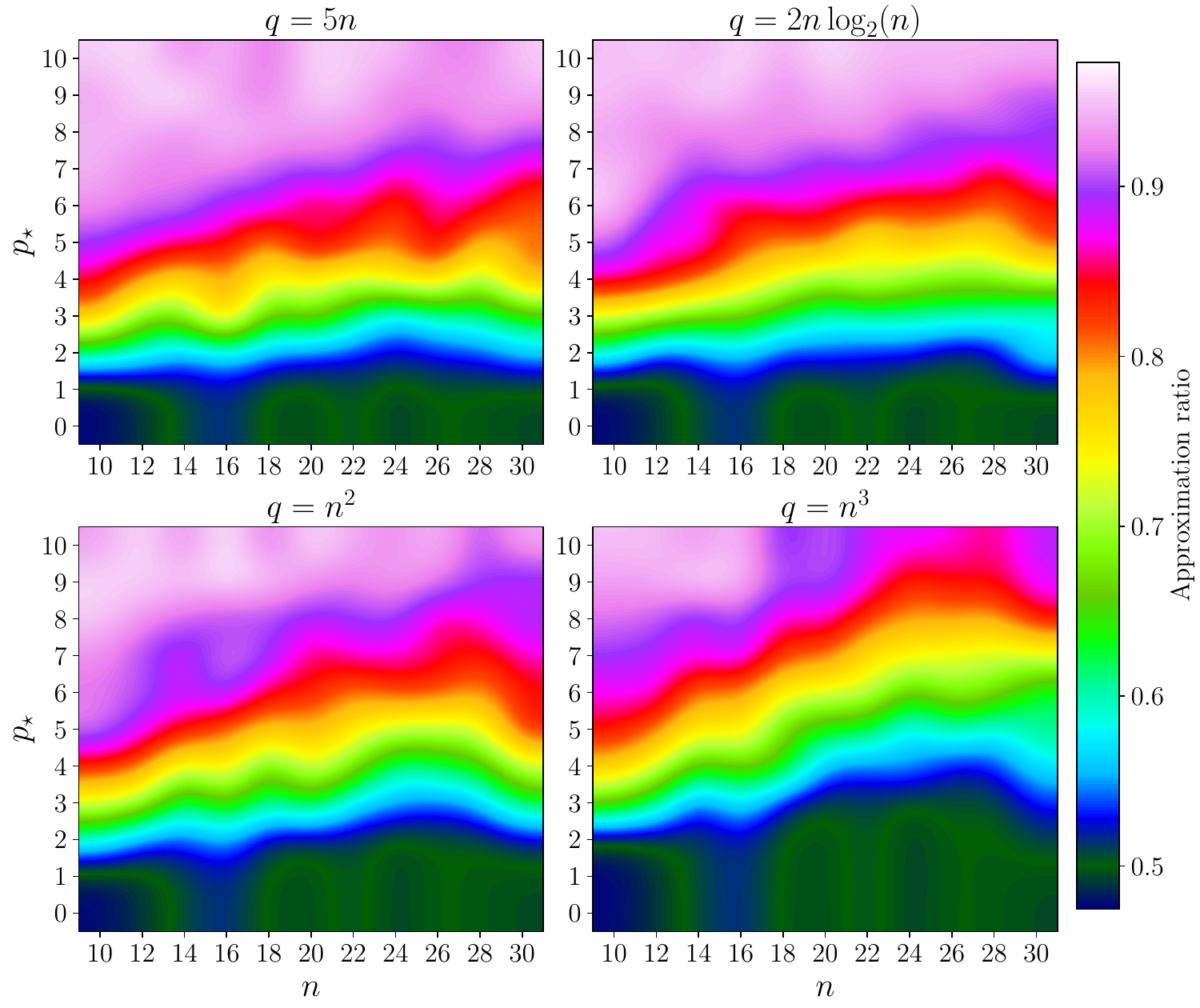}
        \includegraphics[width=.85\linewidth]{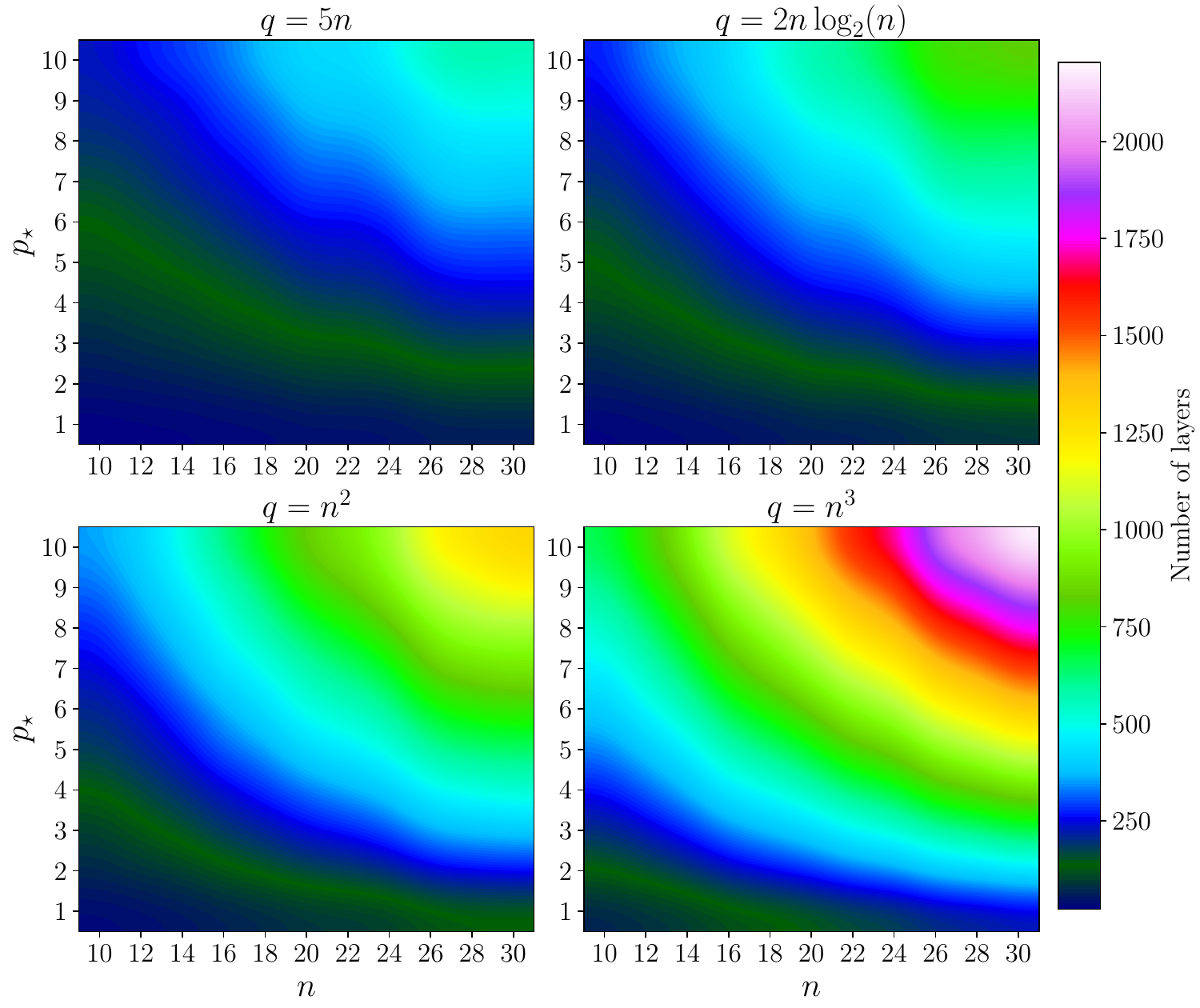}
        \caption{Approximation ratio (top) and number of layers in the circuit (bottom) as a function of the number of qubits $n\in[10,30]$ and the discretization parameter $p_\star\in[1,10]$ for unweighted MaxCut on random Erdős–Rényi graphs with different sampling values $q\in\{5n,2n\log_2(n),n^2,n^3\}$.}
        \label{fig:samba_circuit_maxcut_weighted_n}
\end{figure}

\begin{figure*}
    \centering
	\includegraphics[scale=0.5]{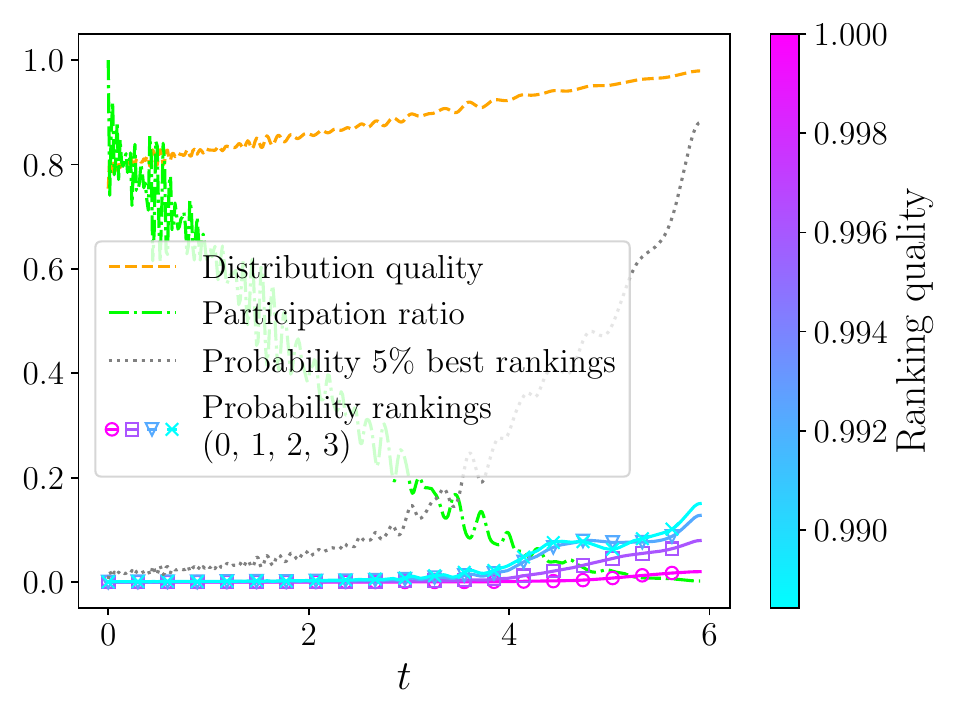}
	\includegraphics[scale=0.5]{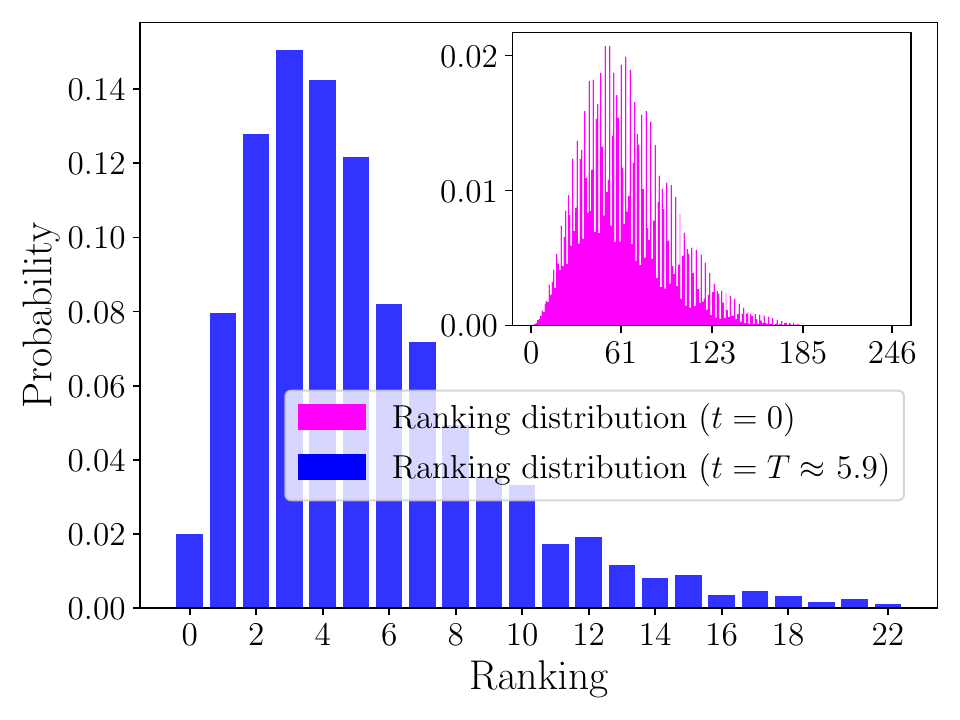}
        \caption{\textsc{SamBa--GQW} Hamiltonian simulations for unweighted MIS on random Erdős–Rényi graphs MIS for $n=20$ qubits with a sampling of $q=n^2$ states.}
    \label{fig:unweighted_mis}
\end{figure*}

We now present the corresponding figures for the weighted MaxCut problem. From Fig.~\ref{fig:samba_circuit_maxcut_weighted_30}, we observe that the number of distinct sampled energies increases smoothly with the sampling parameter $q$. Interestingly, at $p_\star = 10$, the best sampling strategy on average is $q = 5n$. This indicates that, for this problem, such a sampling level provides a sufficient number of sampled energies, whereas larger sampling values require deeper circuits to yield performance improvements. This behavior is further confirmed in Appendix~\ref{app:larger_circuit}, where larger circuits with depths up to $p_\star = 30$ are considered. Figure~\ref{fig:samba_circuit_maxcut_weighted_n} shows that, in contrast to the unweighted case, the weighted MaxCut problem exhibits a smoother and size-dependent improvement with circuit depth. Specifically, the depth required to achieve a given approximation ratio increases with the number of qubits. Increasing the sampling parameter $q$ can partially compensate for this complexity by shifting the onset of high-performance regimes toward smaller depths and enabling approximation ratios above $0.9$ for most system sizes. Nevertheless, these gains are accompanied by a pronounced increase in circuit depth, with the number of layers scaling steeply with both $p_\star$ and $q$, reaching several thousand layers for the largest sampling values. This behavior highlights a fundamental trade-off: larger $q$ improves robustness and solution quality but substantially increases the required computational resources.

\subsubsection{Maximum independent set (MIS)}

\begin{figure*}
    \centering
	\includegraphics[scale=0.5]{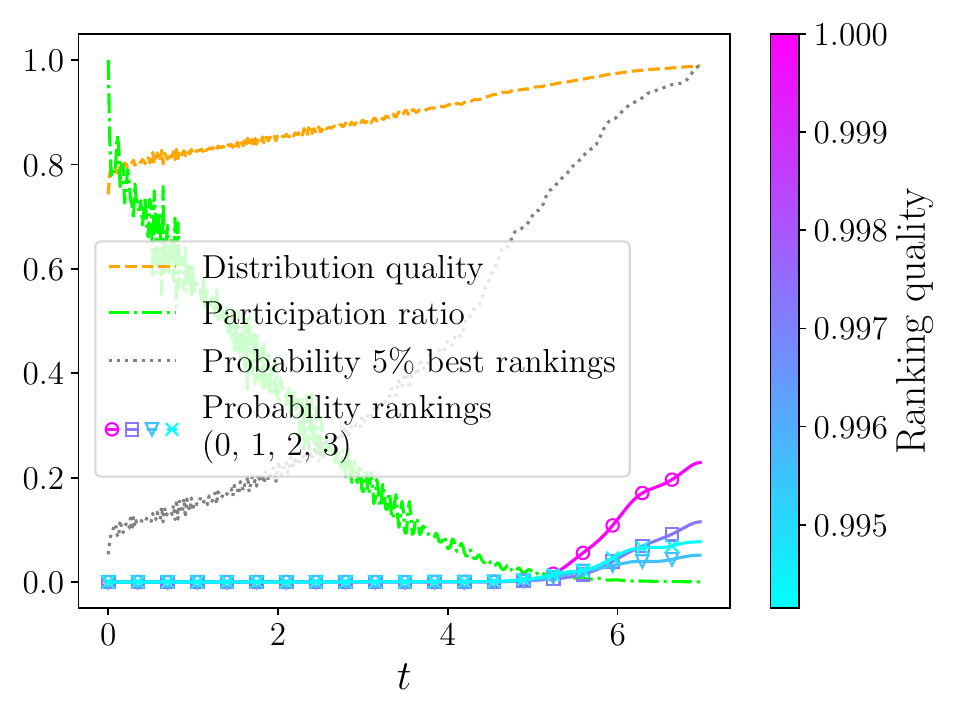}
	\includegraphics[scale=0.5]{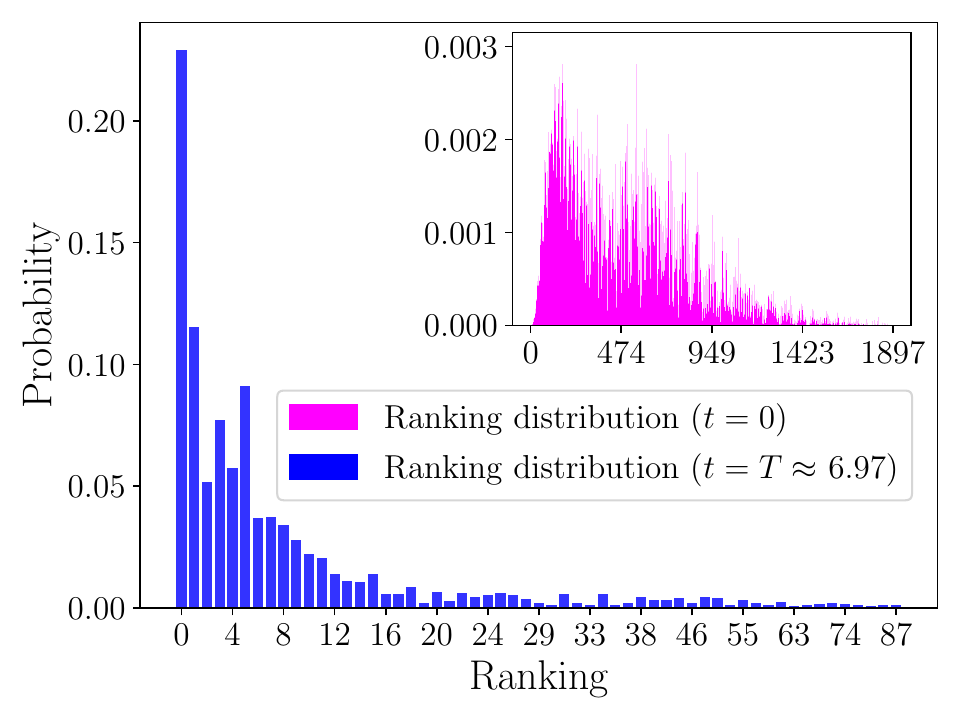}
        \caption{\textsc{SamBa--GQW} Hamiltonian simulations for weighted MIS on Unit-Disk graphs MIS for $n=20$ qubits with a sampling of $q=n^2$ states.}
    \label{fig:weighted_mis}
\end{figure*}

In the case of unweighted MIS, we generate Erdős–Rényi graphs \cite{Erdos1960} of $n=20$ vertices with an edge appearance probability of 1/2, and set $w_i=1$ for all the vertices. Regarding UD-MIS, we generate UD graphs of $n=20$ vertices and we select the weights in the spirit of Ref. \cite{8gjv-ll1f} such that:
\begin{equation}
    w_i=1+\frac{10\cdot C_D(i)}{\max_{j\in V}C_D(j)},
\end{equation}
with $C_D(i)=\text{deg}(i)/(\lvert V\rvert-1)$ the degree centrality of vertex $i$. Furthermore, the procedure we use to generate UD graphs is detailed in the Appendix D of Ref. \cite{8gjv-ll1f}. 

\paragraph{Hamiltonian simulations.}
We present the quality distribution, participation ratio and probabilities per ranking over time for random Erdős–Rényi (resp. UD) graphs on Fig. \ref{fig:unweighted_mis} (resp. \ref{fig:weighted_mis}). In the case of random unweighted graphs, the highest measurement probability is that of ranking 3 with $\mathbb{P}_3[\psi_T]\approx 0.15$, which has a quality close to 0.99. We see that optimal solutions have a relevant success probability $\mathbb{P}_0[\psi_T]\approx 0.02$, considering that the evolution starts in a uniform superposition of $2^n$ states. Moreover, the probability of measuring decisions whose ranking is in the top 5\% of all rankings is around $0.9$. 

Regarding UD graphs, optimal solutions have the highest measurement probability with $\mathbb{P}_0[\psi_T]\approx0.25$. Furthermore, the probability of measuring decisions with the top 5\% rankings almost reaches $1$. In both cases, the participation ratio decreases rapidly, indicating that the quantum state goes from a uniform superposition to a more localized state. Additionally, both random and UD graphs have a high initial distribution quality $\mathbb{E}[\psi_0]\approx 0.8$, suggesting that there are more high quality decisions than low quality ones. Lastly, we note that time evolution is extremely short since it reaches $T\approx 6$ (resp. $T\approx 7$) for random (resp. UD) graphs. Moreover, we see that the ranking distribution is shifted to the left in both cases. However, for UD graphs, the number of different rankings is significantly higher than for random graphs, with $1897$ against $246$. In addition, the distribution for UD graphs seems to be composed of two Gaussians instead of one for random graphs. In both cases, the evolution induced by \textsc{SamBa--GQW} shrinks the initial distribution to a distribution localized around high quality decisions, with high success probability on optimal solutions for UD graphs.

\begin{figure}
\centering
 \includegraphics[width=0.45\linewidth]{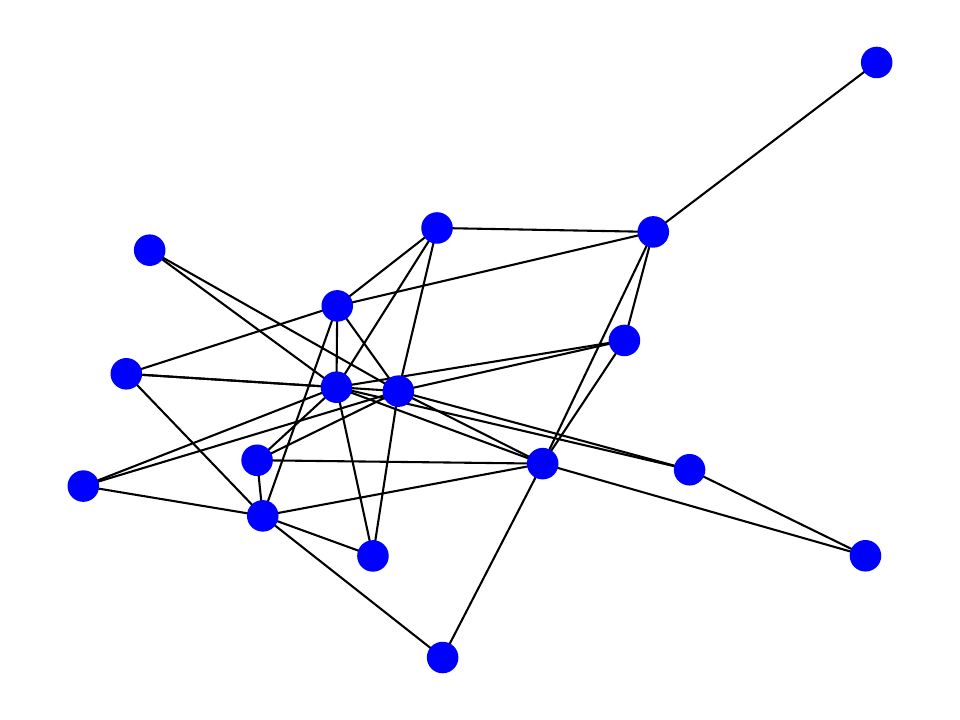}
  \includegraphics[width=0.45\linewidth]{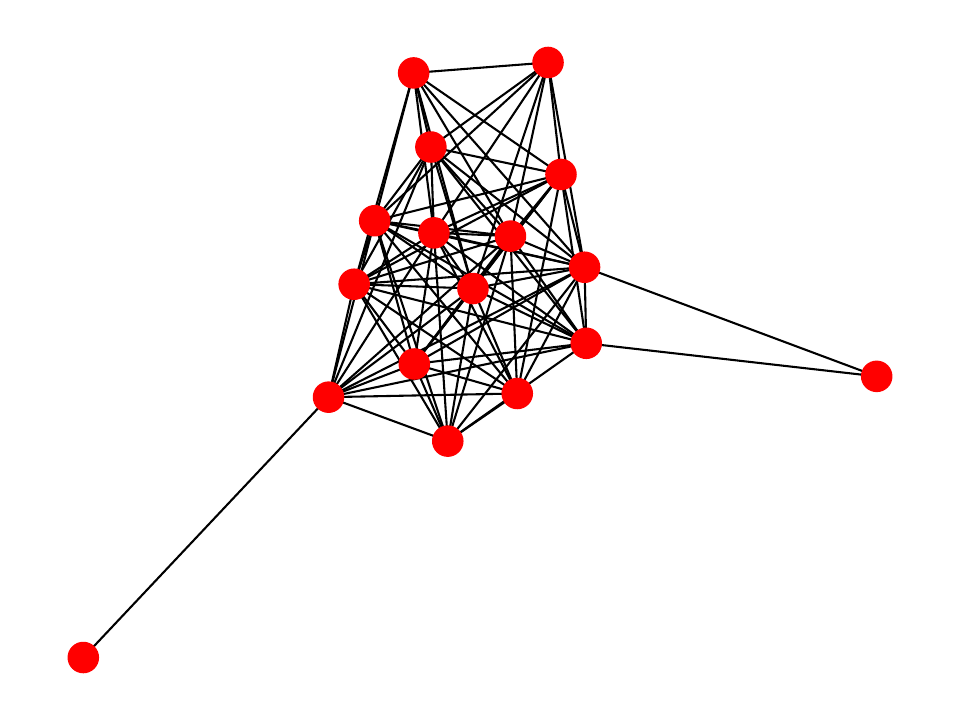}
\caption{Instances \textit{farm} (left) and \textit{mammalia-kangaroo-interactions} (right) of $n=17$ vertices.}
\label{fig:ibm_graphs}
\end{figure}

In addition, we also execute \textsc{SamBa--GQW} on two instances proposed in Ref. \cite{koch2026quantum}, \textit{farm} and \textit{mammalia-kangaroo-interactions} \cite{qoblib2026} of $n=17$ vertices with $\lambda=2$ and $w_i=1$. We display these instances on Fig. \ref{fig:ibm_graphs}. 

\begin{figure*}
	\centering
	\includegraphics[scale=0.5]{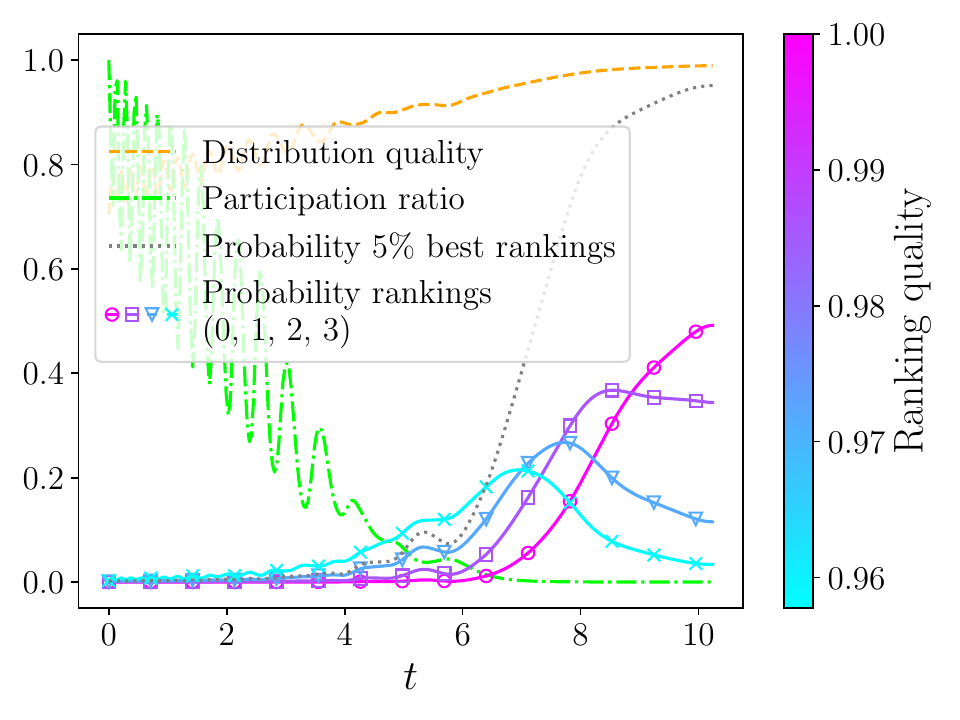}
	\includegraphics[scale=0.5]{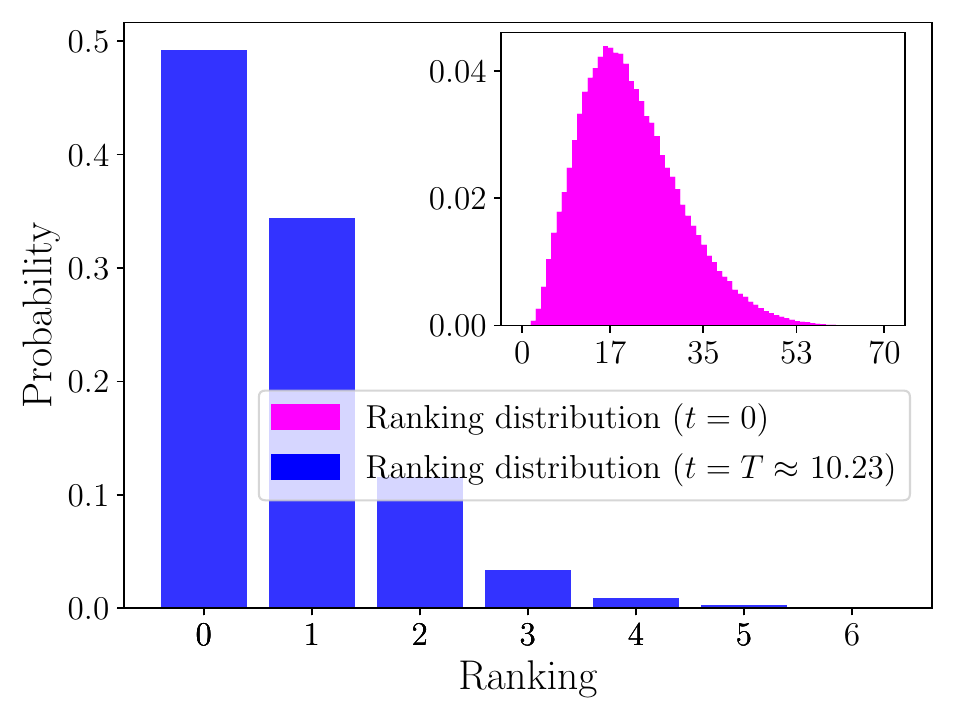}
	\caption{\textsc{SamBa--GQW} Hamiltonian simulations for MIS on \textit{farm} instance from Refs. \cite{koch2026quantum,qoblib2026}, for $n=17$ qubits with a sampling of $q=n^2$ states. }
    \label{fig:ibm_mis_farm}
\end{figure*}

\begin{figure*}
	\centering
	\includegraphics[scale=0.5]{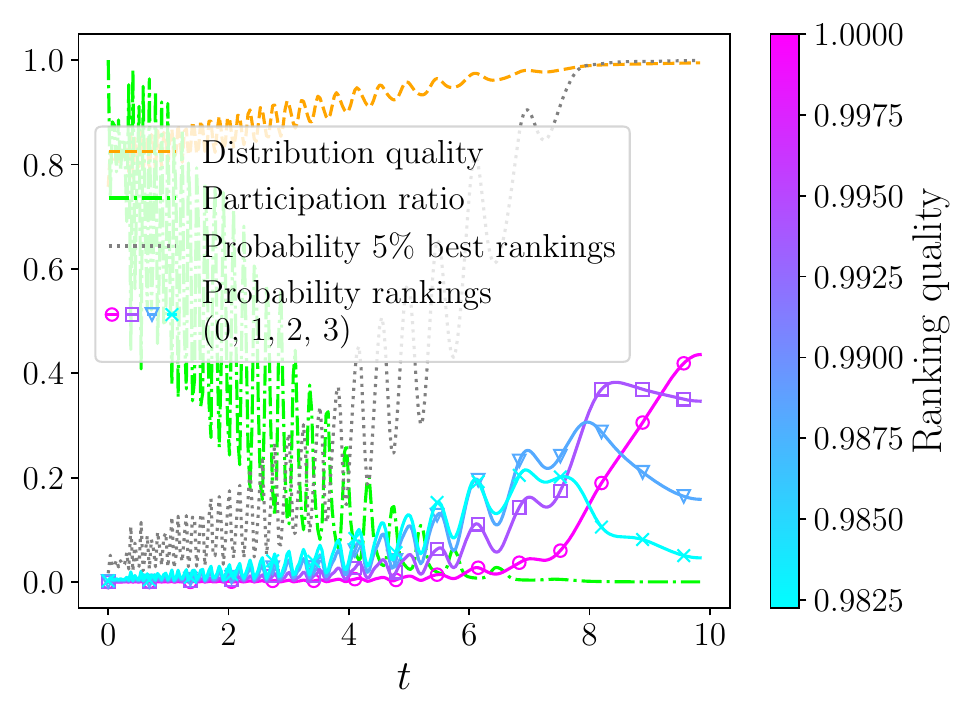}
	\includegraphics[scale=0.5]{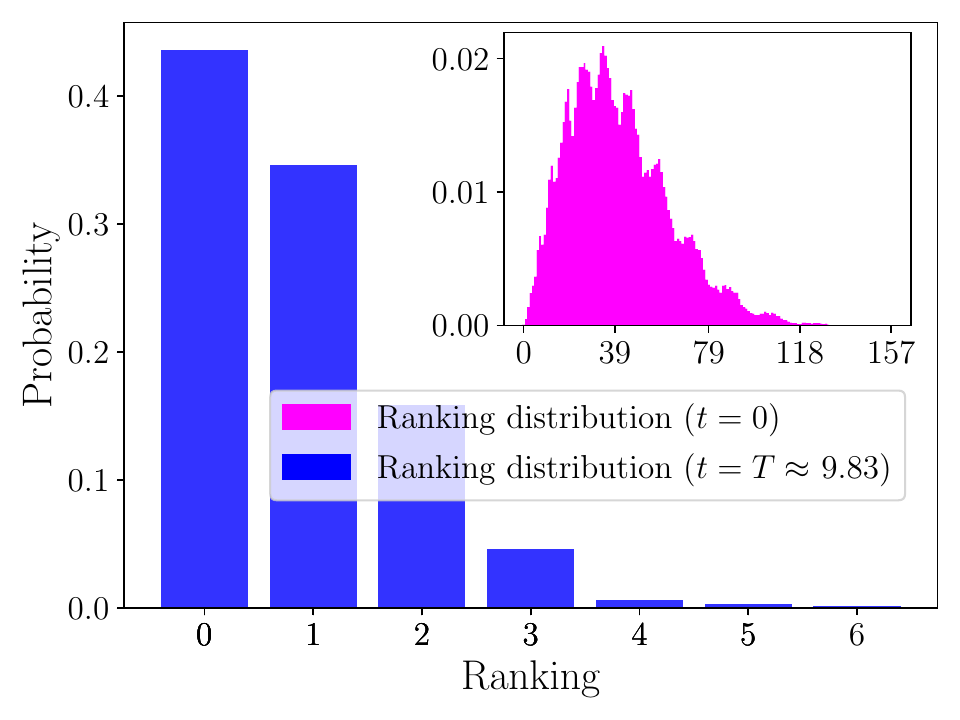}
	\caption{\textsc{SamBa--GQW} Hamiltonian simulations for MIS on \textit{mammalia-kangaroo-interactions} instance from Refs. \cite{koch2026quantum,qoblib2026}, for $n=17$ qubits with a sampling of $q=n^2$ states. }
    \label{fig:ibm_mis_mammalia}
\end{figure*}

We obtain similar results for the two instances, we see on Fig. \ref{fig:ibm_mis_farm} (resp. \ref{fig:ibm_mis_mammalia}) for \textit{farm} (resp. \textit{mammalia-kangaroo-interactions}) that the highest measurement probability is that of the optimal solutions in both cases, with $\mathbb{P}_0[\psi_T]\approx 0.5$ (resp. $\mathbb{P}_0[\psi_T]\approx 0.45$). For \textit{farm}, the probability of measuring decisions with ranking in top 5\% of all rankings reaches approximatively $0.96$, and $1$ for \textit{mammalia-kangaroo-interactions}. The participation ratio decreases up to a localized state quickly and distribution quality is initially high with $\mathbb{E}[\psi_0]\approx 0.8$ for the two instances. The required evolution time is very short since it reaches $T\approx 10$ in both cases. Furthermore, we see that the Gaussian ranking distributions are shifted to higher quality decisions, as suggested by the distribution quality, and instance \textit{mammalia-kangaroo-interactions} owns more different rankings than \textit{farm} with $158$ against $71$ rankings. After the evolution induced by \textsc{SamBa--GQW}, the ranking distributions are almost fully localized on optimal solutions. Moreover, in both cases only six rankings have a measurement probability higher than $10^{-3}$.

\begin{figure*}
    \centering
        \includegraphics[width=0.8\linewidth]{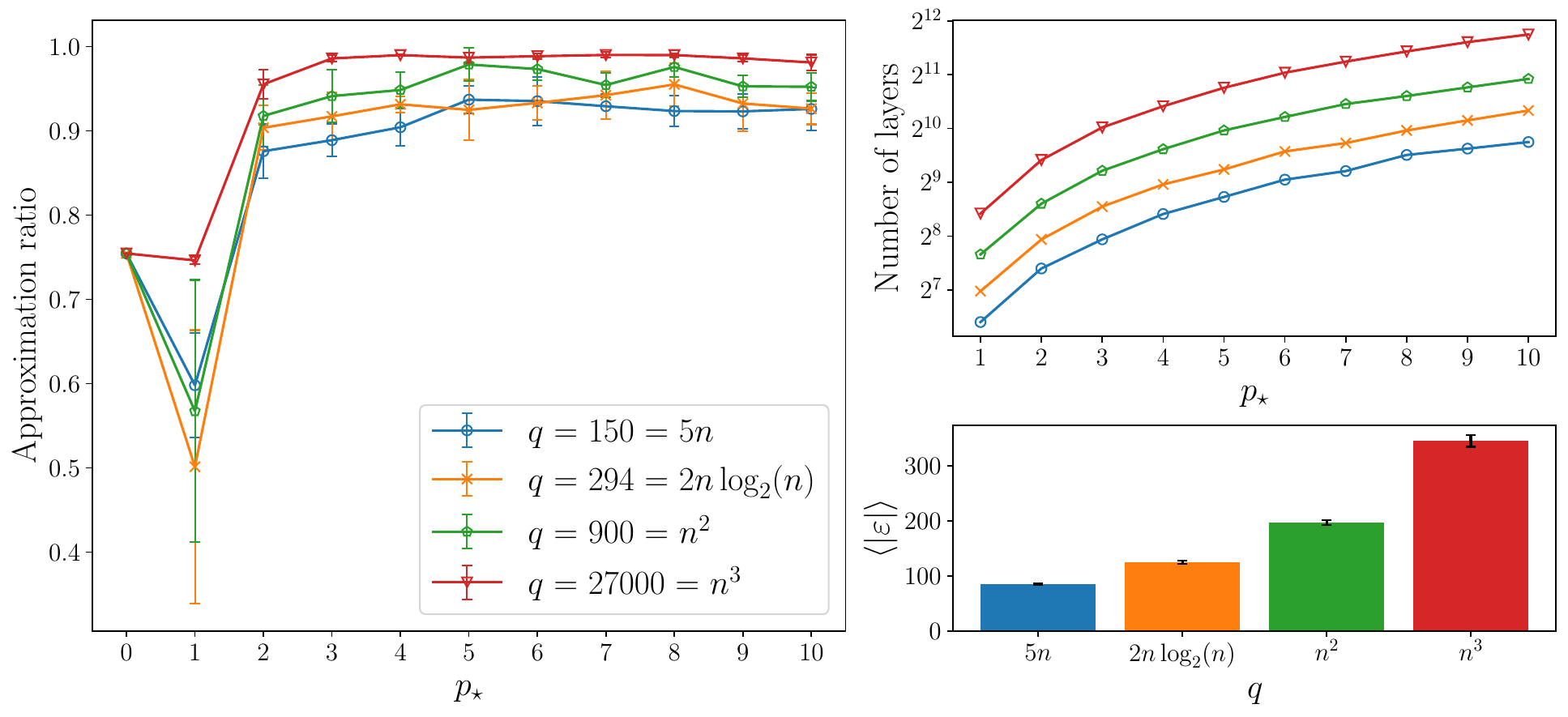}
        \caption{\textsc{SamBa--GQW} quantum circuit simulations for unweighted MIS on random Erdős–Rényi graphs for $n=30$ qubits with different sampling values $q\in\{5n,2n\log_2(n),n^2,n^3\}$.}
        \label{fig:samba_circuit_mis_regular_30}
\end{figure*}

\begin{figure}
    \centering
        \includegraphics[width=0.85\linewidth]{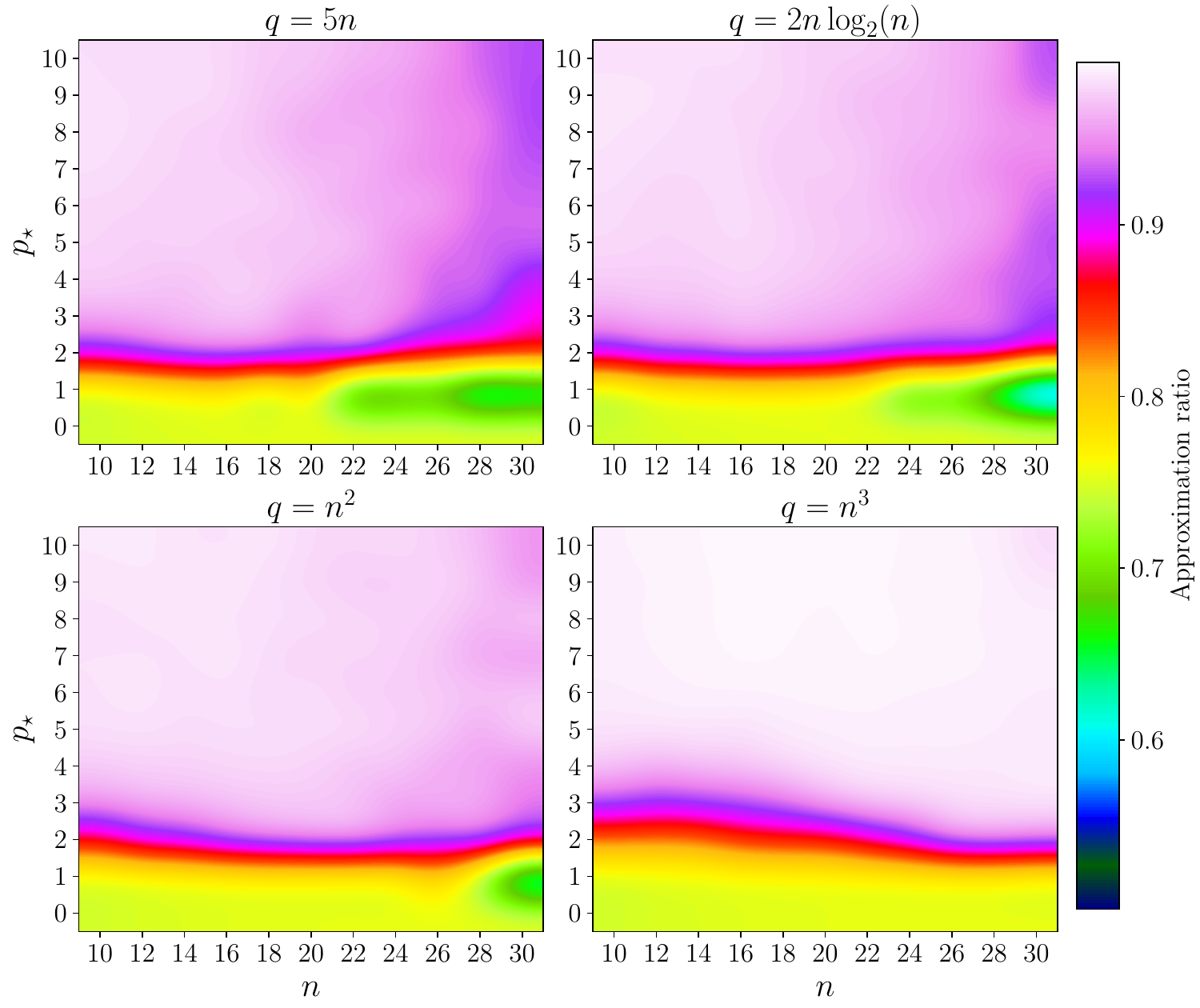}
        \includegraphics[width=0.85\linewidth]{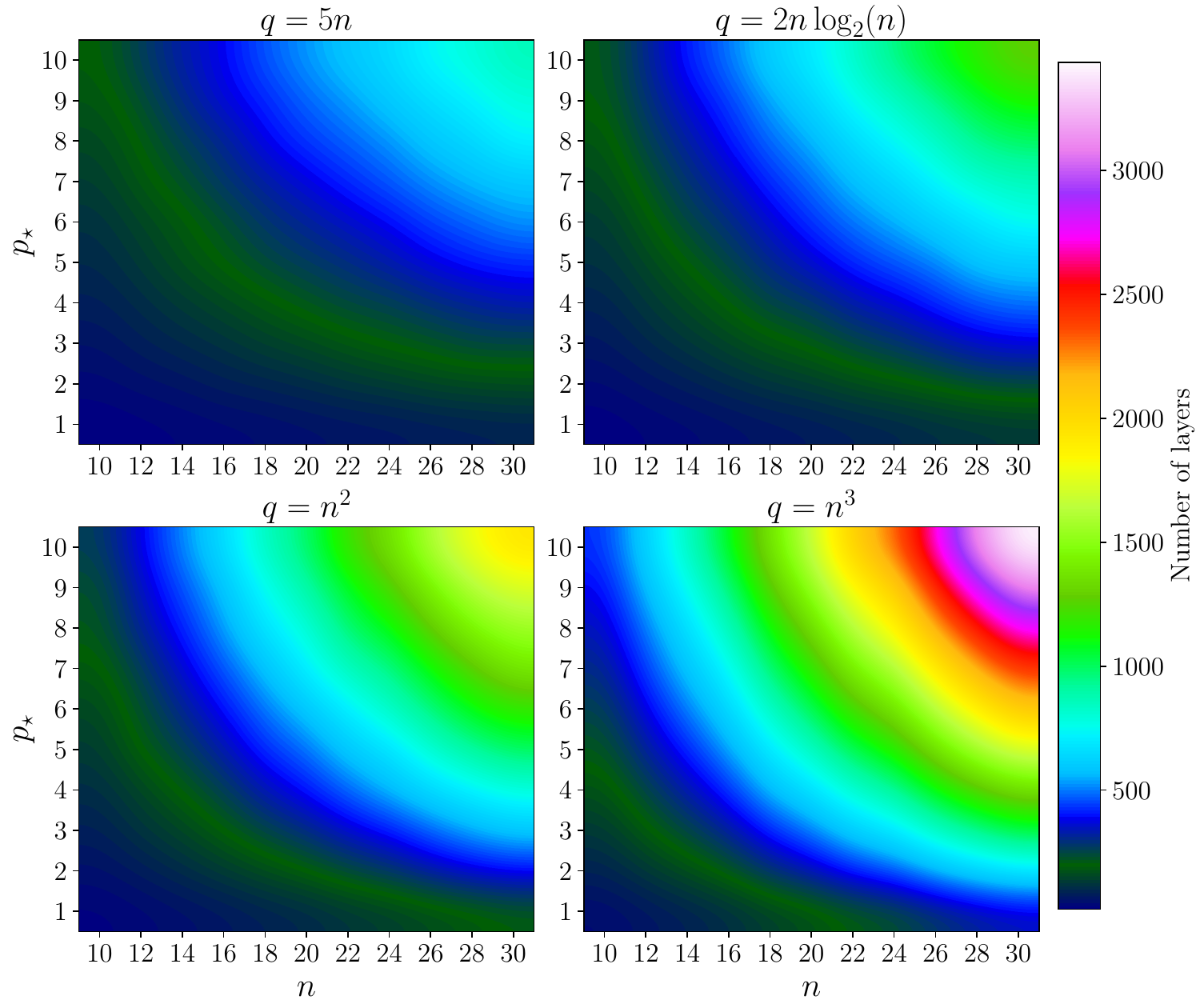}
        \caption{Approximation ratio (top) and number of layers in the circuit (bottom) as a function of the number of qubits $n\in[10,30]$ and the discretization parameter $p_\star\in[1,10]$ for unweighted MIS on random Erdős–Rényi graphs with different sampling values $q\in\{5n,2n\log_2(n),n^2,n^3\}$.}
        \label{fig:samba_circuit_mis_regular_n}
\end{figure}

\paragraph{Quantum circuit simulations.}

For unweighted MIS, the results for $n=30$ are presented in Fig.~\ref{fig:samba_circuit_mis_regular_30}. The approximation ratio rapidly converges to a plateau for $p_\star \geq 2$, indicating that shallow circuits are sufficient to achieve near-optimal solutions. In parallel, the number of layers increases steadily with $p_\star$ and is strongly amplified by larger sampling rates $q$. A broader perspective is provided in Fig.~\ref{fig:samba_circuit_mis_regular_n}, where the algorithm exhibits a distinctive and highly favorable performance profile. Unlike MaxCut, the approximation ratio displays a pronounced non-monotonic dependence on the circuit depth $p_\star$, characterized by a systematic dip at $p_\star=1$ followed by a sharp recovery around $p_\star \simeq 2$. Beyond this point, the performance rapidly saturates close to optimal values and remains largely independent of the system size $n$. This early saturation suggests that the essential combinatorial structure of unweighted MIS instances can be captured with very shallow circuits. Increasing the sampling parameter $q$ primarily enhances the robustness of this high-performance regime rather than significantly improving its asymptotic value, with $q = n^3$ yielding near-optimal approximation ratios across all considered system sizes. The required resources, however, scale steeply with both $p_\star$ and $q$. The number of circuit layers grows rapidly with the circuit depth and system size, and increases dramatically as $q$ is enlarged, reaching several thousand layers for the largest instances. Consequently, while larger values of $q$ provide only marginal improvements in solution quality, they incur a substantial overhead in circuit complexity, highlighting a clear trade-off between robustness and computational cost.

\begin{figure*}
    \centering
        \includegraphics[width=0.8\linewidth]{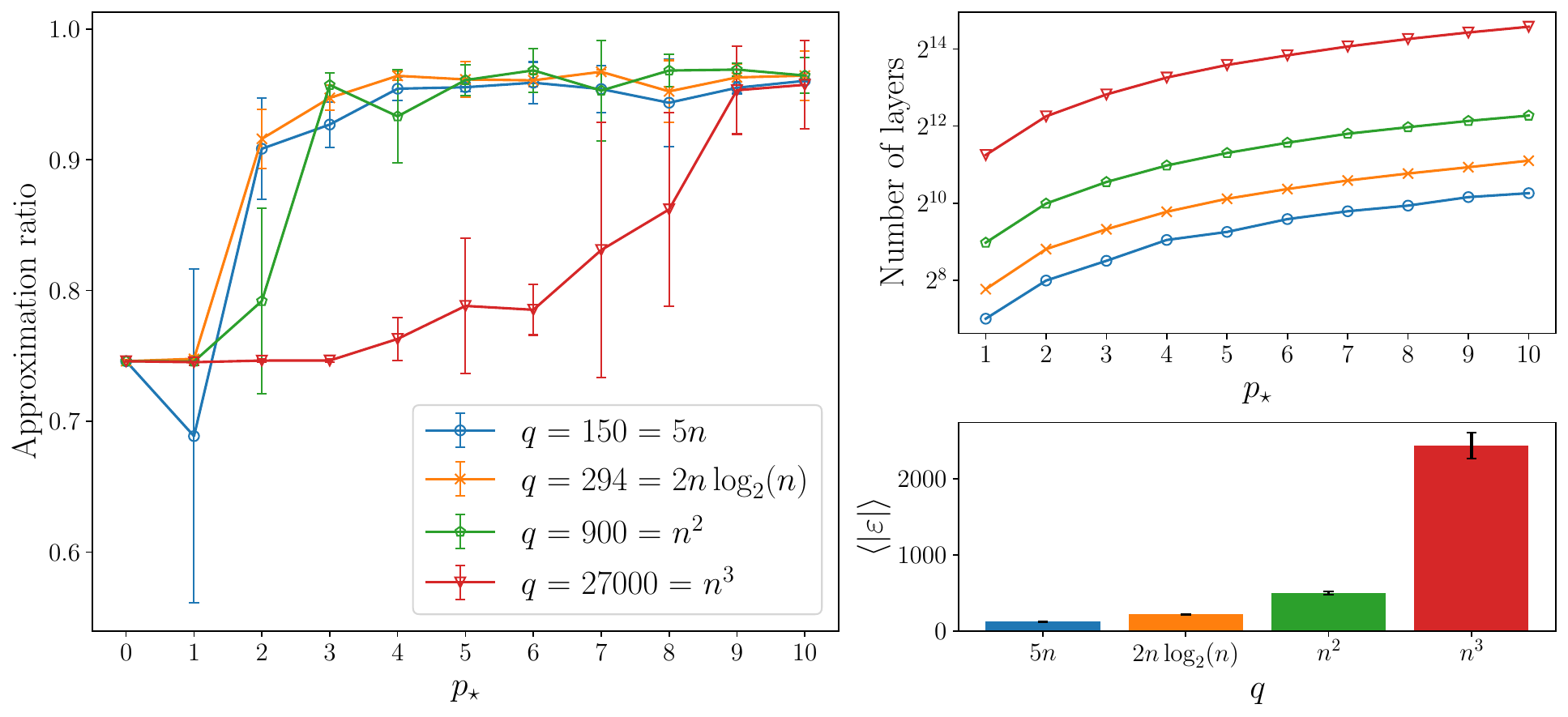}
        \caption{\textsc{SamBa--GQW} quantum circuit simulations for weighted MIS on unit-disk graphs for $n=30$ qubits with different sampling values $q\in\{5n,2n\log_2(n),n^2,n^3\}$.}
        \label{fig:samba_circuit_mis_ud_30}
\end{figure*}

\begin{figure}
    \centering
        \includegraphics[width=0.85\linewidth]{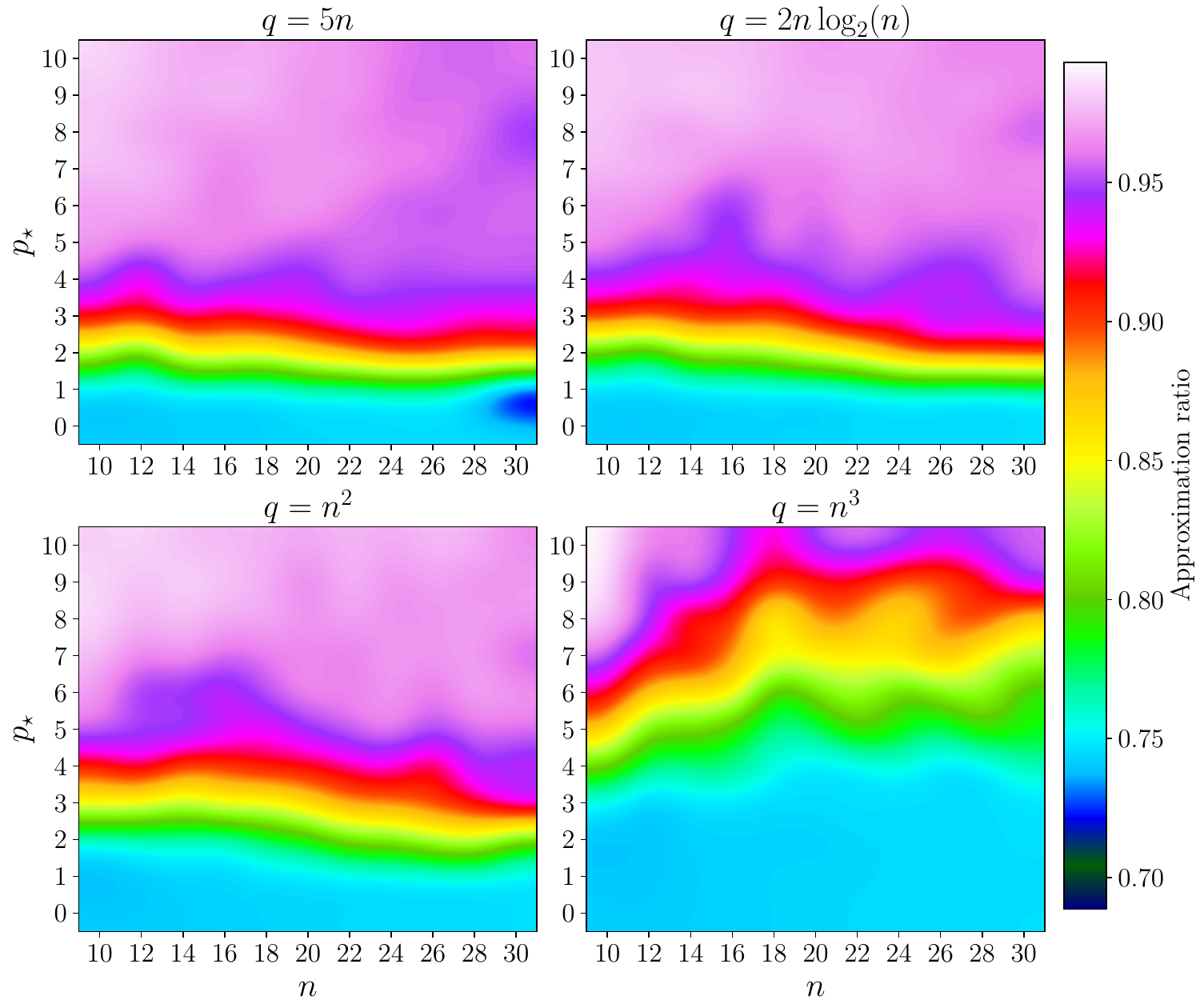}
        \includegraphics[width=0.85\linewidth]{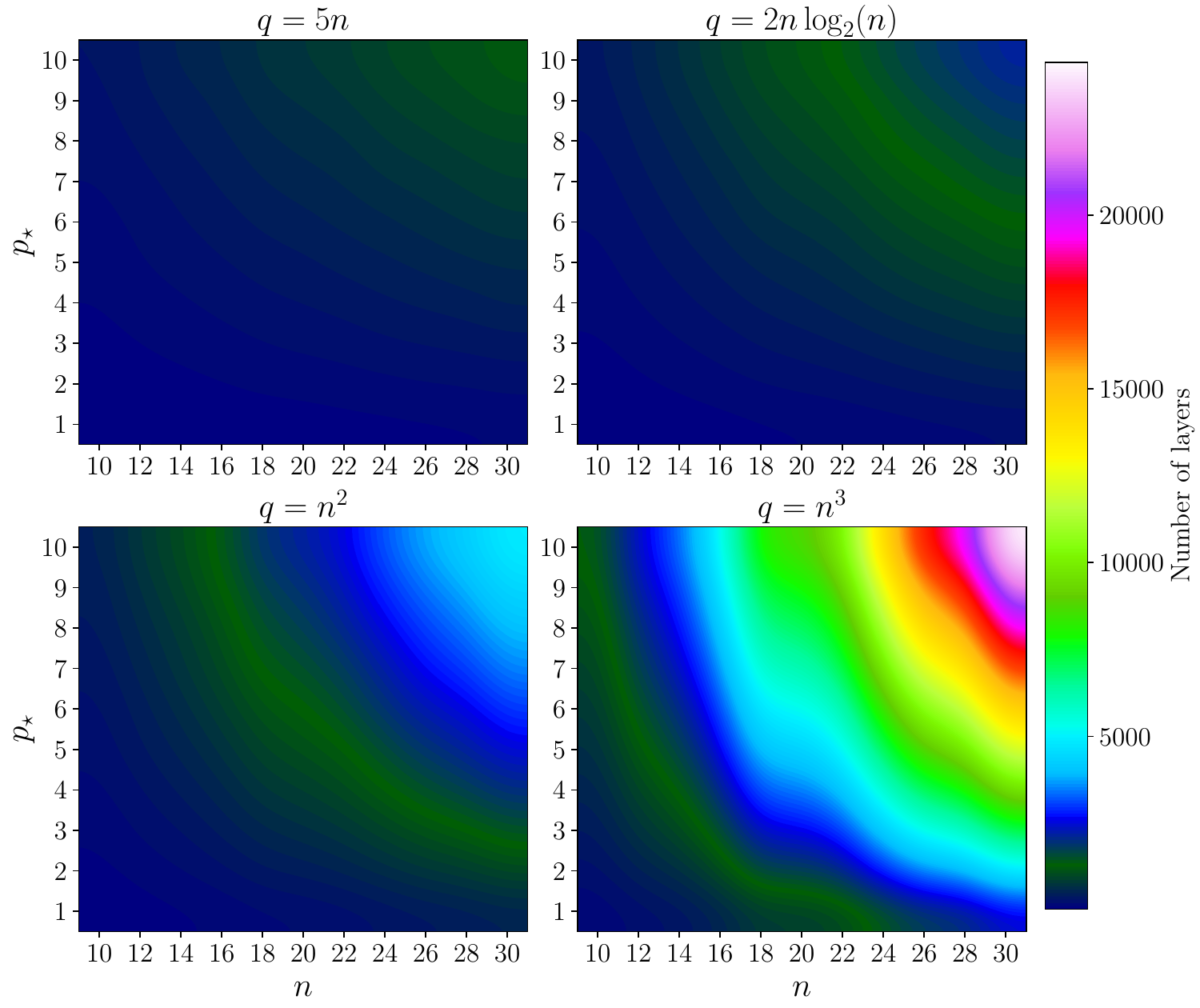}
        \caption{Approximation ratio (top) and number of layers in the circuit (bottom) as a function of the number of qubits $n\in[10,30]$ and the discretization parameter $p_\star\in[1,10]$ for weighted MIS on unit-disk graphs with different sampling values $q\in\{5n,2n\log_2(n),n^2,n^3\}$.}
        \label{fig:samba_circuit_mis_ud_n}
\end{figure}

We now present the results for UD-MIS. Looking at Fig. \ref{fig:samba_circuit_mis_ud_30}, we observe that the number of distinct sampled energies is significantly higher than in the unweighted MaxCut case on $n=30$ qubits, reaching $|\mathcal{E}| = 2000$ for $q = n^3$. Moreover, sampling values $q \in \{5n, 2n\log_2(n), n^2\}$ yield approximately the same approximation ratio, whereas $q = n^3$ requires larger values of $p_\star$ to achieve a noticeable improvement. We also observe that at $p_\star = 10$, all sampling strategies attain similar approximation ratios, with $q = n^3$ not providing any advantage. Considering the number of layers in the circuit, selecting $q = n^3$ is therefore the least favorable strategy, as the approximation ratio converges more slowly while leading to the highest circuit depth. These results highlight a clear trade-off between solution quality and circuit depth: higher sampling values require significantly deeper circuits without necessarily providing better performance.  Figure~\ref{fig:samba_circuit_mis_ud_n} shows that the number of circuit layers increases rapidly with both the discretization parameter $p_\star$ and the sampling value $q$. In particular, for $q = n^3$, the circuit depth grows steeply with the system size, reaching several tens of thousands of layers for the largest instances, whereas smaller sampling values ($q = 5n$ and $q = 2n\log_2 n$) lead to significantly shallower circuits. Furthermore, all sampling strategies achieve high approximation ratios at relatively small discretization parameter values ($p_\star \approx 3$--$4$), with only marginal improvements for larger $p_\star$. Increasing the sampling parameter to $q = n^3$ does not yield a substantial performance advantage and requires considerably deeper circuits to reach comparable approximation ratios.  These results highlight a clear trade-off between solution quality and circuit depth. Moderate sampling values, such as $q = 5n$ or $q = 2n\log_2 n$, provide the most favorable balance, achieving near-optimal approximation ratios while keeping the circuit depth manageable. In contrast, larger sampling values significantly increase the required computational resources without delivering commensurate gains in performance.

\begin{figure*}
	\centering
	\includegraphics[scale=0.5]{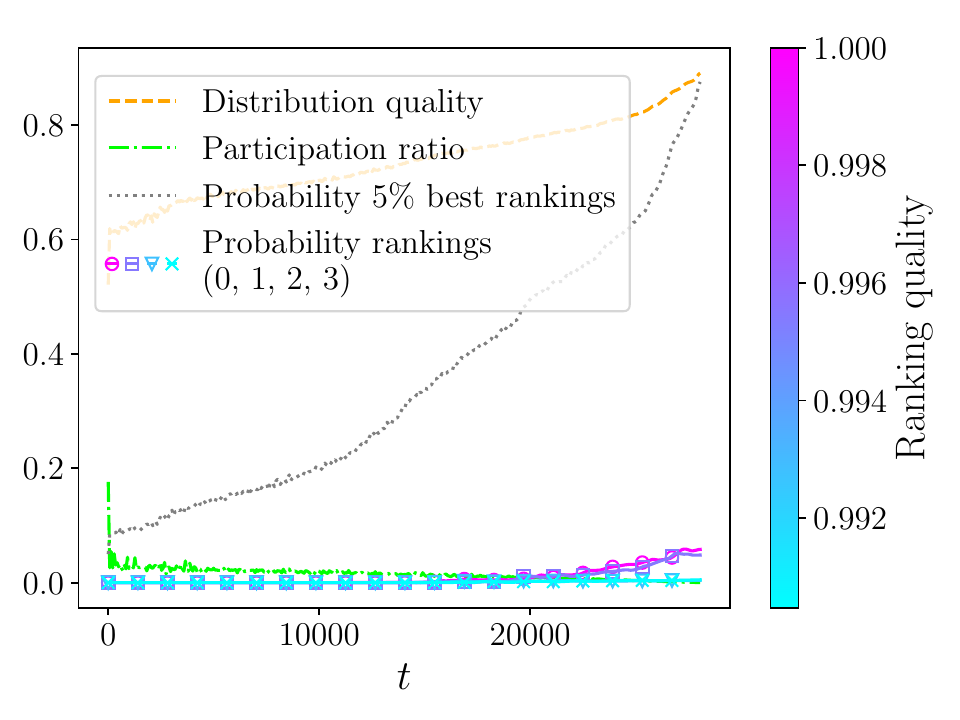}
	\includegraphics[scale=0.5]{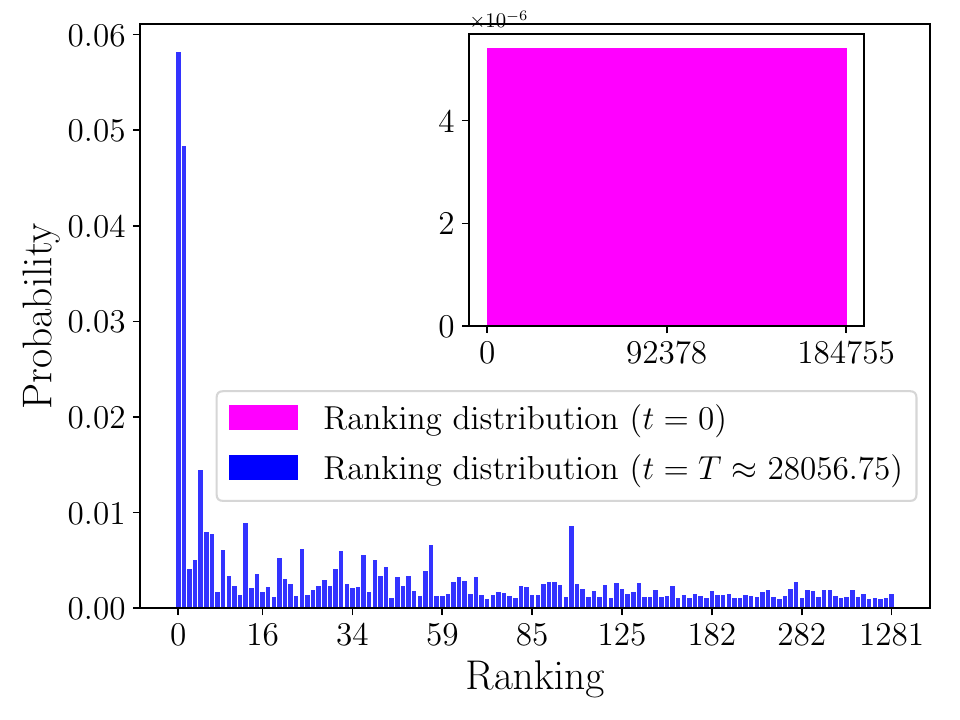}
    \caption{\textsc{SamBa--GQW} Hamiltonian simulations on portfolio-optimization with $XY$-mixers (with ring connectivity) for $n=20$ qubits with a sampling of $q=n^2$ states. The instances are from dataset of Ref. \cite{CZYLOV_2022}.}
    \label{fig:portfolio_optimization}
\end{figure*}

\subsubsection{Portfolio optimization}

We use the file \texttt{JKMP1 500\_1} of dataset \cite{CZYLOV_2022} to generate 10 instances of $n=20$ assets each. Since this file contains portfolio information of $500$ variables, we generate different instances by randomly drawing $n=20$ variables. We set the risk appetite to $\lambda=1/2$ in Eq. \eqref{eq:portfolio} and the number of assets to $k=n/2=10$. Therefore, the initial state is the uniform superposition over basis states of Hamming weight $k=10$. Since we use $XY$-mixers, only the feasible space is explored by the walker, i.e., only the subgraph of Hamming weight $k=10$ is explored. The feasible space contains ${n=20 \choose k=10}=184756$ basis states linked with ring connectivity. In the sampling protocol, we do not restrict ourselves to feasible solutions. Instead, we randomly select any possible solution from the full $2^n$ solution space, including infeasible ones. Additionally, for performance evaluation, non-feasible decisions, i.e., states with Hamming weights not equal to $k$, have a quality of 0. 

We see on Fig. \ref{fig:portfolio_optimization} that the probability of measuring the optimal solution is the highest at the end of the evolution with $\mathbb{P}_0[\psi_T]\approx 0.06$, followed closely by ranking 1 decision with $\mathbb{P}_1[\psi_T]\approx 0.05$. Moreover, the probability of measuring decisions with ranking in the top 5\% of all rankings reaches approximatively $0.85$. The initial distribution quality is $\mathbb{E}[\psi_0]\approx 0.5$, which indicates a good balance between high and low quality solutions. Furthermore, we observe that the time evolution $T\approx 28056$ is significantly higher than it is for MaxCut and MIS since the energy gaps are significantly lower for the portfolio optimization instances, leading to longer transfer time. We can also observe that the initial ranking distribution is uniform for portfolio optimization with $184756$ different rankings for the feasible decisions. The evolution induced by \textsc{SamBa--GQW} leads to a high quality distribution where the first two highest measurement probabilities respectively correspond to ranking $0$ and $1$.

\begin{figure*}
    \centering
	\includegraphics[scale=0.5]{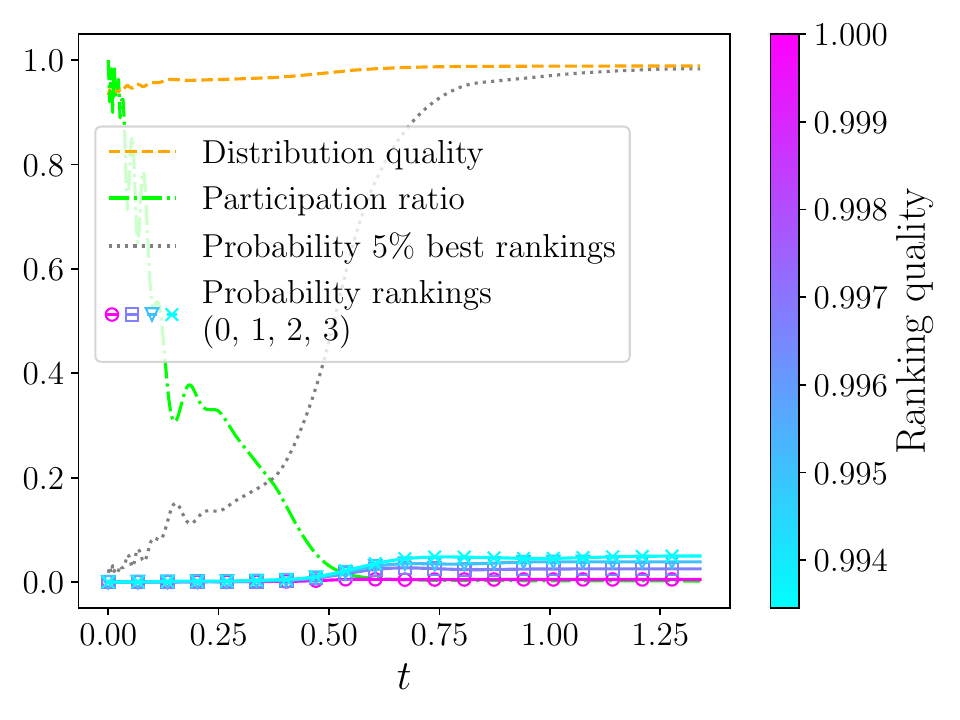}
    	\includegraphics[scale=0.5]{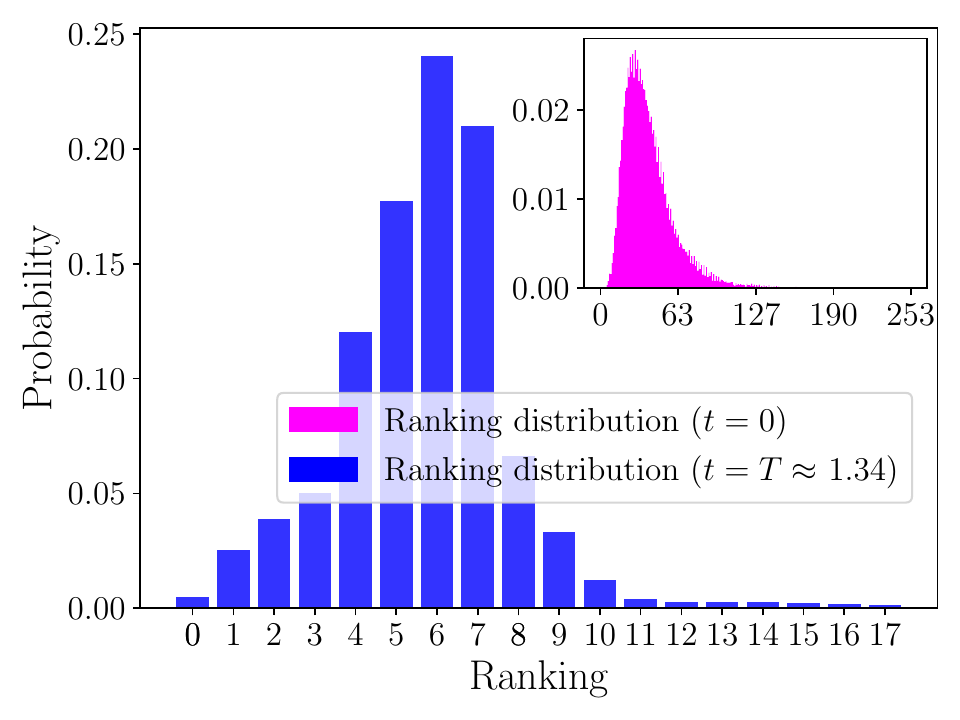}
    \caption{\textsc{SamBa--GQW} Hamiltonian simulations on LABS for $n=20$ qubits with a sampling of $q=n^2$ states.}
    \label{fig:labs}
\end{figure*}

\subsubsection{LABS}

\paragraph{Hamiltonian simulations.}
We present the results for LABS on $n=20$ qubits, where we minimize the energy (instead of maximizing the merit factor). As seen on Fig. \ref{fig:labs}, the initial distribution quality is very high with $\mathbb{E}[\psi_0]\approx 0.95$. This phenomenon indicates that most of the $2^n$ states have a very high quality and may impact the difficulty of solving the problem since it is easy to be stuck on local minima close to the optimal solutions. We observe that after the short time evolution $T\approx 1.4$, the optimal solutions do not have the highest measurement probabilities. However, the probability of measuring states with ranking in the top 5\% of all rankings almost reaches $1$, which indicates good performance of \textsc{SamBa--GQW} since the algorithm finds almost optimal solutions with high probabilities by only sampling $q=n^2$ states. Note that, in general, even restricting the search to the top $5\%$ of rankings may still correspond to an exponentially large search space. However, in the present case, the participation ratio indicates that the quantum state converges toward a well-localized distribution, suggesting that only a small subset of configurations carries a significant fraction of the probability mass.
 Lastly, the participation ratio decreases rapidly as the quantum state localizes during the evolution. We observe this behavior on Fig. \ref{fig:labs} since only $18$ rankings among $254$ have a measurement probability higher than $10^{-3}$ after the evolution induced by the algorithm. The Gaussian ranking distribution is strongly shifted to low rankings, i.e. high quality decisions, as suggested by the initial distribution quality. Furthermore, after the evolution, four high quality rankings have a measurement probability higher than $0.1$ with $\mathbb{P}_6[\psi_T]\approx 0.24$, $\mathbb{P}_7[\psi_T]\approx 0.21$, $\mathbb{P}_5[\psi_T]\approx 0.18$ and $\mathbb{P}_4[\psi_T]\approx 0.12$. The initial measurement probabilities of these rankings are respectively $0.0008, 0.0017, 0.0004,$ and $0.0001$, making the final probabilities a sign of good performance of \textsc{SamBa--GQW}. The cumulated probability of measuring rankings  $r \in \{4,5,6,7\}$ is very high since it reaches $0.75$. The optimal solutions have a significantly lower measurement probability with $\mathbb{P}_0[\psi_T]\approx 0.005$, however at $t=0$ this probability is $\mathbb{P}_0[\psi_0]\approx 7\times 10^{-6}$. Thus, it is fair to consider the final measurement probability a sign a good performance of the algorithm since it multiplied the initial optimal probability by approximately $656$ by sampling only $q=n^2$ states among $2^n$. Moreover, since the quantum state is almost localized at $t=T$ since only $18$ rankings among $254$ have a measurement probability higher than $10^{-3}$, including optimal solutions, one can perform several measurements in the computational basis and classically select the state with lowest energy. This way, one can increase the likeliness of recovering optimal solutions.

\paragraph{Quantum circuit simulations.}

\begin{figure*}
    \centering
        \includegraphics[width=0.8\linewidth]{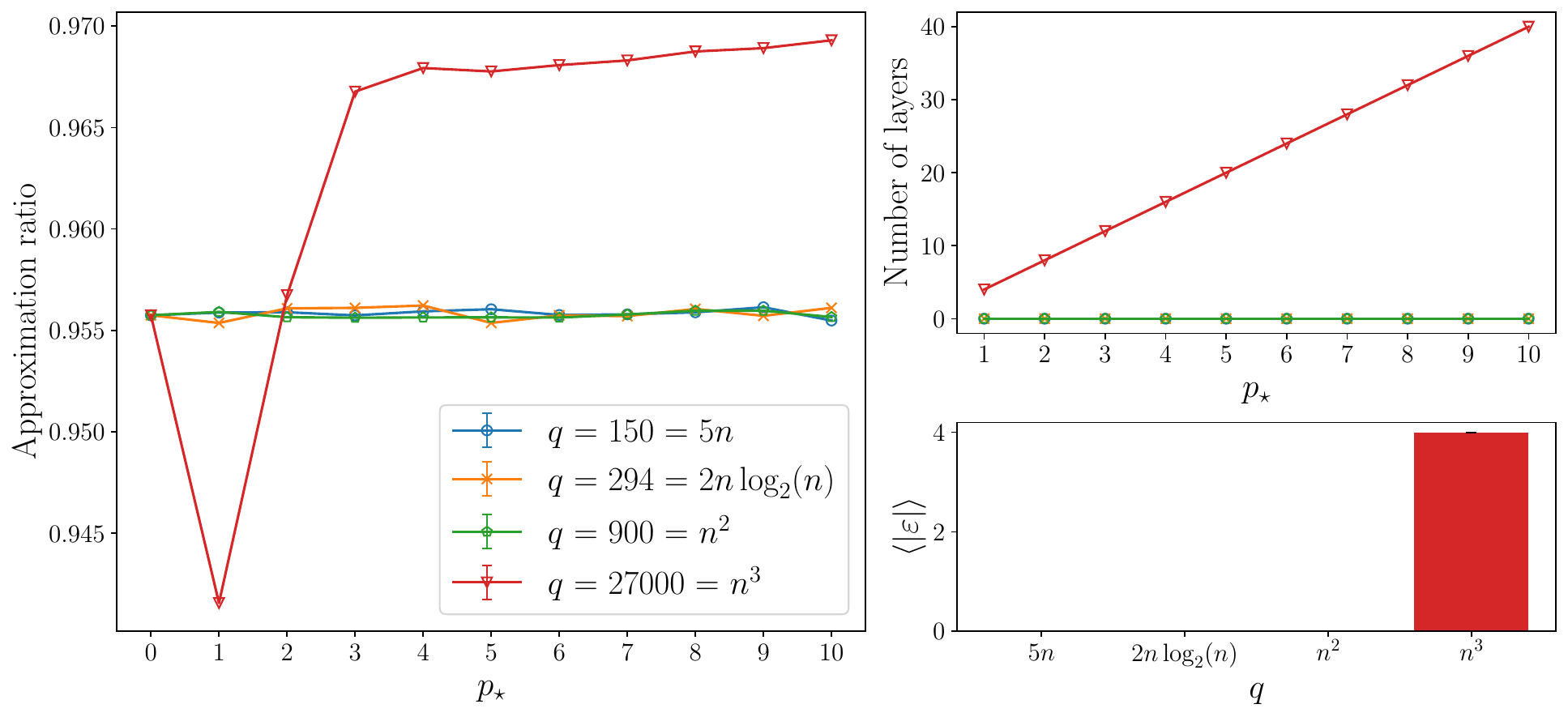}
        \caption{\textsc{SamBa--GQW} quantum circuit simulations for LABS for $n=30$ qubits with different sampling values $q\in\{5n,2n\log_2(n),n^2,n^3\}$.}
        \label{fig:samba_circuit_labs_30}
\end{figure*}

\begin{figure}
    \centering
        \includegraphics[width=0.85\linewidth]{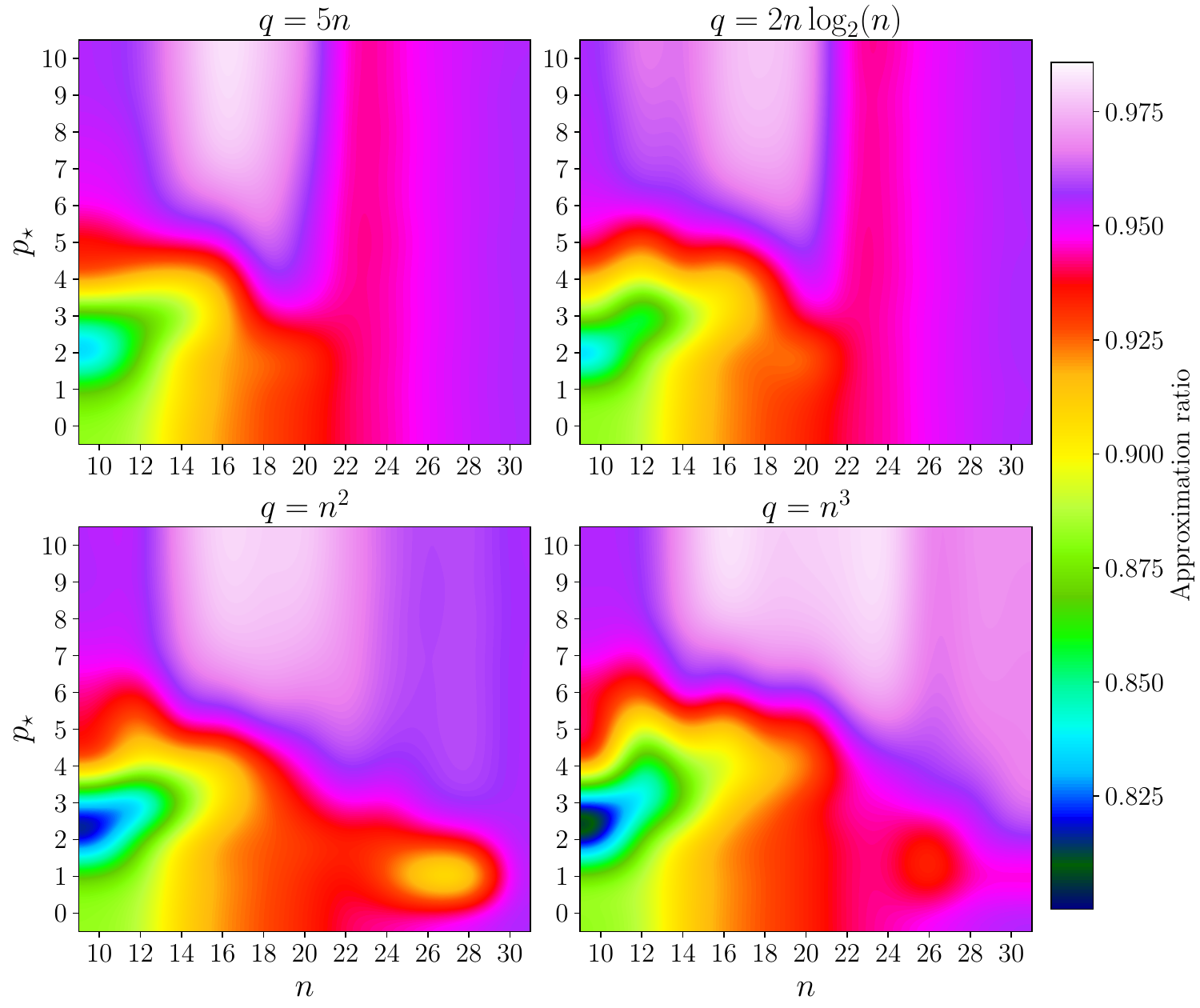}
        \includegraphics[width=0.85\linewidth]{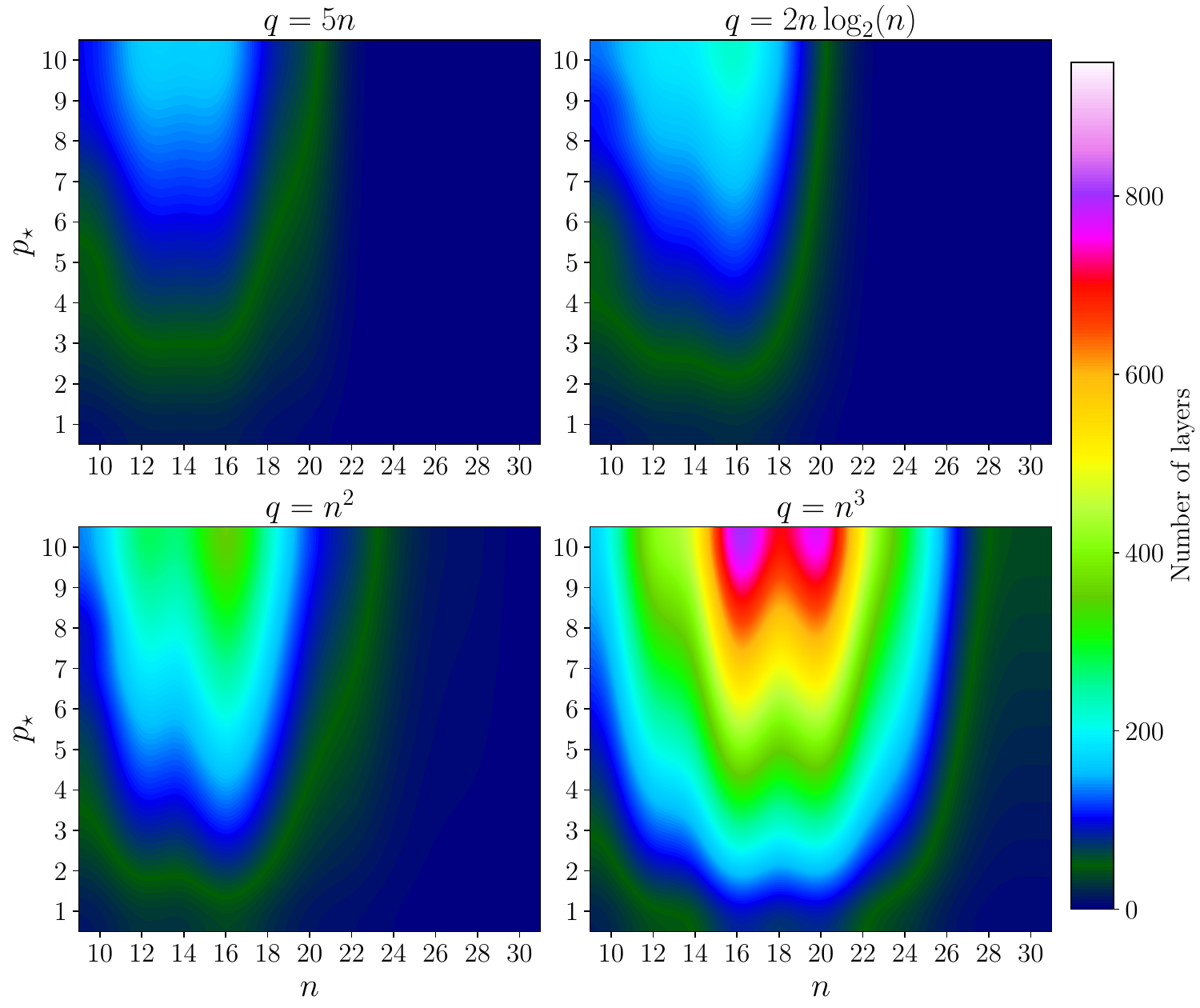}
        \caption{Approximation ratio (top) and number of layers in the circuit (bottom) as a function of the number of qubits $n\in[10,30]$ and the discretization parameter $p_\star\in[1,10]$ for LABS with different sampling values $q\in\{5n,2n\log_2(n),n^2,n^3\}$.}
        \label{fig:samba_circuit_labs_n}
\end{figure}

We present the results for LABS on $n=30$ qubits in Fig.~\ref{fig:samba_circuit_labs_30}. We observe that only the cubic sampling regime, $q=n^3$, leads to variations in the approximation ratio as the circuit depth increases. This is explained by the fact that, for $q\in\{5n,2n\log_2(n),n^2\}$, no energy gaps were sampled, resulting in a quantum circuit composed only of Hadamard gates applied to each qubit. In contrast, the protocol with $q=n^3$ successfully extracted four energy gaps. This behavior arises because the sampling protocol (see Algorithm~\ref{alg:sampling}) considers only positive energy gaps, as shown in Appendix~\ref{app:sampling_absolute}, taking the absolute values of the gaps eliminates this zero-sampled-energy phenomenon. Moreover, we illustrate the scaling of the number of sampled energy gaps as a function of $n$ in Appendix~\ref{app:sampled_energy} for MaxCut, MIS and LABS. For the case $q=n^3$, we observe a slight improvement in the approximation ratio with increasing circuit depth. We also note that the initial approximation ratio is already very high, since the LABS problem possesses a large proportion of high-quality solutions. From Fig.~\ref{fig:samba_circuit_labs_n}, we observe that increasing the number of samples becomes necessary to improve the approximation ratio as $n$ grows.

\subsubsection{MAX-$k$-SAT}

\begin{figure*}[ht]
    \centering

    \begin{subfigure}[b]{0.48\linewidth}
        \centering
        \includegraphics[width=0.49\linewidth]{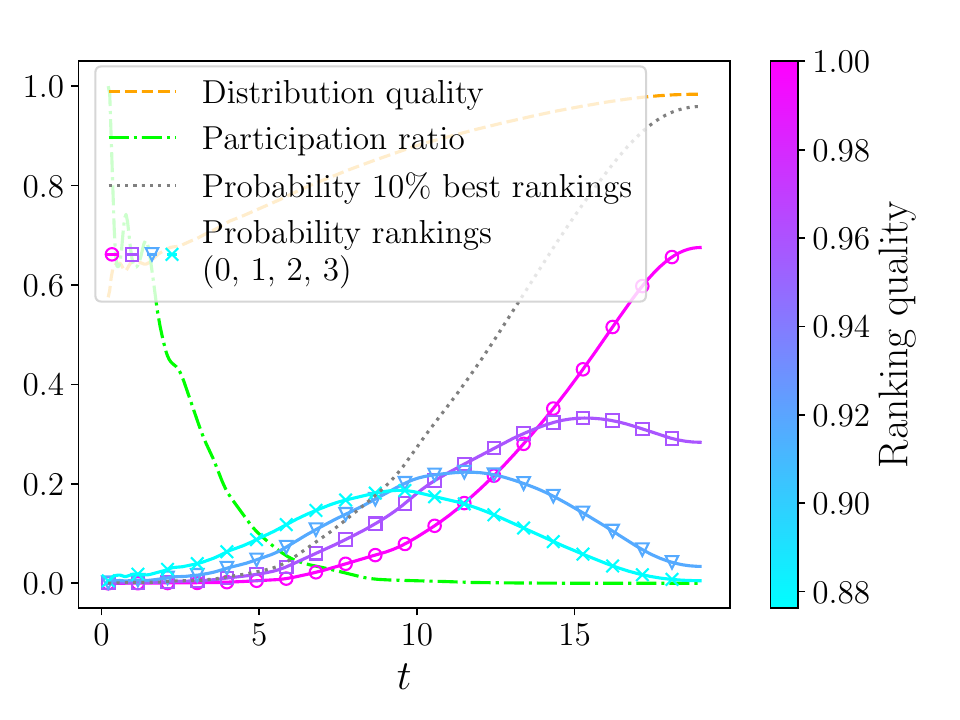}
        \includegraphics[width=0.49\linewidth]{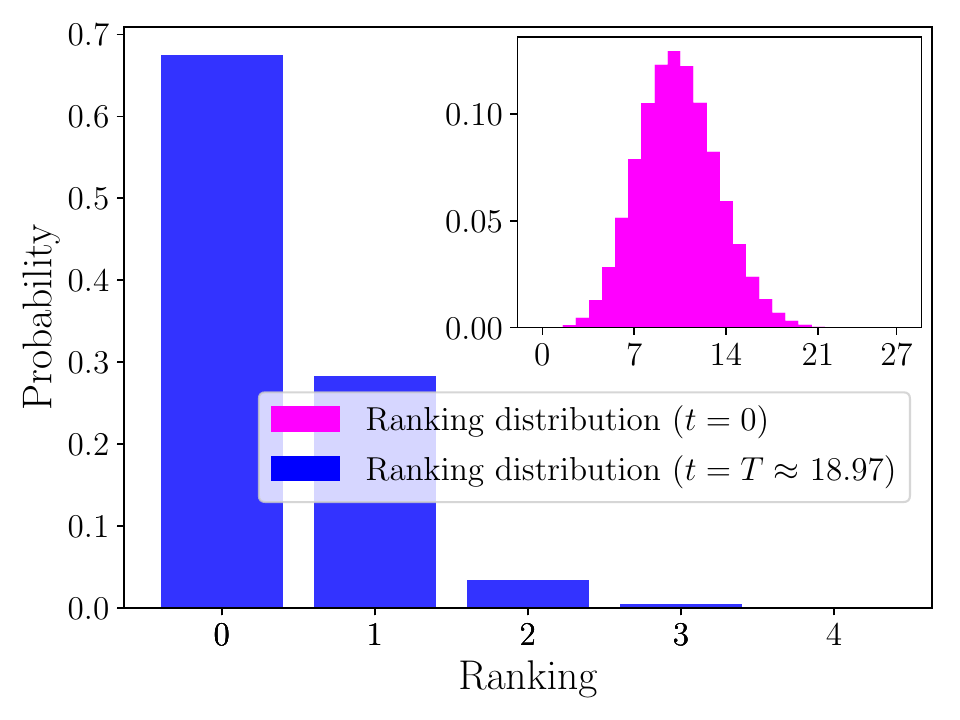}
        \caption{$k=3$}
    \end{subfigure}
    \hfill
    \begin{subfigure}[b]{0.48\linewidth}
        \centering
        \includegraphics[width=0.49\linewidth]{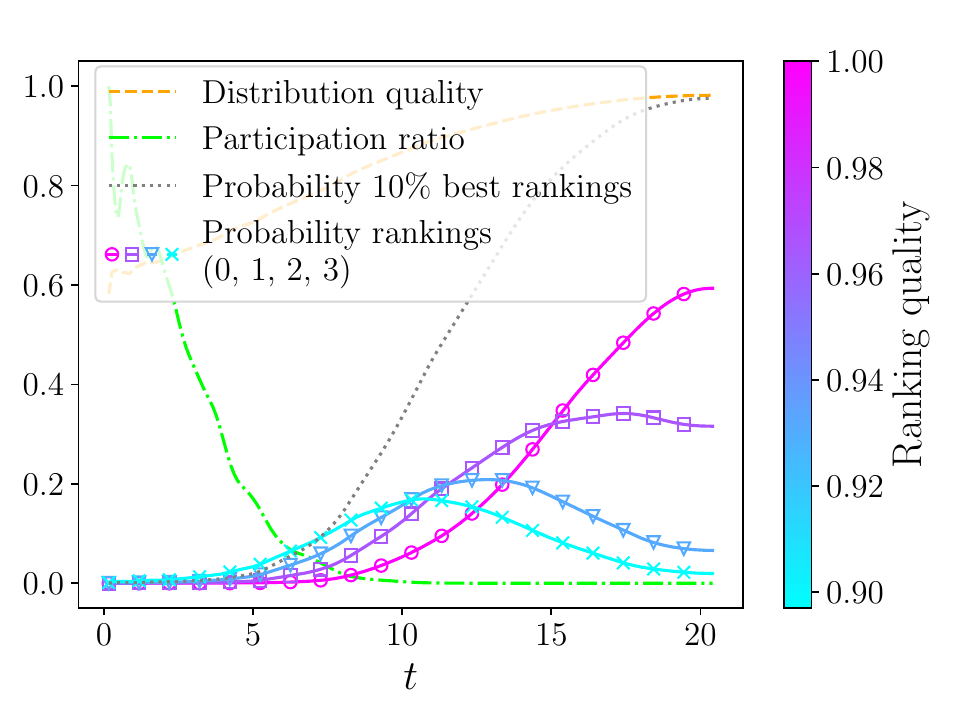}
        \includegraphics[width=0.49\linewidth]{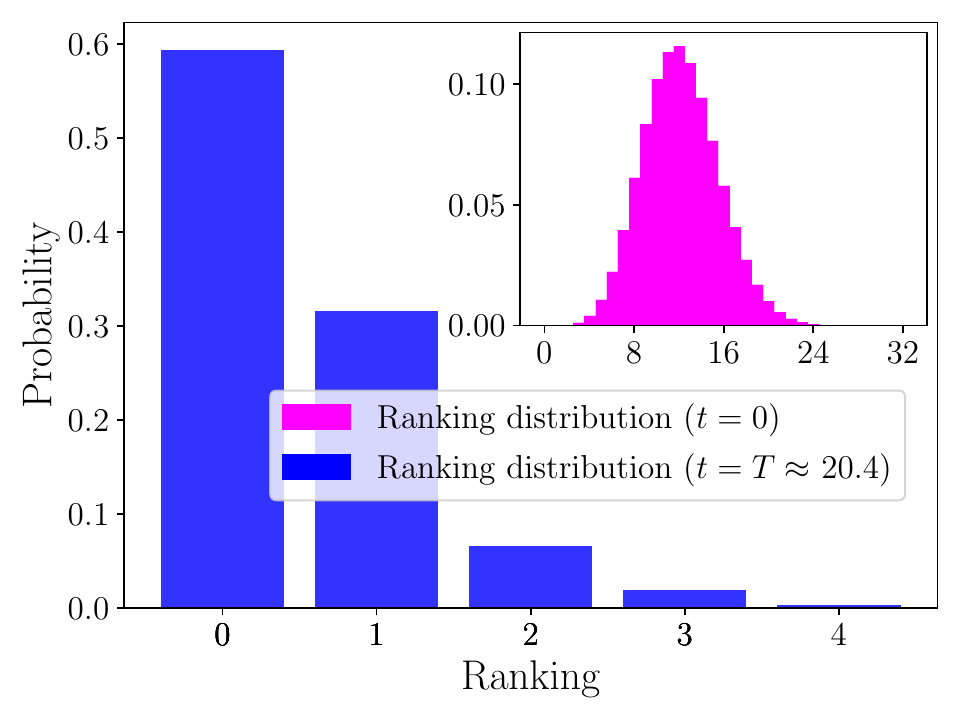}
        \caption{$k=4$}
    \end{subfigure}

    \begin{subfigure}[b]{0.48\linewidth}
        \centering
        \includegraphics[width=0.49\linewidth]{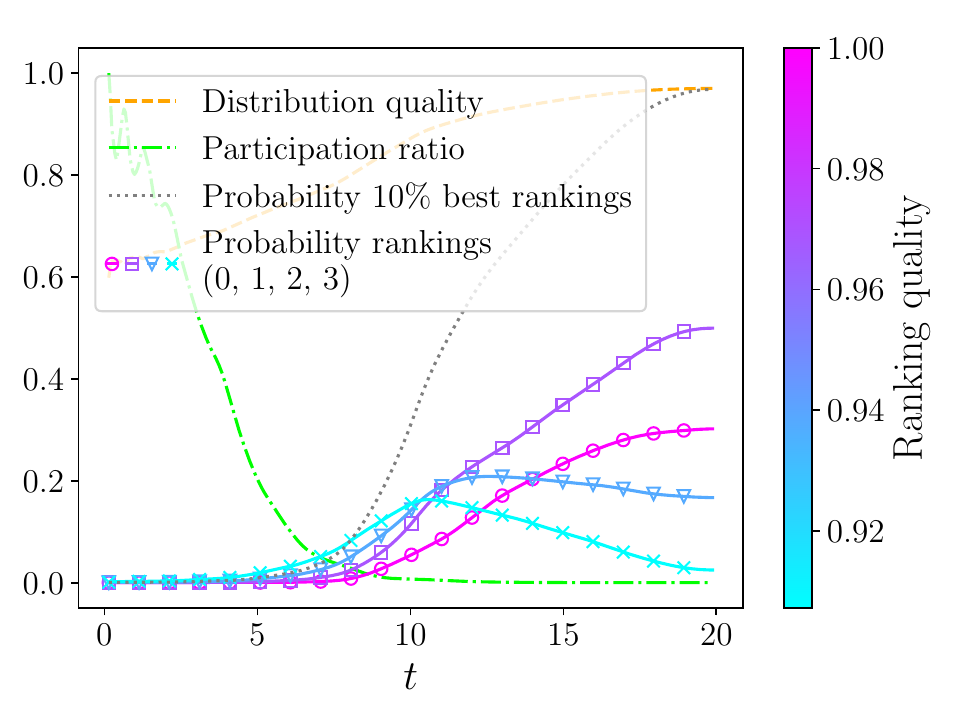}
        \includegraphics[width=0.49\linewidth]{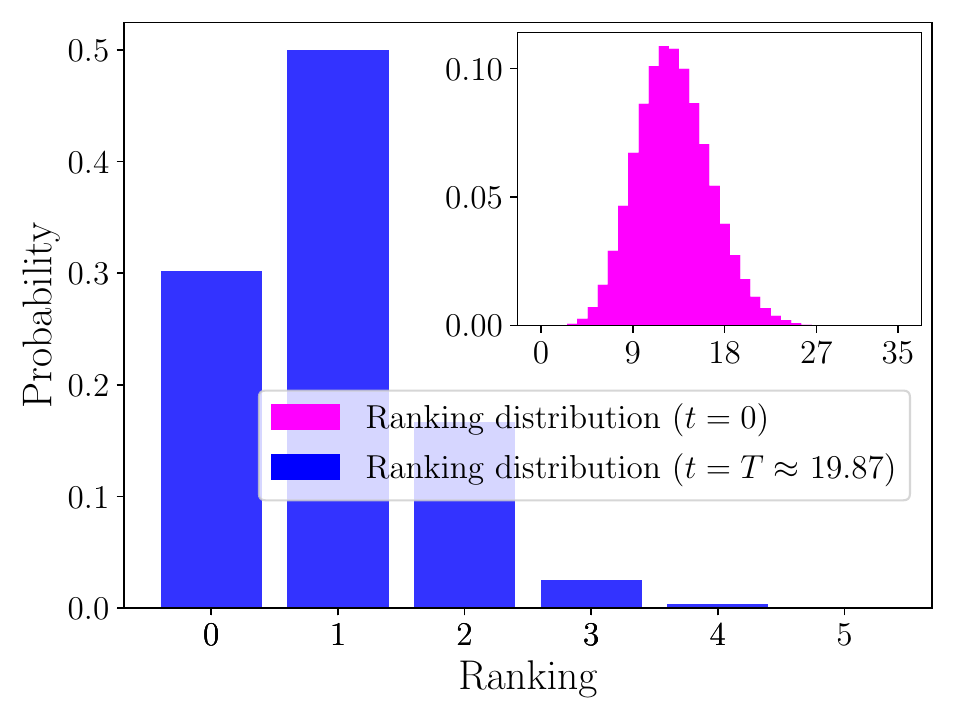}
        \caption{$k=5$}
    \end{subfigure}
    \hfill
    \begin{subfigure}[b]{0.48\linewidth}
        \centering
        \includegraphics[width=0.49\linewidth]{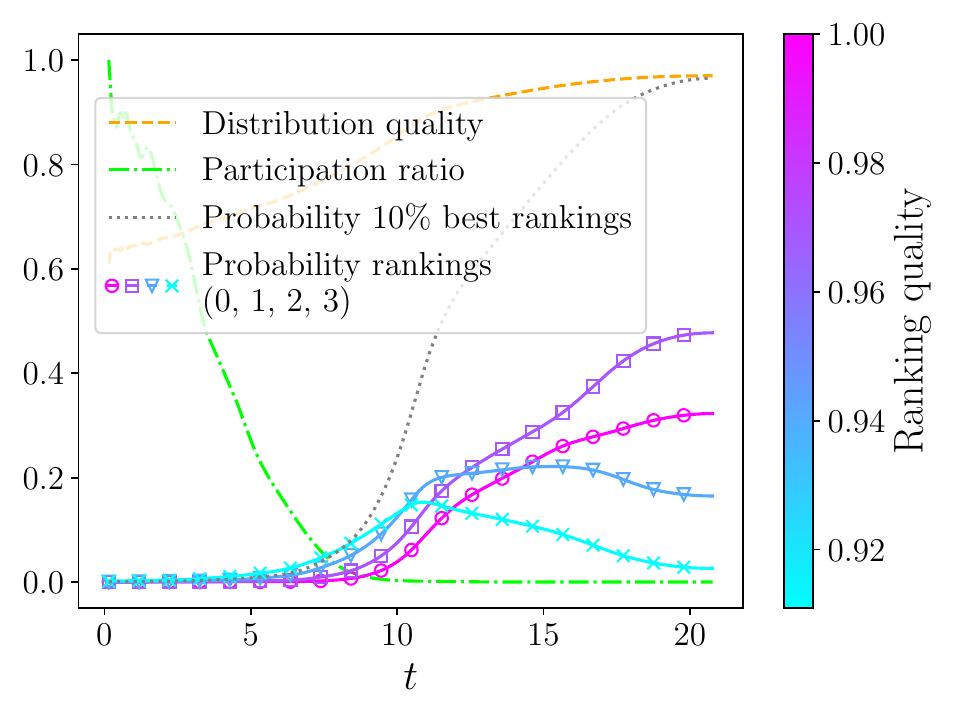}
        \includegraphics[width=0.49\linewidth]{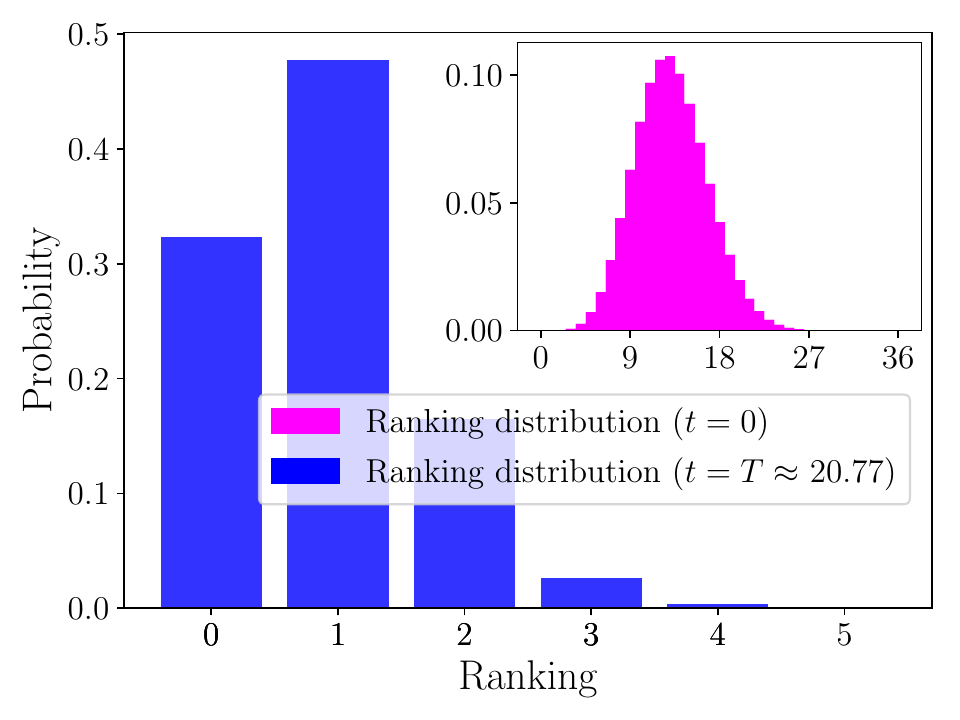}
        \caption{$k=6$}
    \end{subfigure}

    \begin{subfigure}[b]{0.48\linewidth}
        \centering
        \includegraphics[width=0.49\linewidth]{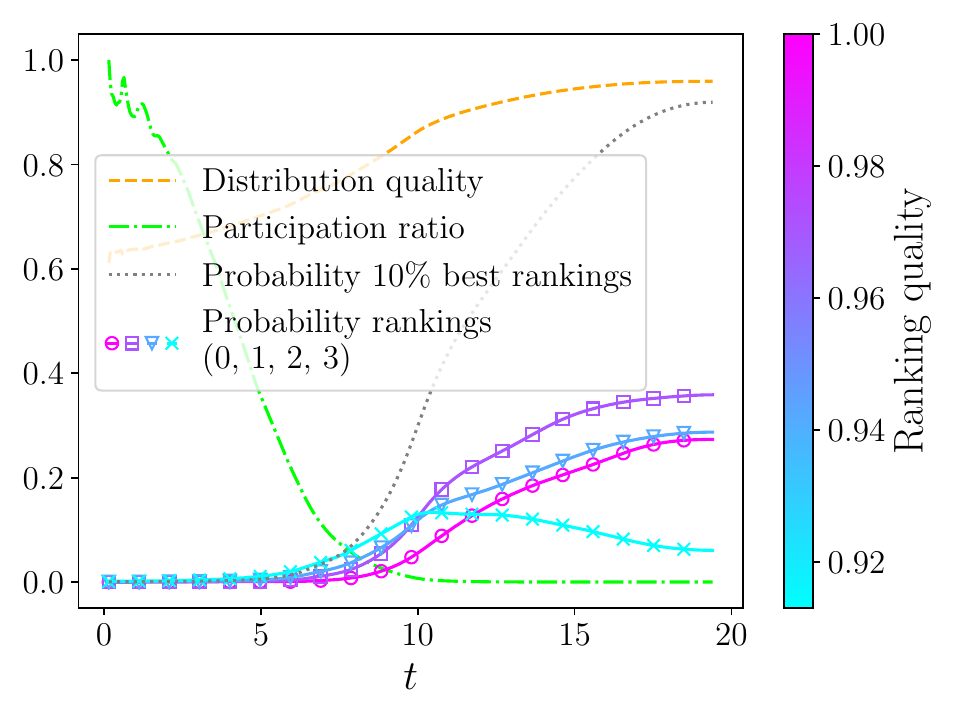}
        \includegraphics[width=0.49\linewidth]{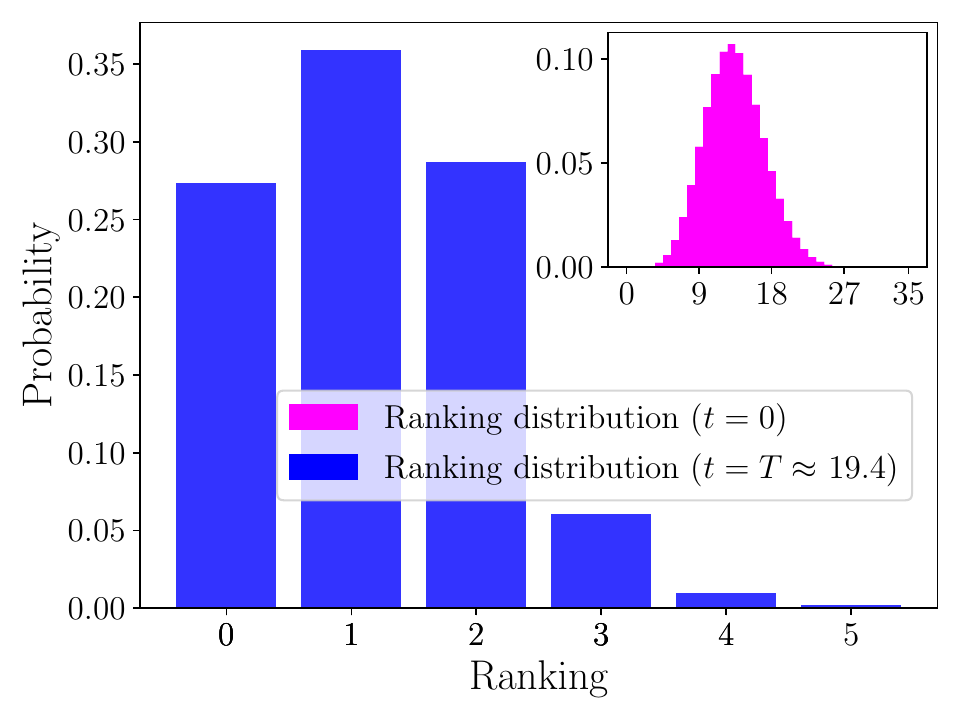}
        \caption{$k=7$}
    \end{subfigure}
    \hfill
    \begin{subfigure}[b]{0.48\linewidth}
        \centering
        \includegraphics[width=0.49\linewidth]{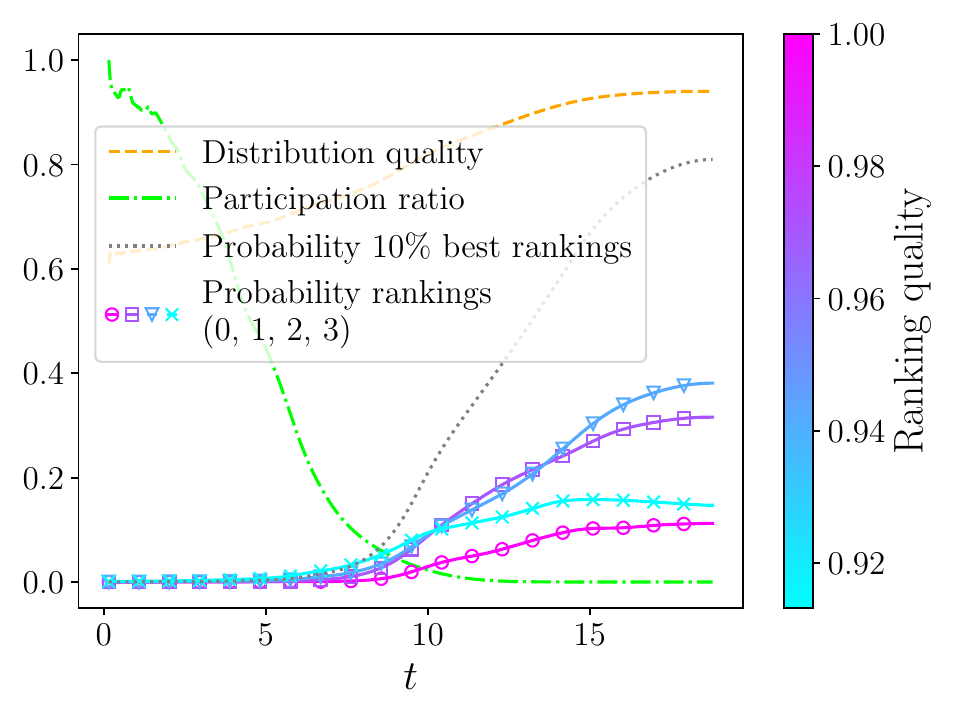}
        \includegraphics[width=0.49\linewidth]{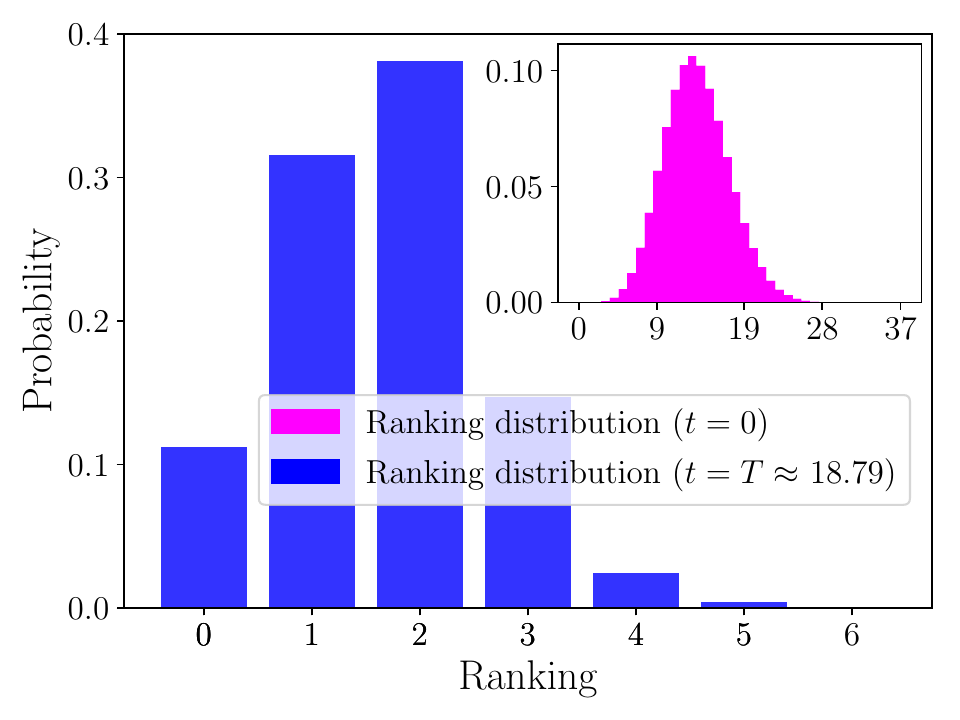}
        \caption{$k=8$}
    \end{subfigure}

    \begin{subfigure}[b]{0.48\linewidth}
        \centering
        \includegraphics[width=0.49\linewidth]{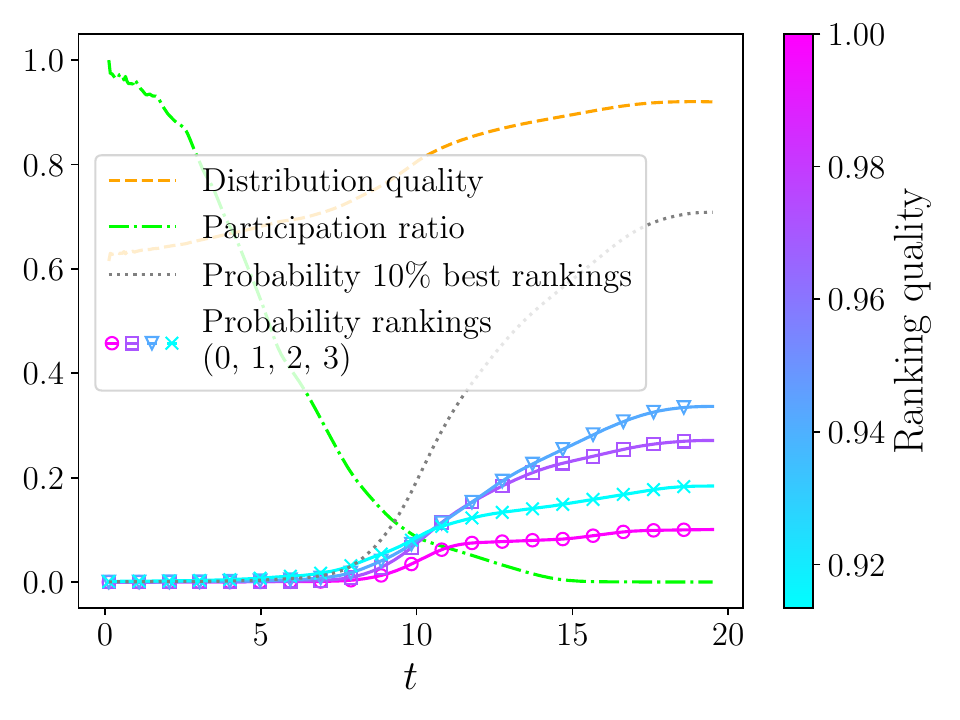}
        \includegraphics[width=0.49\linewidth]{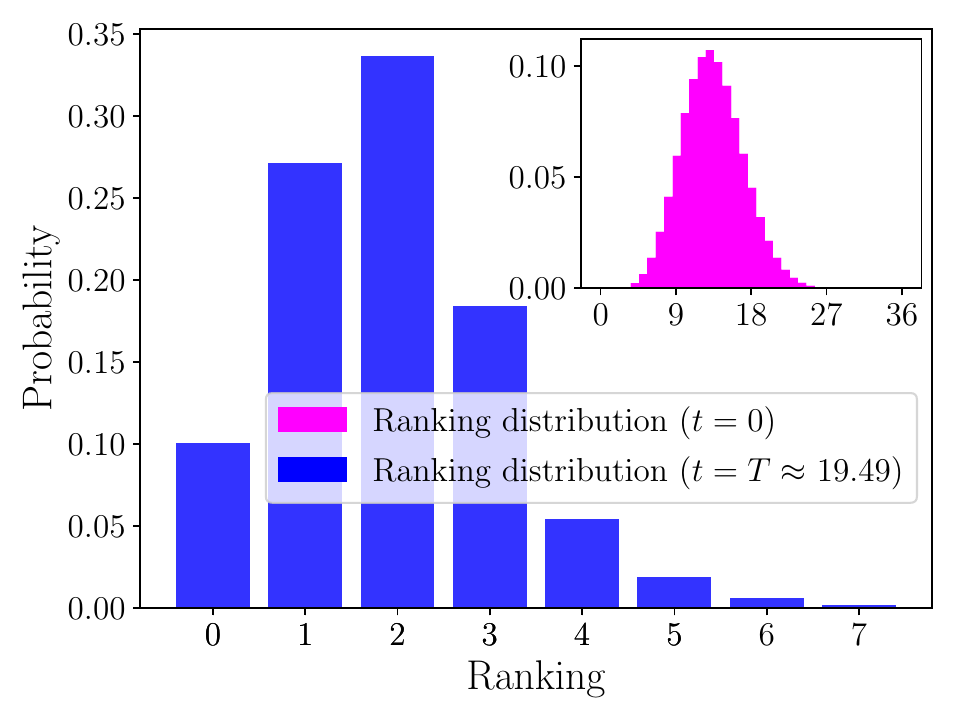}
        \caption{$k=9$}
    \end{subfigure}
    \hfill
    \begin{subfigure}[b]{0.48\linewidth}
        \centering
        \includegraphics[width=0.49\linewidth]{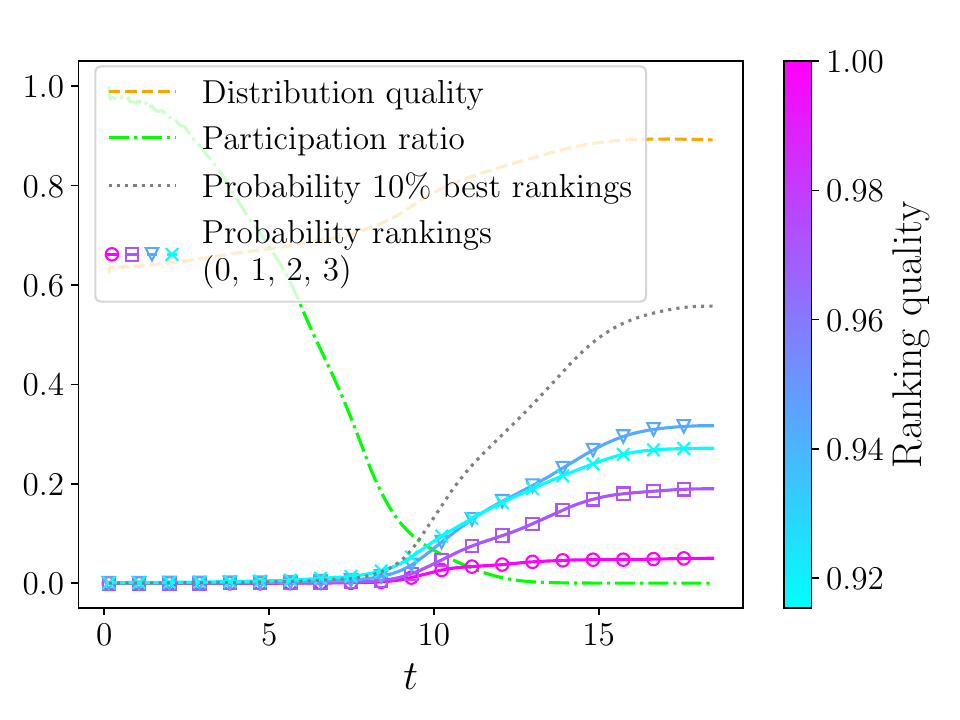}
        \includegraphics[width=0.49\linewidth]{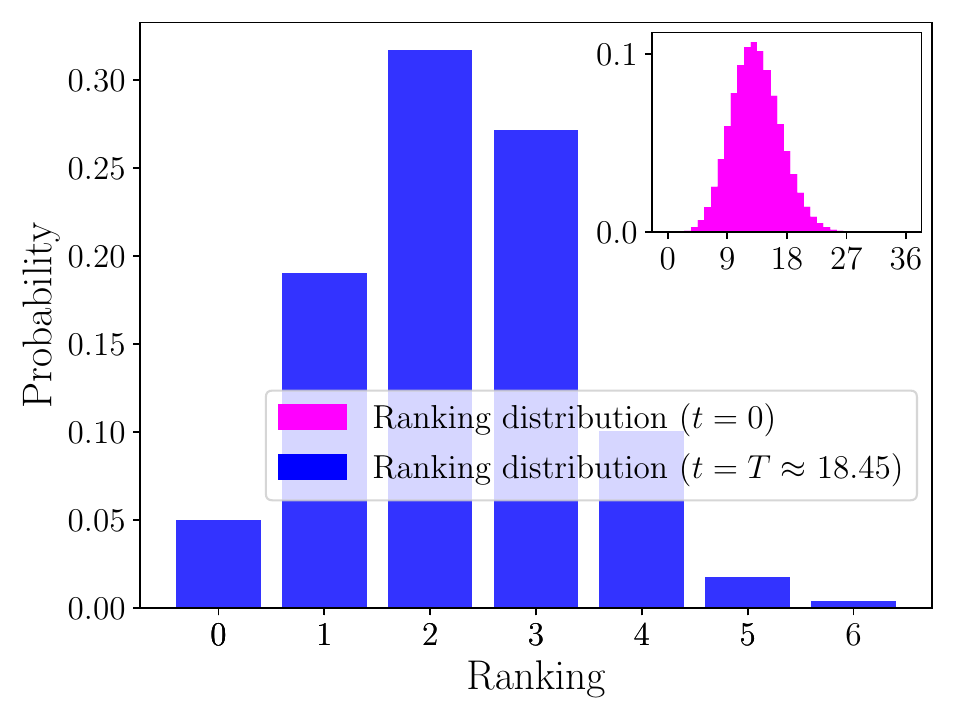}
        \caption{$k=10$}
    \end{subfigure}
    
    \caption{\textsc{SamBa--GQW} Hamiltonian simulations on MAX-$k$-SAT for $n=20$ qubits with a sampling of $q=n^2$ states. We present the results for degrees $k\in\{3,4,5,6,7,8,9,10\}$, where the degree corresponds to the number of literal per clause. The number of clause is set to $m=\lfloor\alpha_k n\rfloor$ where the values of $\alpha_k$ are shown in Table 10 of Ref. \cite{Campbell2019applyingquantum}.}
    \label{fig:maxksat}
\end{figure*}

\begin{figure*}
    \centering
	\includegraphics[scale=0.5]{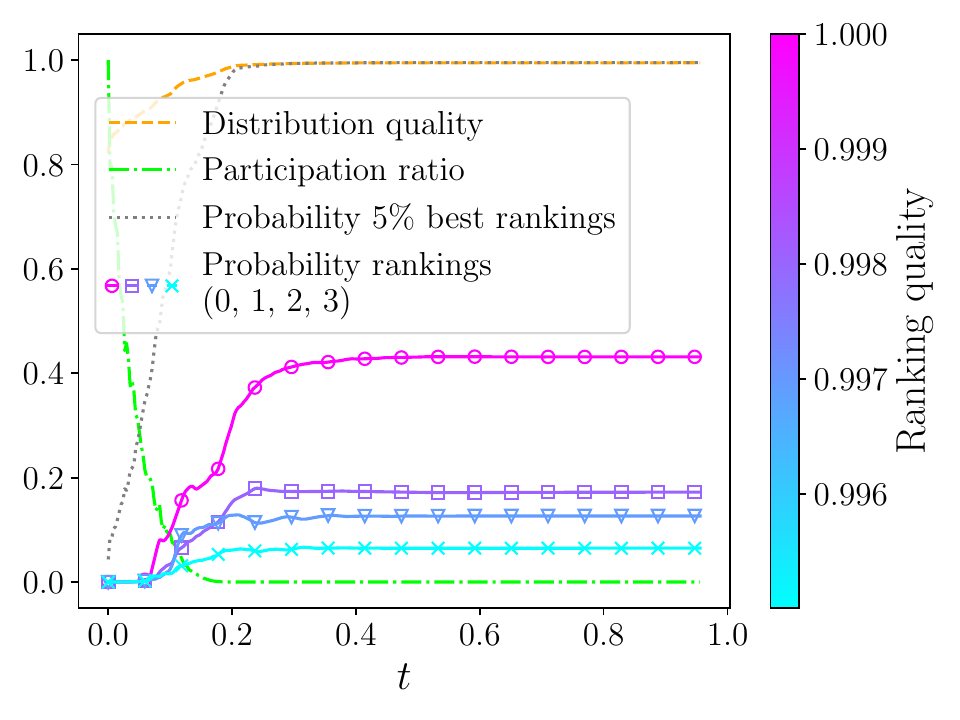}
	\includegraphics[scale=0.5]{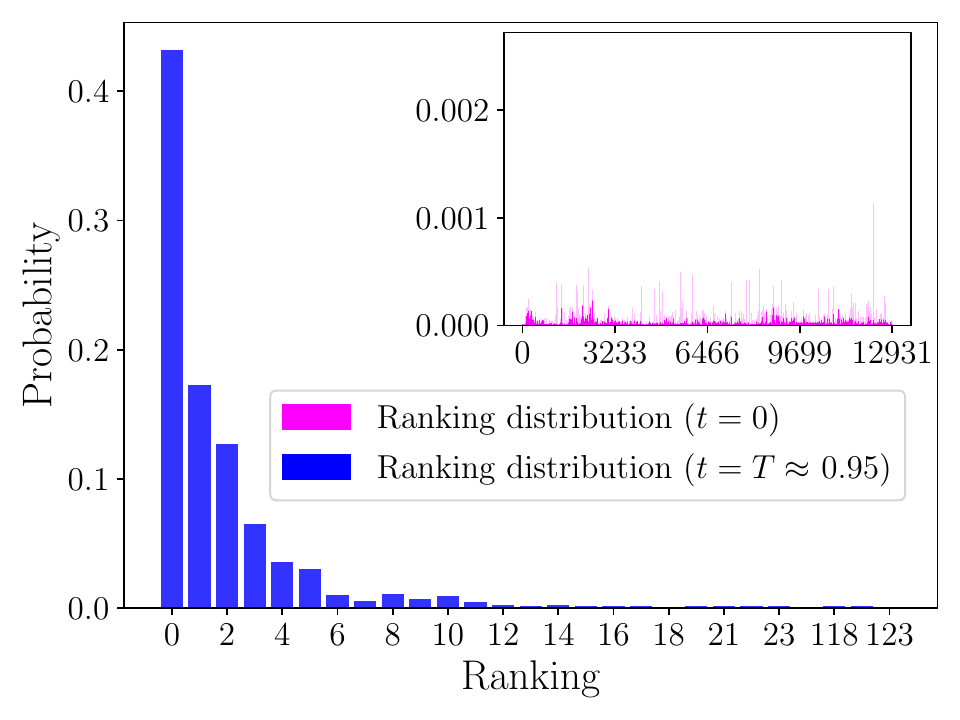}
	\caption{\textsc{SamBa--GQW} Hamiltonian simulations on complete graphs with weight bounded by $1\leq w_{ij}\leq 100$ distances on TSP for $n=18$ qubits with a sampling of $q=n^2$ states.}
    \label{fig:long_tsp}
\end{figure*}

We generate random instances of MAX-$k$-SAT by randomly drawing $m$ clauses uniformly from the set of all possibles clauses $\Phi_{k-\text{SAT}}$ with 
$|\Phi_{k-\text{SAT}}|=2^k {n \choose k}$.
The hardness of MAX-$k$-SAT strongly depends on the clause-to-variable ratio $\alpha=m/n$. There exists a threshold $\alpha_c$, in which no known polynomial-time algorithm can produce an $\epsilon$-approximate solution with high probability~\cite{PhysRevLett.76.3881}. This threshold is hard to determine for any value of $k$ but can be estimated using cavity method \cite{https://doi.org/10.1002/rsa.20090,10.1145/2591796.2591822}. In practice, we run experiments for $k\in[3,10]$ and we set the number of clauses to $m=\lfloor\alpha_k n\rfloor$ where the values of $\alpha_k\equiv \alpha_c$ are shown in Table 10 of Ref. \cite{Campbell2019applyingquantum}. We present the results for MAX-$k$-SAT on $n=20$ qubits for degrees $k\in\{3,4,5,6,7,8,9,10\}$ on Fig. \ref{fig:maxksat}. We observe that the evolution induced by \textsc{SamBa--GQW} leads to a measurement of optimal solutions with highest probabilities only for $k\in \{3,4\}$. As $k$ increases, optimal solutions still have a relevant measurement probability, and in the worst case the highest measurement probability is that of ranking 2, which indicates good performance. We notice that no matter the degree $k$, distribution quality starts around $\mathbb{E}[\psi_0]\approx 0.6$, and it reaches lower values $\mathbb{E}[\psi_T]$ as $k$ increases, which translates an increase in the problem difficulty. We observe that participation ratio decreases less rapidly as $k$ increases, indicating that the state vector requires more evolution time to reach a more localized state with the increase of problem's degree. We notice that evolution time $T$ does not seem to be impacted by the degree $k$ since it is approximatively short and constant with $T\approx 20$. This observation indicates that instances of different degree have energy gaps of similar magnitude. Looking at the initial Gaussian ranking distribution, we see that they are slightly shifted to the left towards higher quality decisions. Furthermore, the number of different rankings does not exceed $38$ which is quite low considering that we work on instances with $2^{20}$ decisions. 

The probability of measuring decisions in the top 10\% of rankings, reaches high values close to $1$ for $k\in\{3,4,5,6,7\}$. We observe a slight drop in performance for $k\in\{8,9,10\}$, since their respective measurement probabilities of the 10\% rankings are approximatively $0.8, 0.7,$ and $0.55$. However, since the number of different rankings does not exceed $38$, measuring the top 10\% of rankings is approximatively equivalent to only measuring the first four rankings. Thus, even if this probability decreases with the increases of the problem's degree $k$, the performance of SamBa are still great considering that in the worst case for $k=10$, the highest measurement probability is that of ranking $2$ with the highest probability $\mathbb{P}_2[\psi_T]\approx 0.32$. For every degree $k$, we observe that a sampling of only $q=n^2$ states for $n=20$ is enough to guide the walker properly to high quality decisions, and even directly to optimal solutions for $k\in\{3,4\}$. As the degree of the problem increases, it becomes harder to find optimal solutions, however, we still measure almost-optimal decisions with high probabilities. In such cases of higher $k$, optimal solutions could be recovered by performing several measurements and by efficiently selecting classically the best measured decision.

\subsubsection{Travelling Salesperson Problem}

\begin{figure*}
    \centering
	\includegraphics[scale=0.5]{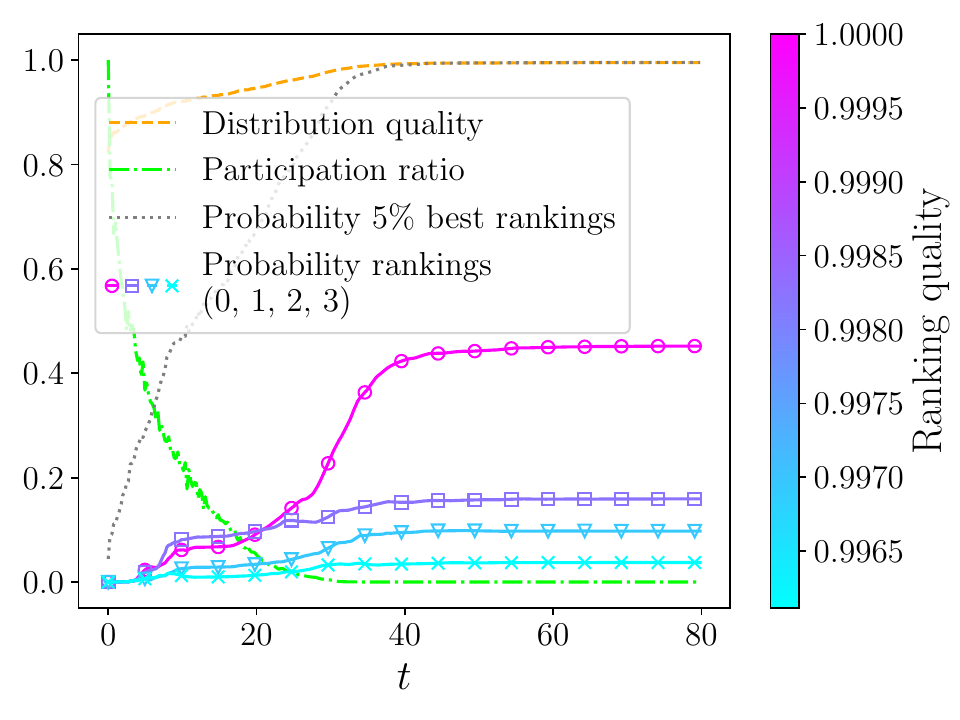}
	\includegraphics[scale=0.5]{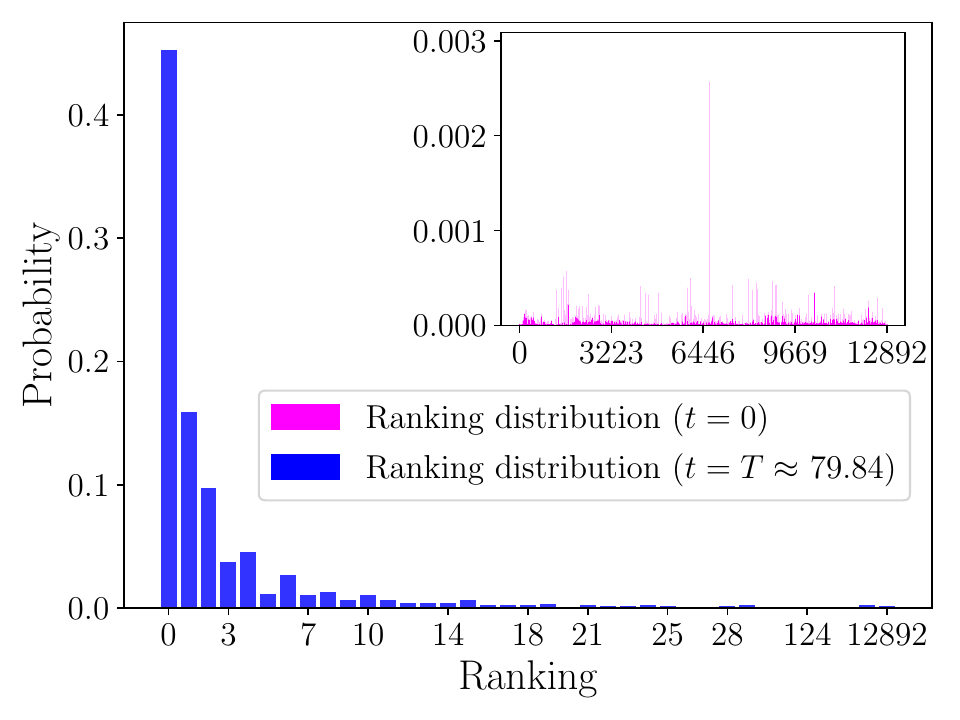}
	\caption{\textsc{SamBa--GQW} Hamiltonian simulations on complete graphs with weight bounded by $0\leq w_{ij}\leq 1$ distances on TSP for $n=18$ qubits with a sampling of $q=n^2$ states.}
    \label{fig:short_tsp}
\end{figure*}

We run TSP on complete graphs of $m=6$ cities (vertices), therefore we use $n=m\lceil\log_2(m)\rceil=18$ qubits for the HUBO encoding. We generate instances with long ($1\leq w_{ij}\leq 100$) and short ($0\leq w_{ij}\leq 1$) distances between the cities. This difference has a direct impact on the values of the energy gaps. With long distances we have short evolution times since gaps are quite high and with short distances the gaps get a lot smaller making the evolution time longer. We set the parameters of Eq. \eqref{eq:tsp} to $\mu=1$ and $\lambda=\gamma=2\mu\max_{(i,j)\in E}w_{ij}$. We obtain similar results for long and short distance instances with a significant difference in the evolution time as shown on Figs. \ref{fig:long_tsp} and \ref{fig:short_tsp}. The long distance instances require an evolution time $T\approx 1$ against $T\approx 80$ for the short distance instances. This comes from the difference in energy gaps which are lower in the second case, leading to longer evolution time. For both long and short distance instances, the optimal solutions have the highest measurements probabilities, reaching $\mathbb{P}_0[\psi_T]\approx 0.4$ with slightly better results for short distances. The probability of measuring the top 5\% rankings reaches $1$ at $t\approx 0.2$ (resp. $t\approx 40$) for long (resp. short) distance instances, and the participation ratio decreases fast to approach $1/2^n$, indicating a localized state early in the evolution. The initial ranking distribution shown on Figs. \ref{fig:long_tsp} and \ref{fig:short_tsp} are spread among a higher amount of rankings than the previous studied problems. Long (resp. short) distance instances have $12932$ (resp. $12893$) different rankings, and the evolution induced by \textsc{SamBa--GQW} leads the walker to the optimal solutions with high success probability even if the TSP problem is encoded as a HUBO.

\section{Comparison with the state of the art}\label{sec:sota}

\subsection{Guided quantum walk}

The authors of Ref. \cite{PhysRevResearch.6.013312} introduced the guided quantum walk (GQW) as an heuristic for combinatorial optimization. They follow the time evolution of Eq. \eqref{eq:hamiltonian} and use a classical optimizer to learn the optimal hopping rate $\Gamma$. Knowing that this function must be monotonically decreasing and positive \cite{PhysRevResearch.6.013312}, they reduce the number of learning parameters by using cubic Bézier curves. Such curves are defined by four control points $p_i=(x_i,y_i)^{\top}$ with $i\in \{0,1,2,3\}$. Their coordinates reads:
\begin{equation}
    \begin{split}
        x(\tau) &= x_0(1-\tau)^3+3x_1(1-\tau)^2\tau+3x_2(1-\tau)\tau^2\\&+x_3\tau^3 \\
        y(\tau) &= y_0(1-\tau)^3+3y_1(1-\tau)^2\tau+3y_2(1-\tau)\tau^2\\&+y_3\tau^3.
    \end{split}
\end{equation}
To ensure strict decrease conditions, they set $p_0=(0,1)$ and $p_1=(1,0)$, i.e. $x_0=y_3=0$ and $y_0=x_3=1$. Moreover, they define the hopping rate function as:
\begin{equation}
    \Gamma(t)=y(t)\cdot 10^{a\cdot\alpha}+(1-y(t))\cdot 10^{b\cdot\beta},
\end{equation}
with $(\alpha,\beta)\in[0,1]^2$ and they set $a=2$ and $b=-3$. Thus, the parameters $\alpha$ and $\beta$ control the extreme values of $\Gamma$, and the coordinates of points $p_1$ and $p_2$ control the shape of the curve. The authors argue that if $x_0<x_1$ and $x_2<x_3$, inserting the solution of $t=x(\tau)$ (for $\tau(t)$) in $y$ gives a mapping from two to one dimension. Therefore, one has to fine-tune the six hyperparameters $\vec{\theta}=(x_1,y_1,x_2,y_2,\alpha,\beta)$ to approach the optimal hopping rate $\Gamma$. However, the limitation of this method is the emergence of exponential scalings in the classical learning of $\vec{\theta}$ \cite{PhysRevResearch.6.013312}.

We compare GQW and \textsc{SamBa--GQW} on unweighted and weighted MaxCut, as well as on unweighted MIS and LABS. The problem instances, described in Sec.~\ref{subsec:results}, have even sizes $n \in [10, 20]$. As in the previous section, we proceed by Hamiltonian simulation. The classical optimization of Bézier parameters $\vec{\theta}$ is done with $100$ iterations of Nelder-Mead method with bounds $\vec{\theta}\in [0,1]^6$. We run the GQW evolution for the same evolution time $T$ as \textsc{SamBa--GQW}, and we select $q=n^2$ for the sampling protocol. 

\begin{figure}
    \centering
        \includegraphics[width=1.\linewidth]{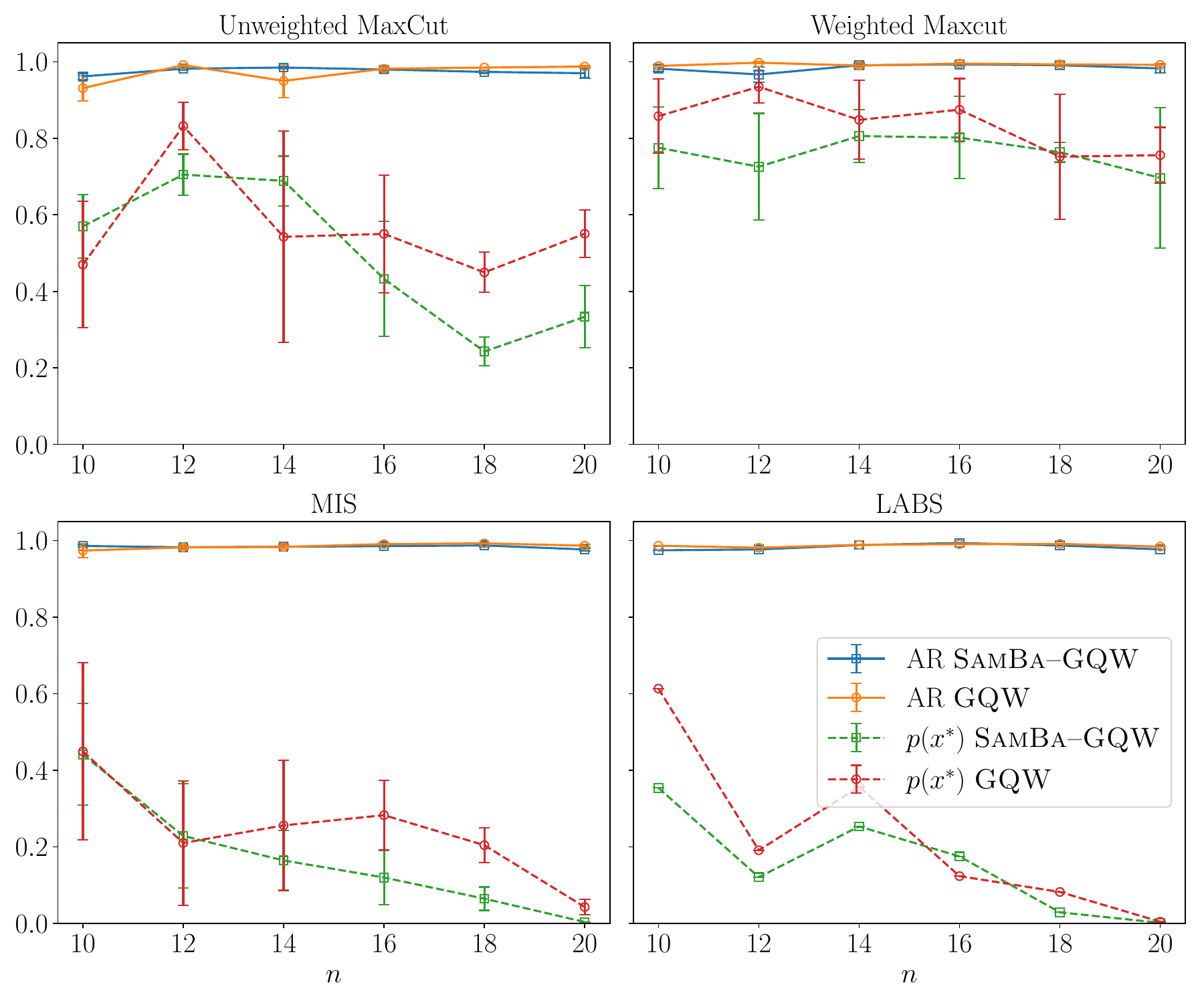}
        \caption{Approximation ratio (AR) and probability of measuring optimal solutions as a function of the number of qubits for \textsc{SamBa--GQW} and GQW.}
        \label{fig:gqw_ratios_samba}
\end{figure}

The results are presented in Fig.~\ref{fig:gqw_ratios_samba}. For all considered problems, both methods achieve very high approximation ratios, with GQW consistently attaining slightly better values. Regarding the probability of measuring optimal solutions, the variational GQW also yields higher probabilities, although the differences remain relatively small. This advantage can be attributed to the use of a classical optimizer in GQW to determine the hopping-rate schedule that steers the evolution toward high-quality states, whereas \textsc{SamBa--GQW} does not rely on such optimization. The probability of measuring optimal solutions is notably high for both unweighted and weighted MaxCut across all instance sizes. However, for MIS and LABS, it decreases significantly and becomes close to zero as the system size approaches $n=20$.

\begin{figure}
    \centering
        \includegraphics[width=1.\linewidth]{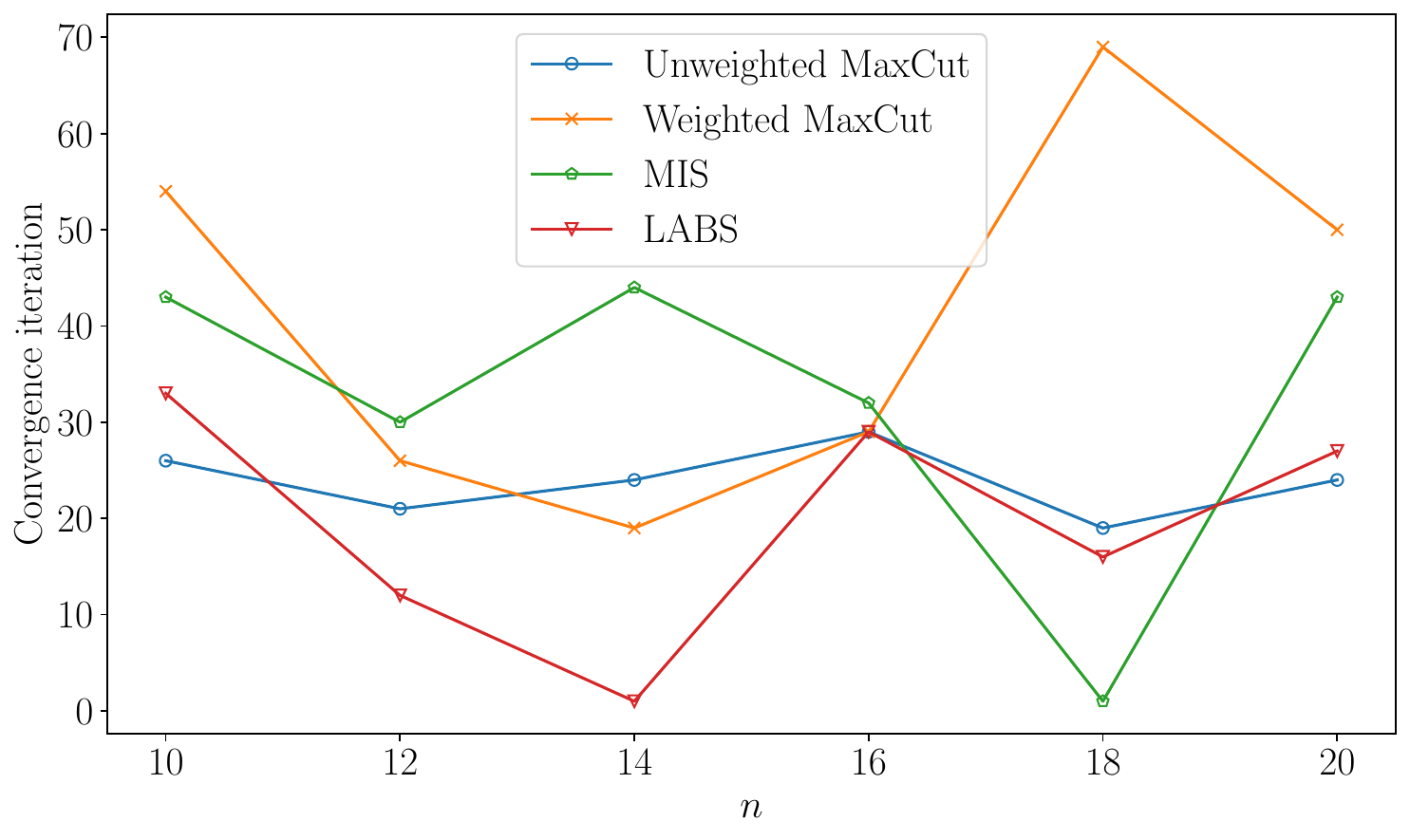}
        \caption{Number of optimization iterations required to reach convergence as a function of the number of qubits $n$ for GQW.}
        \label{fig:gqw_conv}
\end{figure}

\begin{figure}
    \centering
        \includegraphics[width=1.\linewidth]{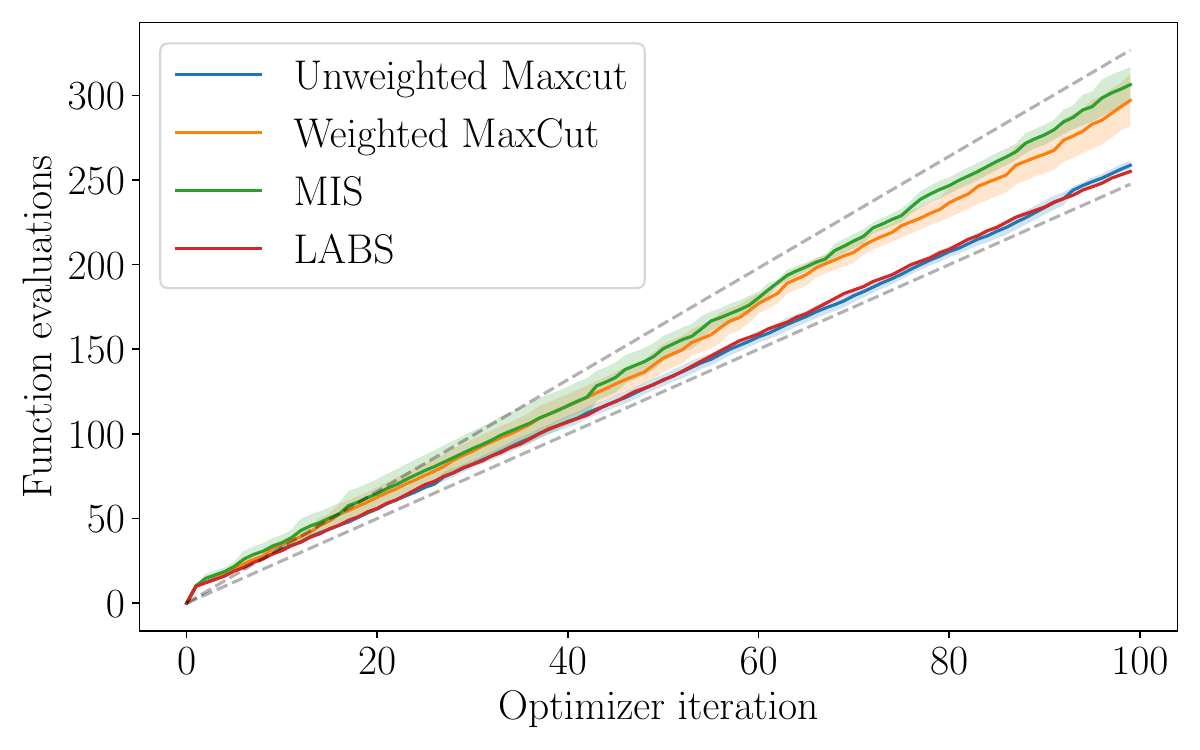}
        \caption{Number of quantum circuit evaluations as a function of the number of iterations of Nelder-Mead for \textsc{GQW} for $n=20$ qubits. The dotted lines respectively denote $f(x)=2.5x$ and $f(x)=3.3x$.}
        \label{fig:gqw_evals}
\end{figure}

The number of iterations required for GQW to reach convergence is shown in Fig.~\ref{fig:gqw_conv}. Fewer than 100 iterations are sufficient for all considered problems. In particular, unweighted MaxCut and LABS exhibit a similar convergence behavior, requiring approximately 25 iterations, largely independent of the instance size. Weighted MaxCut and MIS require slightly more iterations to converge. Furthermore, Fig.~\ref{fig:gqw_evals} illustrates the number of quantum circuit evaluations as a function of the number of optimizer iterations for GQW. For all problems, a linear relationship is observed between the number of circuit evaluations and the number of optimization iterations. In contrast, \textsc{SamBa--GQW} requires only a single quantum circuit evaluation, as it operates as a one-shot method.

In summary, although GQW achieves slightly better performance than \textsc{SamBa--GQW}, the difference remains relatively small. Considering that variational GQW relies on classical optimization, which necessitates repeated quantum circuit evaluations, whereas \textsc{SamBa--GQW} requires only a single evaluation, the latter offers a good trade-off between solution quality and computational cost compared to the variational approach.

\subsection{QAOA}

\begin{figure*}
    \centering
        \includegraphics[width=0.8\linewidth]{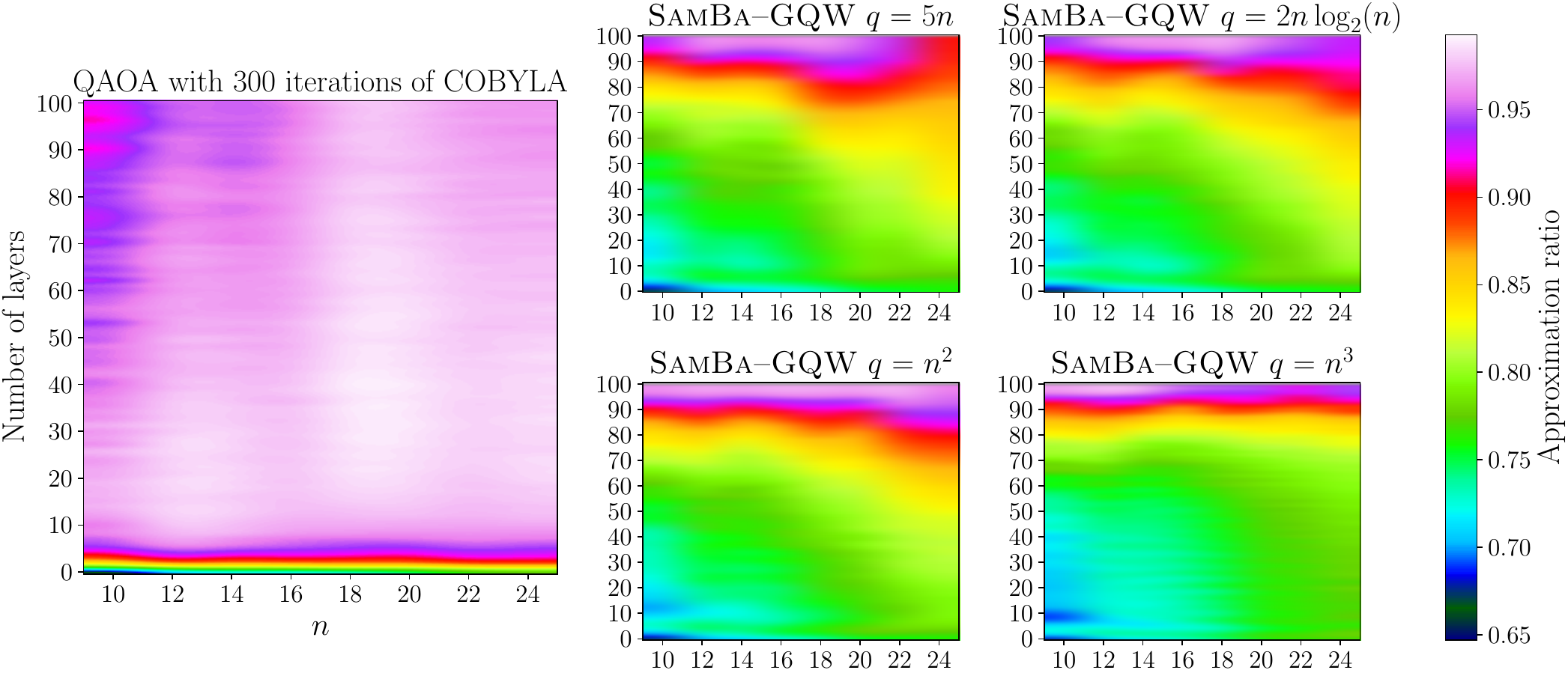}
        \caption{Approximation ratio obtained with QAOA (left) and \textsc{SamBa--GQW} (right) for unweighted MaxCut on random Erdős–Rényi graphs as a function of the number of qubits number of layers of the circuits.}
        \label{fig:samba_qaoa_unweighted_maxcut_ratio}
\end{figure*}

We present now results comparing \textsc{Samba--GQW} in its circuit form with QAOA, which is one of the most studied variational approach. This heuristic can be interpreted as a Trotterized version of quantum annealing, where the continuous-time evolution under a time-dependent Hamiltonian is replaced by a sequence of alternating unitaries generated by the problem and mixer Hamiltonians. The QAOA evolution is parametrized by a number of layers $p$ that controls the precision of the computation. The resulting state reads:
\begin{equation}\label{eq:qaoa}
    \ket{\psi_{\vec{\theta}_p}}=\left(\prod_{k=1}^p\mathrm{e}^{-i\beta_kH_M}\mathrm{e}^{-i\gamma_kH_C}\right)\ket{+}^{\otimes n},
\end{equation}
and we set $\vec{\theta}_p=(\gamma_1,\beta_1,\dots, \gamma_p,\beta_p)$ and $(\gamma_k,\beta_k)\in \mathbb{R}^2$. The cost and mixer Hamiltonian are respectively defined in Eq. \eqref{eq:cost_hamiltonian} and \eqref{eq:mixer}. However, one is not restricted to the use of $X$-mixers since a lot of different families of mixer can be of interest \cite{hadfield2019quantum}. After the quantum evolution of Eq. \eqref{eq:qaoa}, the expected cost value $F(\psi_{\vec{\theta}_p})=\bra{\psi_{\vec{\theta}_p}}H_C\ket{\psi_{\vec{\theta}_p}}$ is minimized classically. Then, this variational process is iterated with the updated $\vec{\theta}_p$ as long as the expectation value is above a given threshold. In the context of minimization problems, the performance of QAOA are usually measured with the approximation ratio:
\begin{equation}
    r=\frac{C_{\min}}{F(\psi_{\vec{\theta}_p})}.
\end{equation}
Since we aim at minimizing a cost function with negative values, we slightly rescale this ratio as:
\begin{equation}
    \tilde{r}=\frac{C_{\max}-F(\psi_{\vec{\theta}_p})}{C_{\max}-C_{\min}}.
\end{equation}
To go further, a detailed review of QAOA and its variant can be found in Ref. \cite{BLEKOS20241}.

We compare QAOA with \textsc{Samba--GQW} on unweighted and weighted MaxCut, as well as on unweighted MIS and LABS. The problem instances, described in Sec.~\ref{subsec:results}, have even sizes $n \in [10, 24]$. QAOA is optimized using 300 COBYLA iterations, and the maximum number of layers is fixed to $p = 100$ for both methods\footnote{Here, $p$ denotes the number of circuit layers, as in standard QAOA, whereas in previous sections $p_\star$ referred to the uniform discretization parameter of \textsc{Samba--GQW}. The latter is still Trotterized as described in Sec.~\ref{subsec:circuit}, with the maximum number of layers fixed to 100. Thus, QAOA and \textsc{SamBa--GQW} are evaluated using circuits with the same number of layers, ensuring a fair comparison.}. Different values of the sampling parameter $q \in \{5n, 2n\log_2(n), n^2, n^3\}$ are explored for \textsc{Samba--GQW}.

The INTERP initialization strategy of Zhou et al.~\cite{Zhou2020} is used for the MaxCut and MIS problems, treating optimal parameters as a smooth schedule in the layer index and initializing deeper circuits via interpolation of previously optimized angles. We use a deterministic initialization for $p=1$, setting \(\beta_1=\pi/8\) and \(\gamma_1=0.4/w_{\mathrm{scale}}\), where \(w_{\mathrm{scale}}\) is the maximum absolute adjacency matrix entry.

For the LABS problem, parameters are initialized using the dataset \cite{ruslan_shaydulin_2023_8190275} from Ref.~\cite{doi:10.1126/sciadv.adm6761}, which provides near-optimal angles for fixed circuit depths. When the values for a given depth $p$ are unavailable, the initialization is obtained by interpolating the parameters from depth $p-1$.

\subsubsection{MaxCut}
\begin{figure*}
    \centering
        \includegraphics[width=0.8\linewidth]{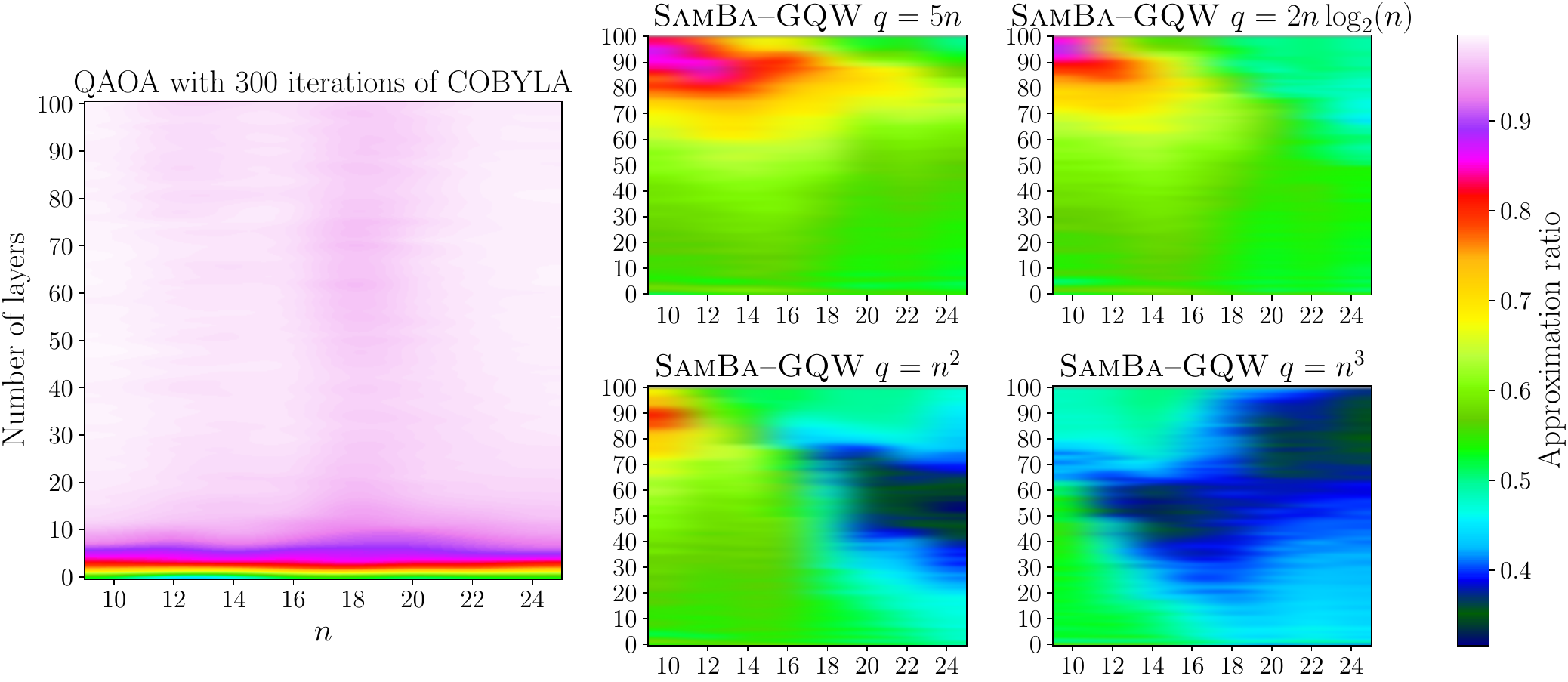}
        \caption{Approximation ratio obtained with QAOA (left) and \textsc{SamBa--GQW} (right) for weighted MaxCut on random Erdős–Rényi graphs as a function of the number of qubits number of layers of the circuits.}
        \label{fig:samba_qaoa_weighted_maxcut_ratio}
\end{figure*}

The MaxCut cost Hamiltonian reads:
\begin{equation}
	H_C = -\frac{1}{2}\sum_{(i,j)\in\mathcal{E}}w_{ij}\left(\mathbb{I}-\sigma^z_i\sigma^z_j\right),
\end{equation}
where $w_{ij}=1$ for unweighted instances. The results for the unweighted and weighted MaxCut problems are summarized in Figs.~\ref{fig:samba_qaoa_unweighted_maxcut_ratio} and \ref{fig:samba_qaoa_weighted_maxcut_ratio}. Overall, QAOA achieves high approximation ratios across all system sizes, with performance improving as the number of layers increases and rapidly saturating at shallow depths ($p < 10$) for both unweighted and weighted instances. For the unweighted MaxCut instances, \textsc{Samba--GQW} attains approximation ratios close to those of QAOA only at significantly larger depths ($p > 90$), with similar behavior observed across all sampling values. In contrast, for the weighted instances, increasing the sampling parameter $q$ leads to a decrease in approximation ratios. This trend suggests that larger sampling values require substantially deeper circuits, likely well beyond the considered limit of $p =100$ to achieve comparable performance.

\subsubsection{Maximum independent set}
\begin{figure*}
    \centering
        \includegraphics[width=0.75\linewidth]{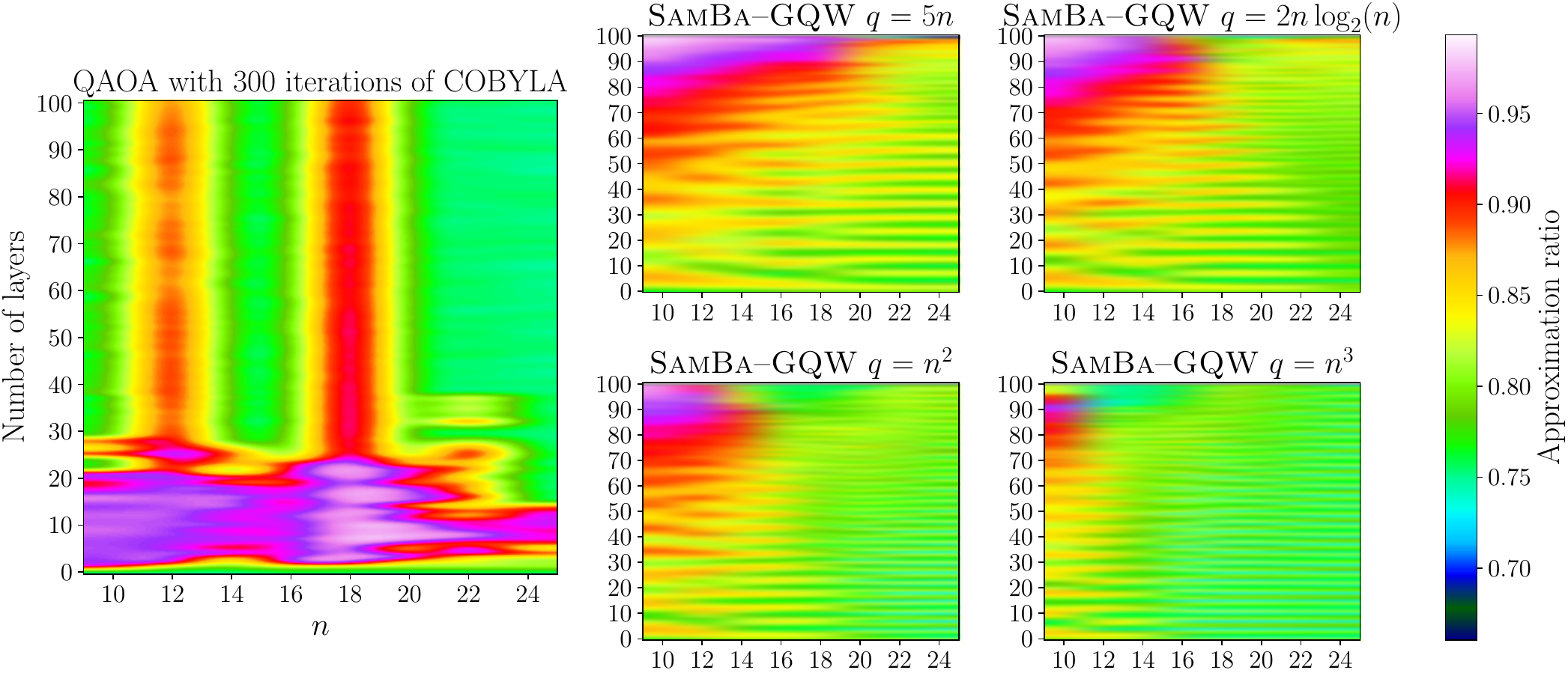}
        \vskip-0.175cm
        \caption{Approximation ratio obtained with QAOA (left) and \textsc{SamBa--GQW} (right) for unweighted MIS on random Erdős–Rényi graphs as a function of the number of qubits number of layers of the circuits.}
        \label{fig:samba_qaoa_mis_ratio}
\end{figure*}

The unweighted MIS cost Hamiltonian is defined as:
\begin{equation}
	\begin{split}
	H_C &= \frac{1}{2}\sum_{i\in \mathcal{V}}\sigma^z_i-\frac{\lambda}{4}\sum_{(i,j)\in\mathcal{E}}\left(\sigma^z_i+\sigma^z_j\right)\\
	&+\frac{\lambda}{4}\sum_{(i,j)\in\mathcal{E}}\sigma^z_i\sigma^z_j.
	\end{split}
\end{equation}
The results for the unweighted MIS problem are summarized in Fig.~\ref{fig:samba_qaoa_mis_ratio}. QAOA achieves high approximation ratios at shallow circuit depths ($p < 10$), with some variations across system sizes. At larger depths ($p > 20$), the approximation ratio decreases, indicating difficulties for the optimizer in effectively tuning deeper circuits. In contrast, \textsc{Samba--GQW} exhibits distinct behaviors depending on the sampling parameter $q$. Sampling values $q \in \{5n, 2n\log_2(n), n^2\}$ yield competitive approximation ratios that approach those of QAOA at sufficiently large depths ($p > 90$). However, further increasing the sampling value leads to a noticeable degradation in performance as the instance size grows.

\subsubsection{LABS}
\begin{figure*}
    \centering \vskip-0.3cm
        \includegraphics[width=0.75\linewidth]{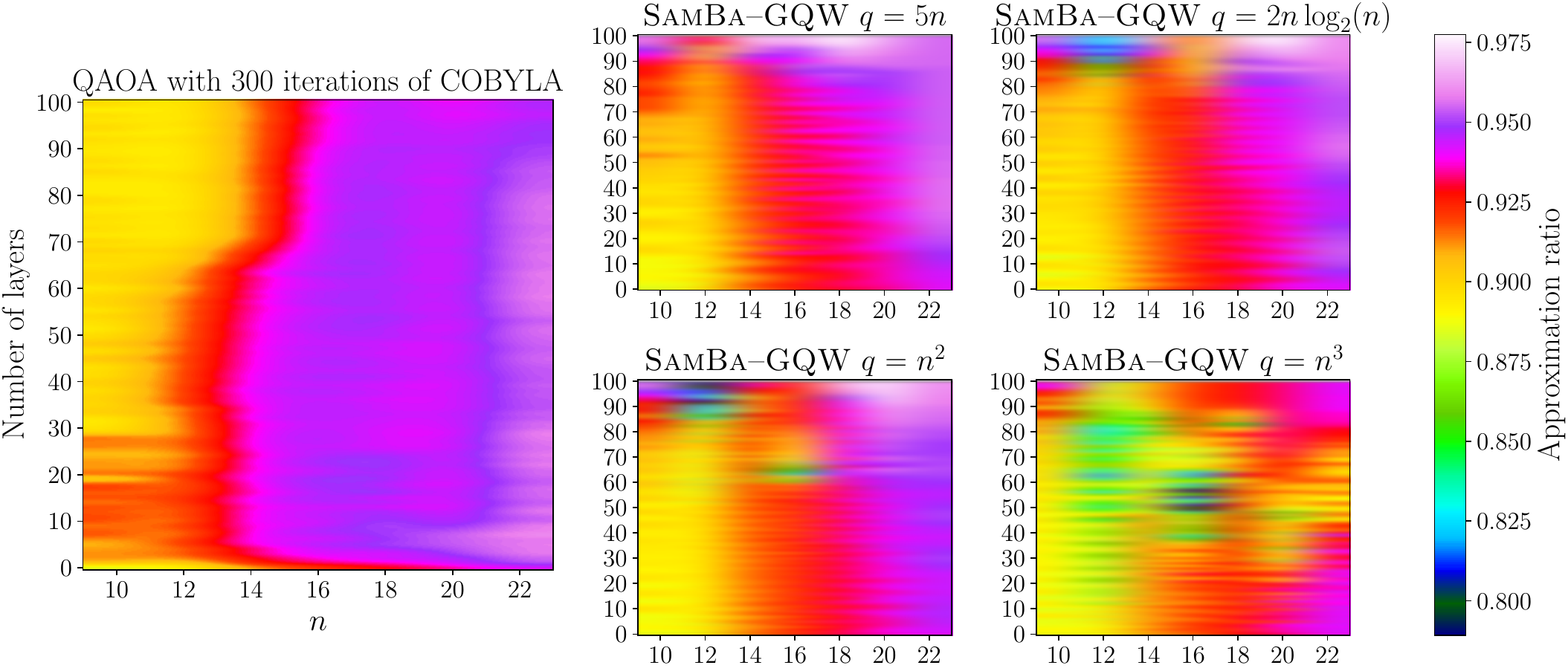}
        \vskip-0.175cm
        \caption{Approximation ratio obtained with QAOA (left) and \textsc{SamBa--GQW} (right) for LABS as a function of the number of qubits and the number of layers of the circuits.}
        \label{fig:samba_qaoa_labs_ratio}
\end{figure*}

The LABS cost Hamiltonian reads \cite{doi:10.1126/sciadv.adm6761}:
\begin{equation}
	\begin{split}
	H_C &= 2\sum_{i=1}^{n-3}\sigma^z_i\sum_{t=1}^{\lfloor\frac{n-i-1}{2}\rfloor}\sum_{k=t+1}^{n-i-t}\sigma^z_{i+t}\sigma^z_{i+k}\sigma^z_{i+k+t}\\ &+\sum_{i=1}^{n-2}\sigma^z_i\sum_{k=1}^{\lfloor\frac{n-i}{2}\rfloor}\sigma^z_{i+2k}.
	\end{split}
\end{equation}
The results for the LABS problem are summarized in Fig.~\ref{fig:samba_qaoa_labs_ratio}. The LABS problem differs from MaxCut and MIS in that it possesses a large number of high-quality solutions, making it difficult to distinguish the optimal solution from the rest. As a result, the approximation ratio is already very high at $p = 0$, when the circuit consists only of a layer of Hadamard gates. QAOA achieves high approximation ratios across all system sizes, with performance steadily improving as the number of layers increases. In comparison, \textsc{Samba--GQW} exhibits a strong dependence on the sampling parameter $q$. Sampling values $q\in\{5n,2n\log_2 (n),n^2\}$ yield competitive results, approaching QAOA performance at large depths. Moreover, we notice that QAOA performs substantially better at shallow depths for $14 \leq n \leq 20$, whereas for larger system sizes ($n \gtrsim 20$), \textsc{Samba--GQW} achieves similar performance with fewer layers. This trend may also be explained by the growing number of high-quality solutions as $n$ increases, which naturally raises the approximation ratio. Furthermore, increasing the sampling to $q=n^3$ leads to a noticeable degradation in approximation ratios, particularly for intermediate depths.

\begin{figure*}
    \centering \vskip-0.3cm
        \includegraphics[width=0.75\linewidth]{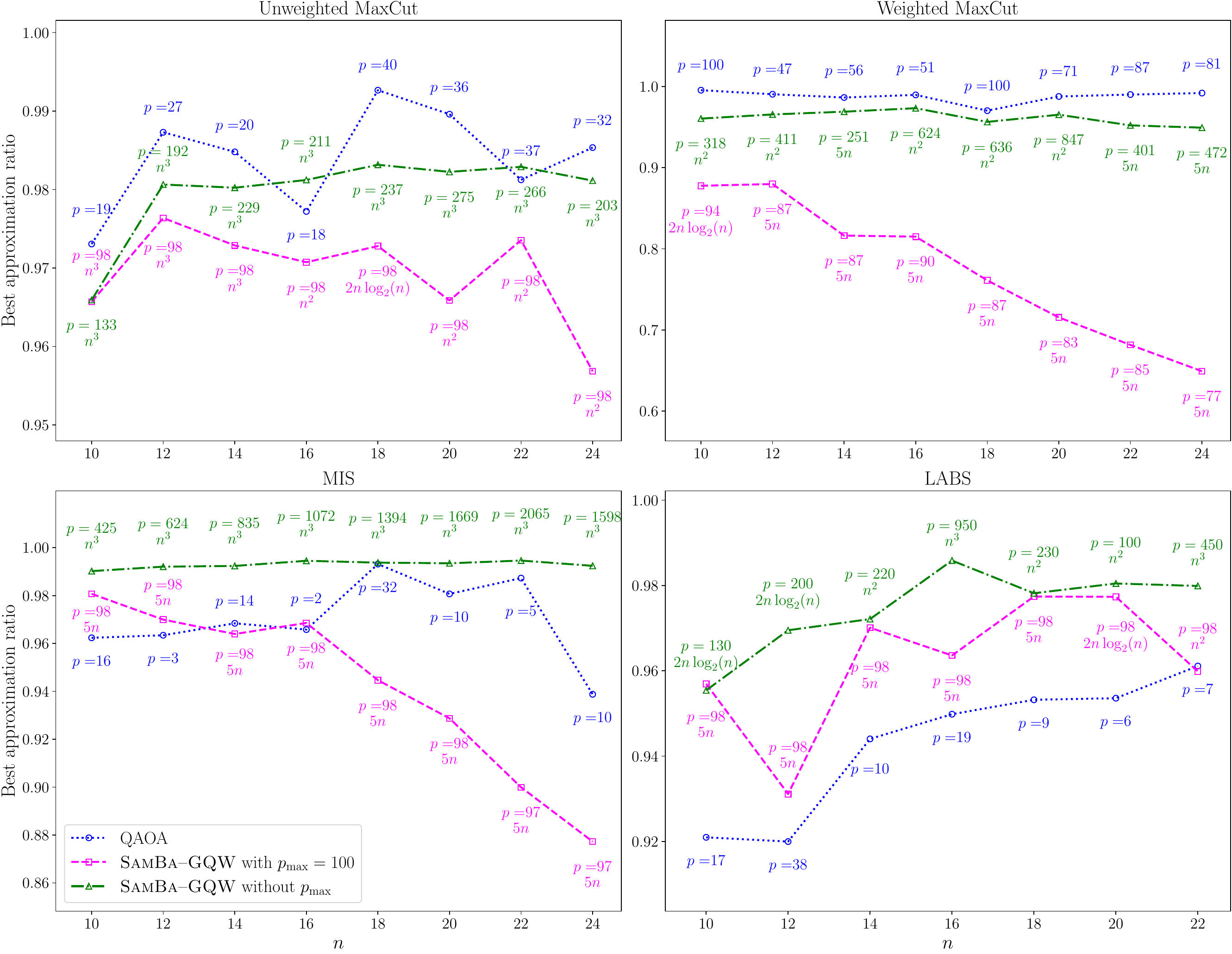}
        \vskip-0.175cm
        \caption{Best approximation ratio obtained by QAOA and \textsc{SamBa--GQW} as a function of the number of qubits. Results for \textsc{SamBa--GQW} are shown both with a restricted number of layers ($p_{\max}=100$) and with an unrestricted layer budget. We indicate the number of layers $p$ at which the highest approximation is reached, we also display the sampling parameter $q$ that led to this performance for \textsc{SamBa--GQW}.}
        \label{fig:qaoa_overall}
\end{figure*}
\subsection{Overall comparison}
We illustrate in Fig.~\ref{fig:qaoa_overall} the best approximation ratios obtained over quantum circuits with number of layers $p \in [1,100]$ for QAOA and \textsc{SamBa--GQW}, as a function of the number of qubits. For unweighted MaxCut, QAOA consistently achieves slightly higher approximation ratios across all $n$. This trend becomes more pronounced for weighted MaxCut, where QAOA maintains near-optimal ratios even at large $n$, whereas \textsc{SamBa--GQW} exhibits a marked degradation as $n$ increases. We also observe that the best approximation ratios for QAOA are typically obtained at relatively large depths ($p \gtrsim 70$). However, as shown in Fig.~\ref{fig:samba_qaoa_weighted_maxcut_ratio}, high-quality solutions are already reached at shallow depths ($p \approx 10$). Moreover, in contrast with the unweighted case where the optimal sampling parameter is most often $q = n^2$ or $q = n^3$, for weighted MaxCut the best performance is consistently obtained with $q = 5n$, except for $n=10$.  A similar gap is observed for MIS, where QAOA reaches higher-quality solutions at moderate depths, while \textsc{SamBa--GQW} again deteriorates significantly despite using maximal circuit depths ($p = 98$). In contrast, for LABS, \textsc{SamBa--GQW} behaves better than QAOA for $n < 22$, albeit at the cost of much deeper circuits. Overall, fixing the maximal number of layers to $p_{\max} = 100$ leads to QAOA performing generally better than \textsc{SamBa--GQW}, achieving higher approximation ratios with shallower circuits. It is important to note, however, that QAOA relies on a classical optimization routine to determine its parameters, which contributes to the quality of the solutions, whereas \textsc{SamBa--GQW} only relies on an efficient sampling protocol on the classical side. These results highlight that \textsc{SamBa--GQW} is less suited to NISQ settings, as it requires  deeper circuits. Finally, Fig.~\ref{fig:qaoa_overall}, also depicts the best approximation ratio achieved by \textsc{SamBa--GQW}, with unrestricted layer budget. We see that beyond the NISQ setting, \textsc{SamBa--GQW} is well in par with QAOA.

\section{Conclusion}

In this work, we propose a new quantum algorithm for approximating the solution of polynomial combinatorial optimization problems without using classical optimizers: \textsc{SamBa--GQW}. The algorithm is based on a classical offline sampler and a quantum walk. The quantum walk is guided via a time-dependent hopping rate, which is determined a priori by the sampler. We present numerical results on several hard combinatorial problems, showcasing the high performance, consistency, and simplicity of \textsc{SamBa--GQW}. Our algorithm performs also very favorably with respect to the variational guided quantum walk, which requires a classical optimizer to determine the time-dependent hopping rate. Compared to QAOA, \textsc{SamBa--GQW} behaves very similarly for approximation ratios, as long as its layer count is unrestricted (beyond NISQ regime), and less so, when forced to use shallow circuits. The latter is to be expected, since QAOA runs a classical optimizer, while \textsc{SamBa--GQW} does not.

\textsc{Samba--GQW} strong performance makes it a promising heuristic candidate for solving large-scale combinatorial problems as quantum hardware continues to improve. Our approach is not intended to provide a polynomial-time solution to NP-hard problems; instead, numerical evidence suggests that polynomial resources may be sufficient to effectively guide the search toward high-quality states.

\section*{Data availability}

Hamiltonian simulations were carried out with Dynamiqs \cite{guilmin2025dynamiqs} and quantum circuits simulations with Qiskit \cite{qiskit2024}. Quantum circuit diagrams were generated with quantikz package \cite{kay2018tutorial}, and graphs with NetworkX \cite{hagberg2008exploring}. The code and examples of use are available at \cite{samba_code}.

\section*{Author contributions}

U. Nzongani conceived the presented idea and carried out the simulations. D. Laplace Mermoud developed the quantum circuit discretization. U. Nzongani and D. Laplace Mermoud wrote the manuscript with support from A. Simonetto. A. Simonetto supervised the project. G. Di Molfetta has participated to several discussions, and provided his insight on the final manuscript.

\section*{Acknowledgements}

This work is supported by the PEPR EPiQ ANR-22-PETQ-0007, by the ANR JCJC DisQC ANR-22-CE47-0002-01, and by the ANR project HQI-ANR-22-PNCQ-0002.

\bibliographystyle{unsrtnat}
\bibliography{ref}

\newpage
\appendix

\begin{center}
\Large \sf  Appendices
\end{center}

\section{Hopping rate as a function of energy}\label{app:hopping_rate}

\begin{figure}[h]
    \centering
\includegraphics[width=\linewidth]{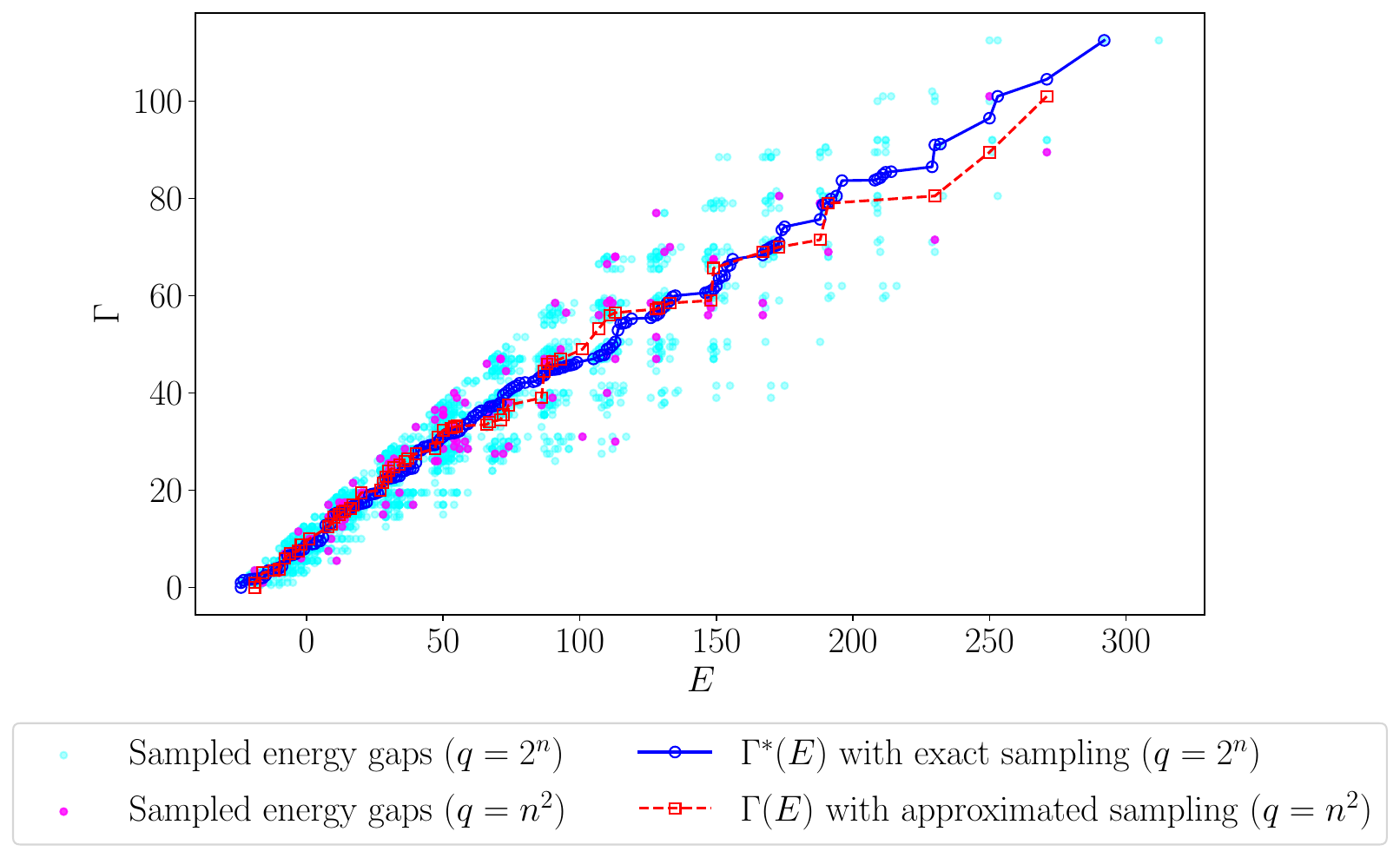}   
 \caption{Hopping rate $\Gamma$ as a function of energy $E$ for $n=11$ qubits with $\Gamma(E)=\frac{\langle\Delta^C\rangle(E)}{2}$ for UD-MIS instance.}
    \label{fig:hopping_energy}
\end{figure}

The hopping rate $\Gamma$ is initially a function of energy, which corresponds to the output of the cost function, which is the reason why we can consider “negative energies”. We make it a function of time to define a QW on the solution space, whose connectivity is defined by the structure of the mixer. Since each energy requires a specific transfer time (see Eq. \eqref{eq:time}), the original function of energy $\Gamma(E)$ is rescaled to obtain the time-dependent hopping rate $\Gamma(t)$. We show the exact ($q=2^n$) and approximated ($q=n^2$) original function of energy on Fig. \ref{fig:hopping_energy} for $n=11$ qubits. Unlike the time-dependent hopping rates displayed on Figs. \ref{fig:interpolation_exact} and \ref{fig:interpolation_approx}, it is clear that when it depends on energy, and are therefore not rescaled, the exact and approximate hopping rates are very similar.

\section{Uniform discretization of $\Gamma$}\label{app:constant_discretisation}
\begin{figure}
    \centering
        \includegraphics[width=1.\linewidth]{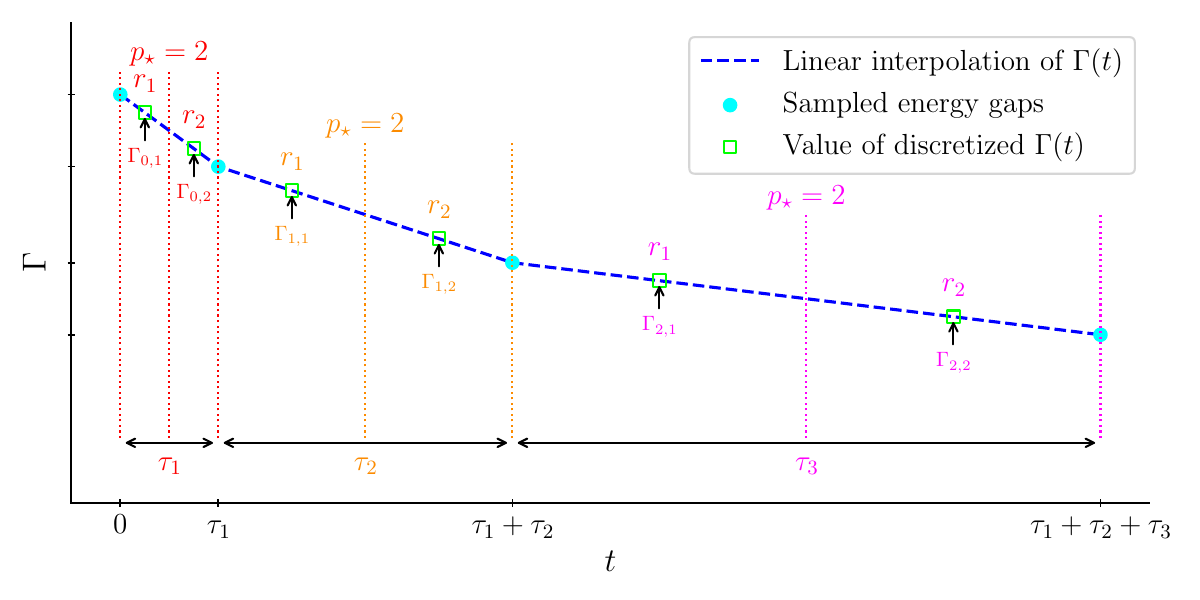}
        \caption{Uniform discretization of the hopping rate $\Gamma$.}
        \label{fig:uniform_discretization}
\end{figure}

In our numerical simulations we split each linear sub-interval into $p_\star$ cuts, we illustrate such a discretization scheme with $p_\star=2$ in Fig. \ref{fig:uniform_discretization}.

\section{Sampled energy scaling with $n$}\label{app:sampled_energy}

\begin{figure}
    \centering
        \includegraphics[width=1.\linewidth]{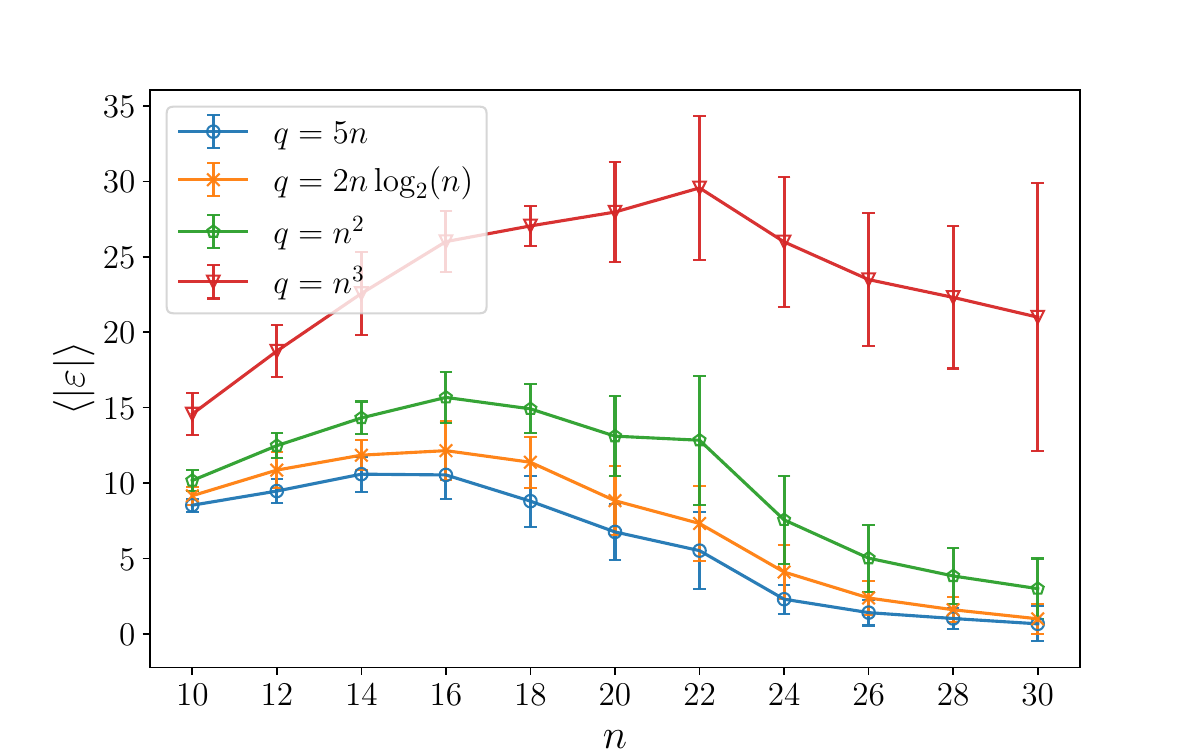}
        \includegraphics[width=1.\linewidth]{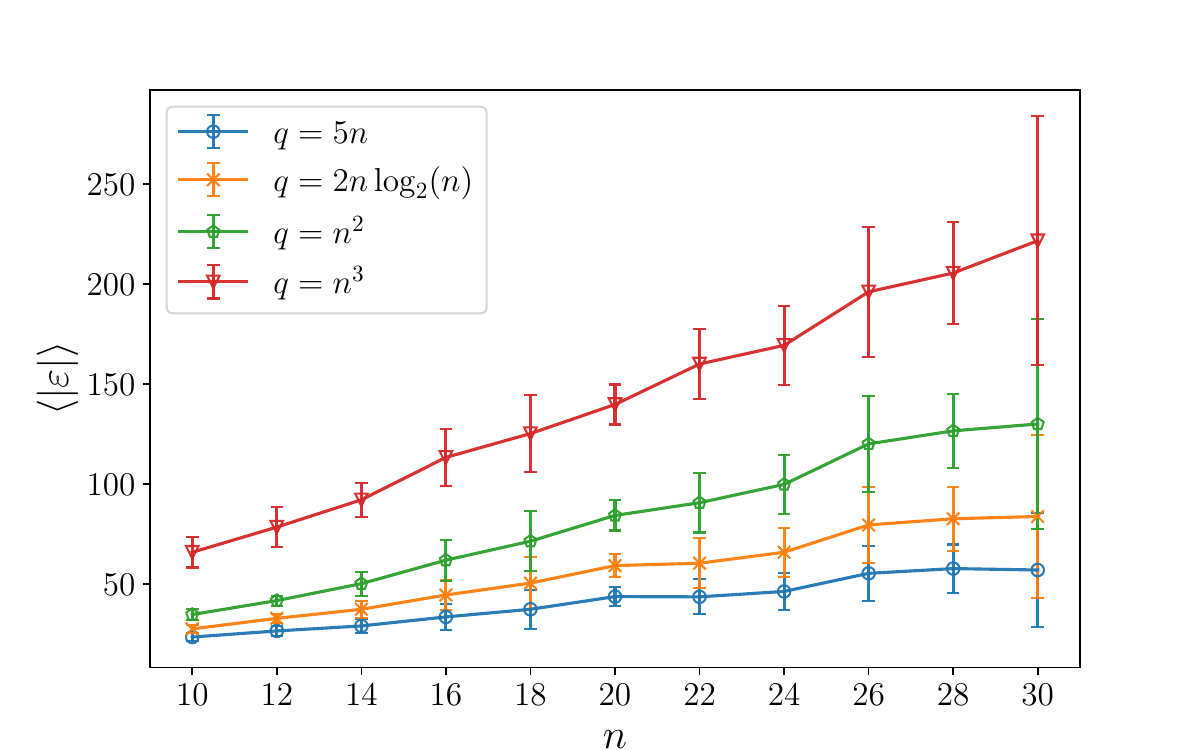}
        \caption{Number of different sampled energies for unweighted (top) and weighted (bottom) MaxCut on random Erdős–Rényi graphs for $n\in[10,30]$ qubits with different sampling values $q\in\{5n,2n\log_2(n),n^2,n^3\}$.}
        \label{fig:samba_circuit_maxcut_samples_n}
\end{figure}

\begin{figure}
    \centering
        \includegraphics[width=1.\linewidth]{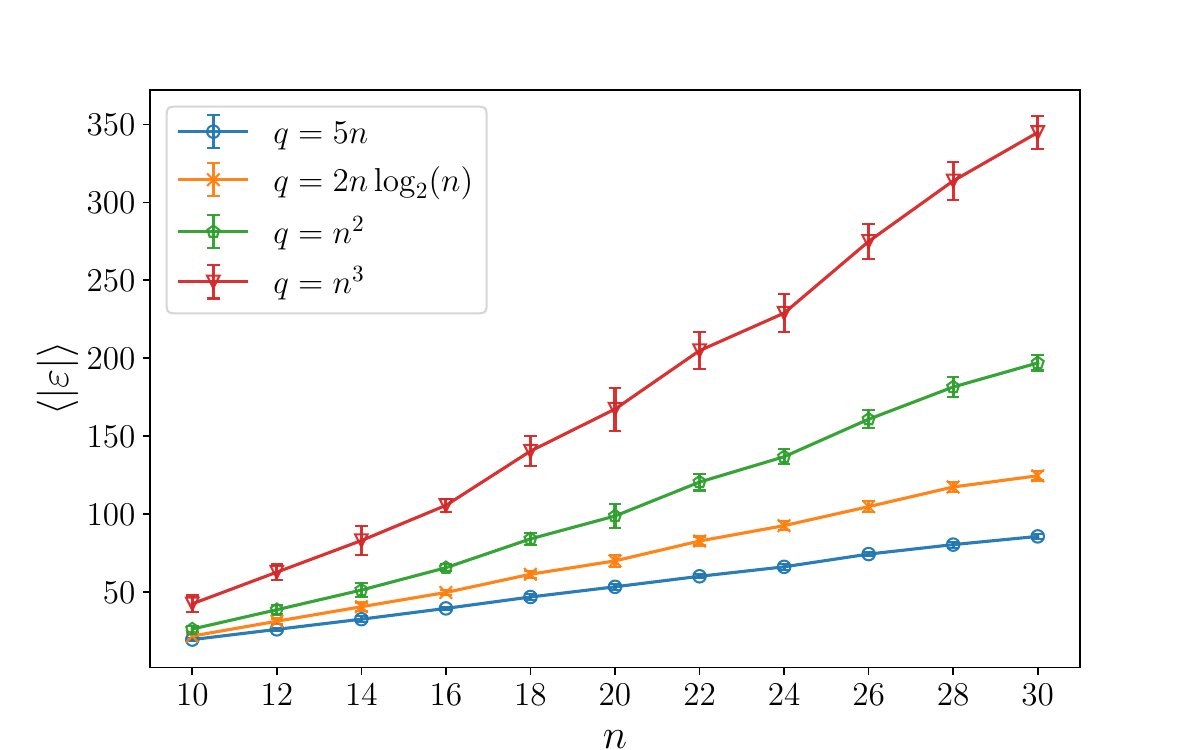}
        \includegraphics[width=1.\linewidth]{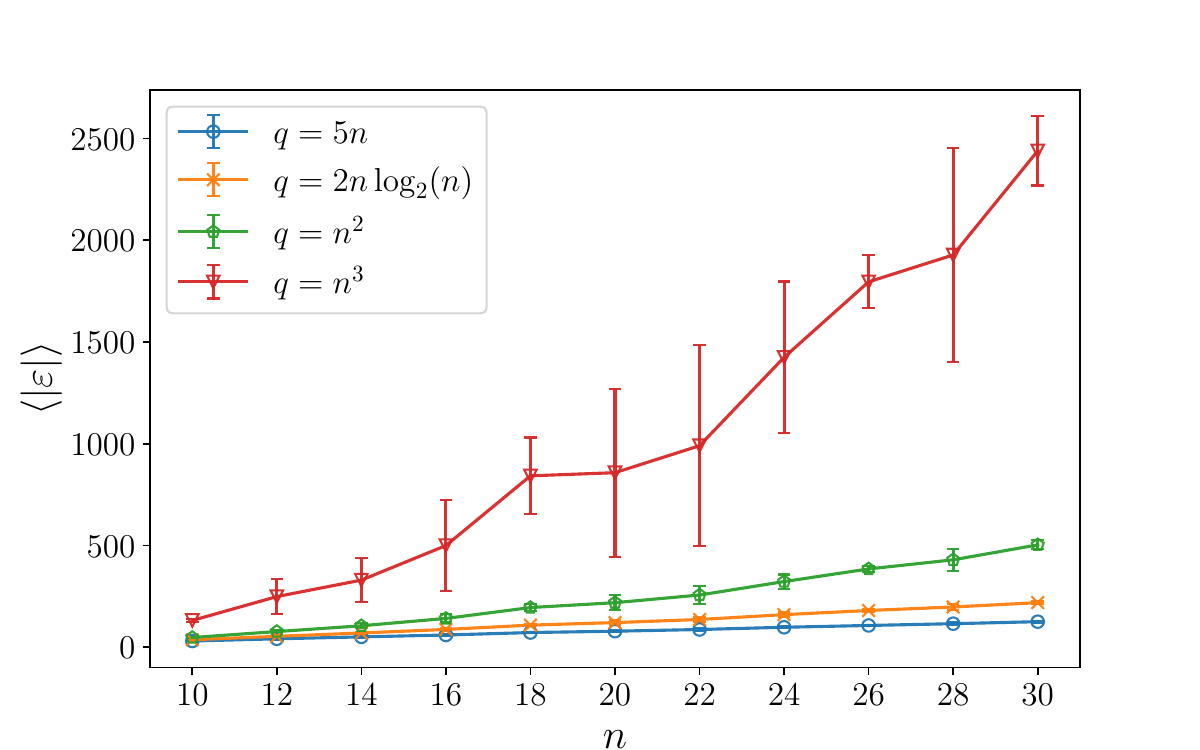}
        \caption{Number of different sampled energies for unweighted MIS on random Erdős–Rényi graphs (top) and weighted MIS on unit-disk graphs (bottom) for $n\in[10,30]$ qubits with different sampling values $q\in\{5n,2n\log_2(n),n^2,n^3\}$.}
        \label{fig:samba_circuit_mis_samples_n}
\end{figure}

\begin{figure}
    \centering
        \includegraphics[width=1.\linewidth]{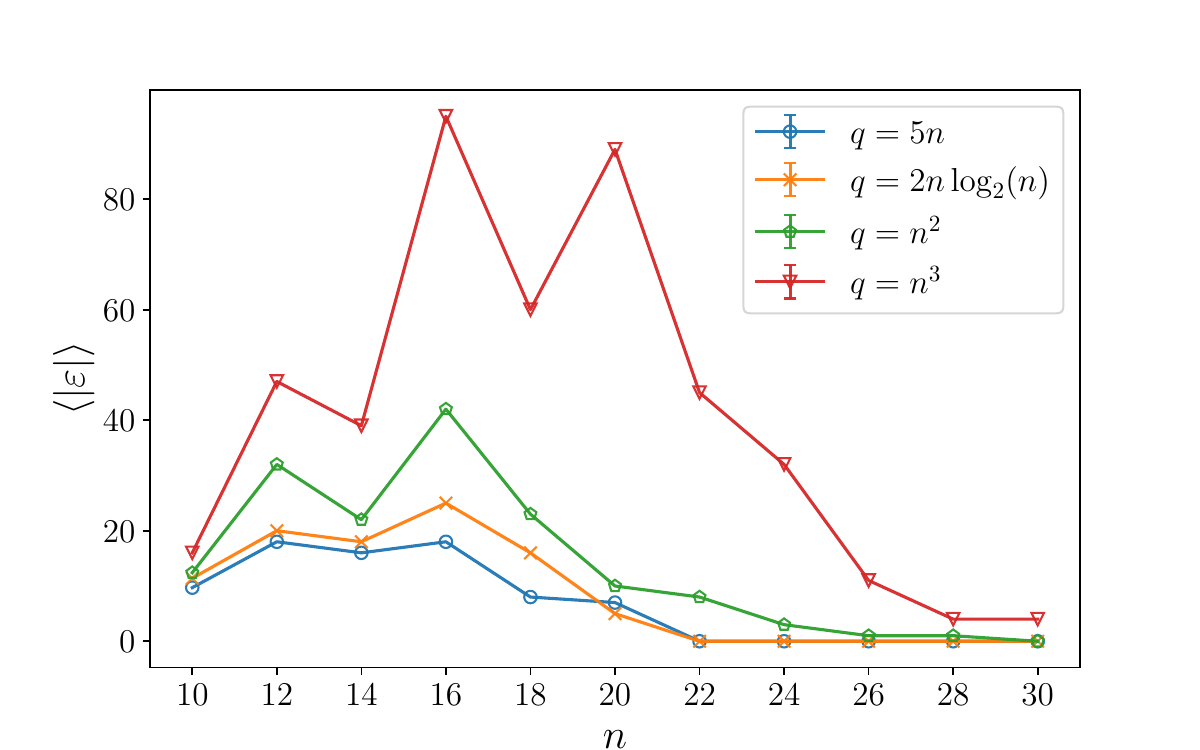}
        \caption{Number of different sampled energies for LABS for $n\in[10,30]$ qubits with different sampling values $q\in\{5n,2n\log_2(n),n^2,n^3\}$.}
        \label{fig:samba_circuit_labs_samples_n}
\end{figure}

In this section, we show the number of different sampled energies $|\mathcal{E}|$ as a function of the number of qubits for MaxCut (Fig. \ref{fig:samba_circuit_maxcut_samples_n}), MIS (Fig.~\ref{fig:samba_circuit_mis_samples_n}) and LABS (Fig.~\ref{fig:samba_circuit_labs_samples_n}). As $n$ increases, this quantity approaches zero for LABS, explaining the behavior observed in Fig.~\ref{fig:samba_circuit_labs_30}. In Appendix~\ref{app:sampling_absolute}, we show how to prevent this vanishing behavior.

\section{Deeper quantum circuits for \textsc{SamBa--GQW}}\label{app:larger_circuit}

\begin{figure*}
    \centering
        \includegraphics[width=1.\linewidth]{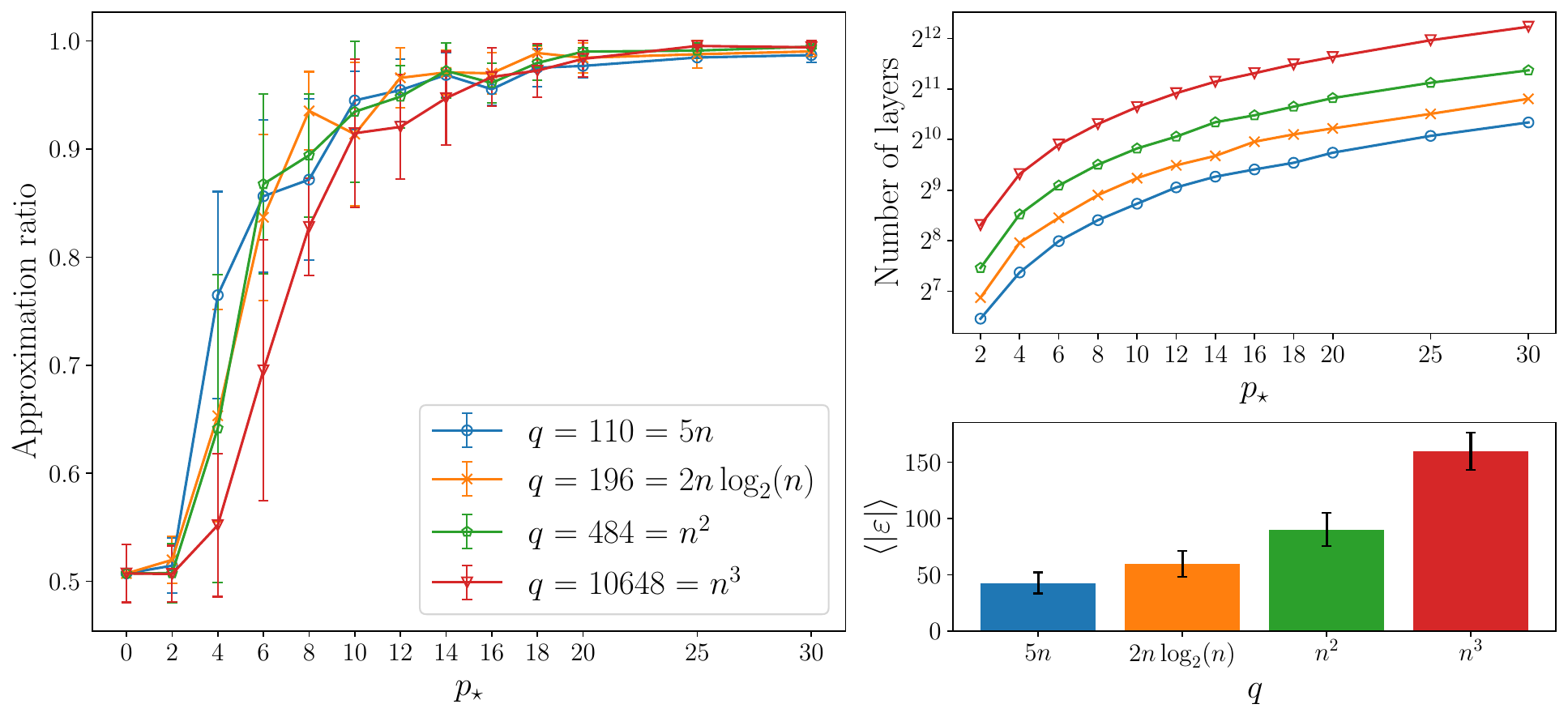}
        \caption{\textsc{SamBa--GQW} quantum circuit simulations for weighted MaxCut on random Erdős–Rényi graphs for $n=22$ qubits with different sampling values $q\in\{5n,2n\log_2(n),n^2,n^3\}$. The discretization parameter $p_\star$ takes values in interval $[0,30]$.}
        \label{fig:samba_circuit_maxcut_weighted_22_large}
\end{figure*}

\begin{figure*}
    \centering
        \includegraphics[width=1.\linewidth]{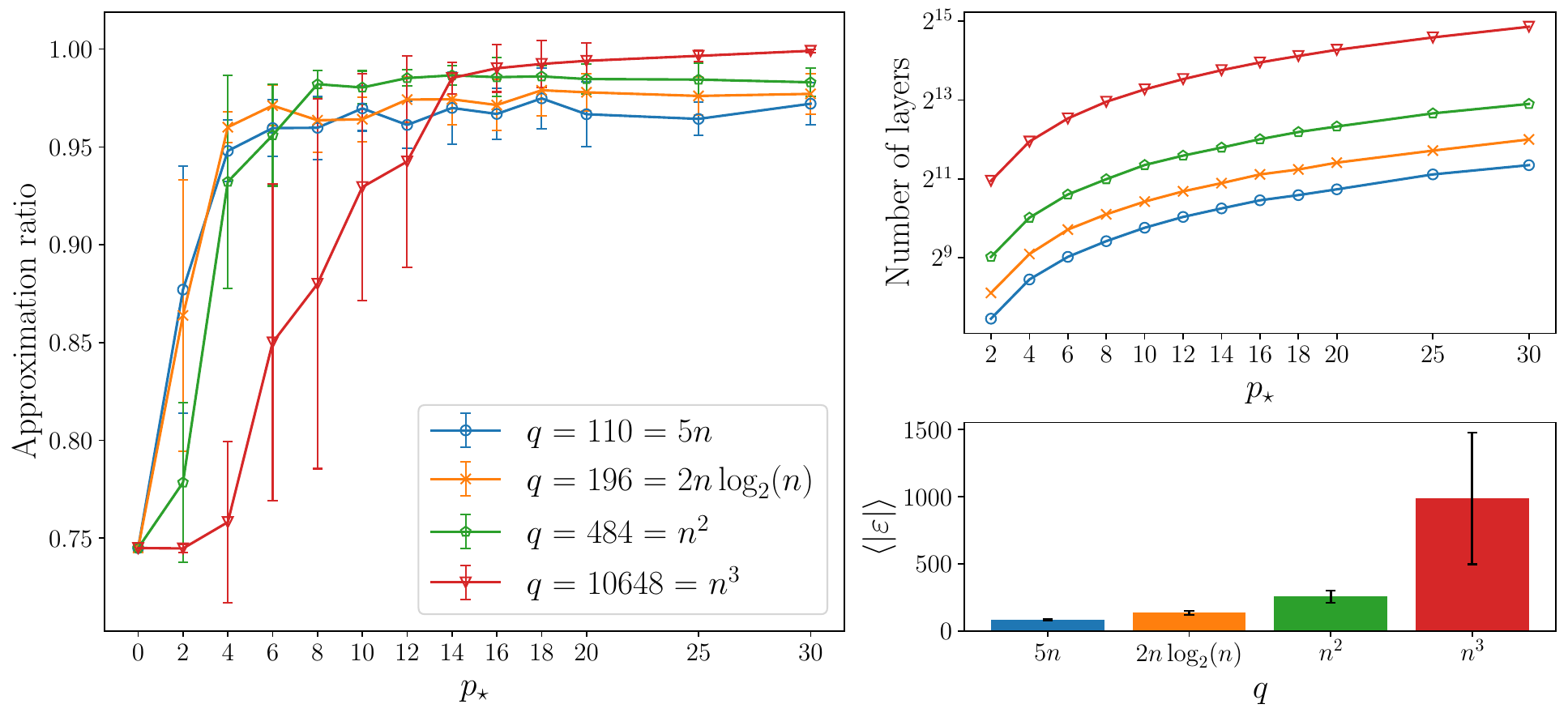}
        \caption{\textsc{SamBa--GQW} quantum circuit simulations for weighted MIS on unit-disk graphs for $n=22$ qubits with different sampling values $q\in\{5n,2n\log_2(n),n^2,n^3\}$. The discretization parameter $p_\star$ takes values in interval $[0,30]$.}
        \label{fig:samba_circuit_mis_ud_22_large}
\end{figure*}

In this section, we illustrate the effect of using higher-depth circuits for weighted MaxCut and UD-MIS on \( n = 22 \) qubits by extending the discretization parameter to the range \( p_\star \in [0, 30] \). As shown in Figs.~\ref{fig:samba_circuit_maxcut_weighted_22_large} and \ref{fig:samba_circuit_mis_ud_22_large}, increasing the circuit depth leads to the best performance when sampling with \( q = n^3 \), although for the weighted MaxCut problem the results are nearly identical to those obtained with other sampling strategies. However, despite the improvement in the approximation ratio, the number of circuit layers increases significantly, making this approach inefficient. Therefore, it does not appear necessary to use large values of the discretization parameter \( p_\star \), as the gain in solution quality is negligible compared to the substantial increase in circuit depth.

\begin{figure*}
    \centering
        \includegraphics[width=1.\linewidth]{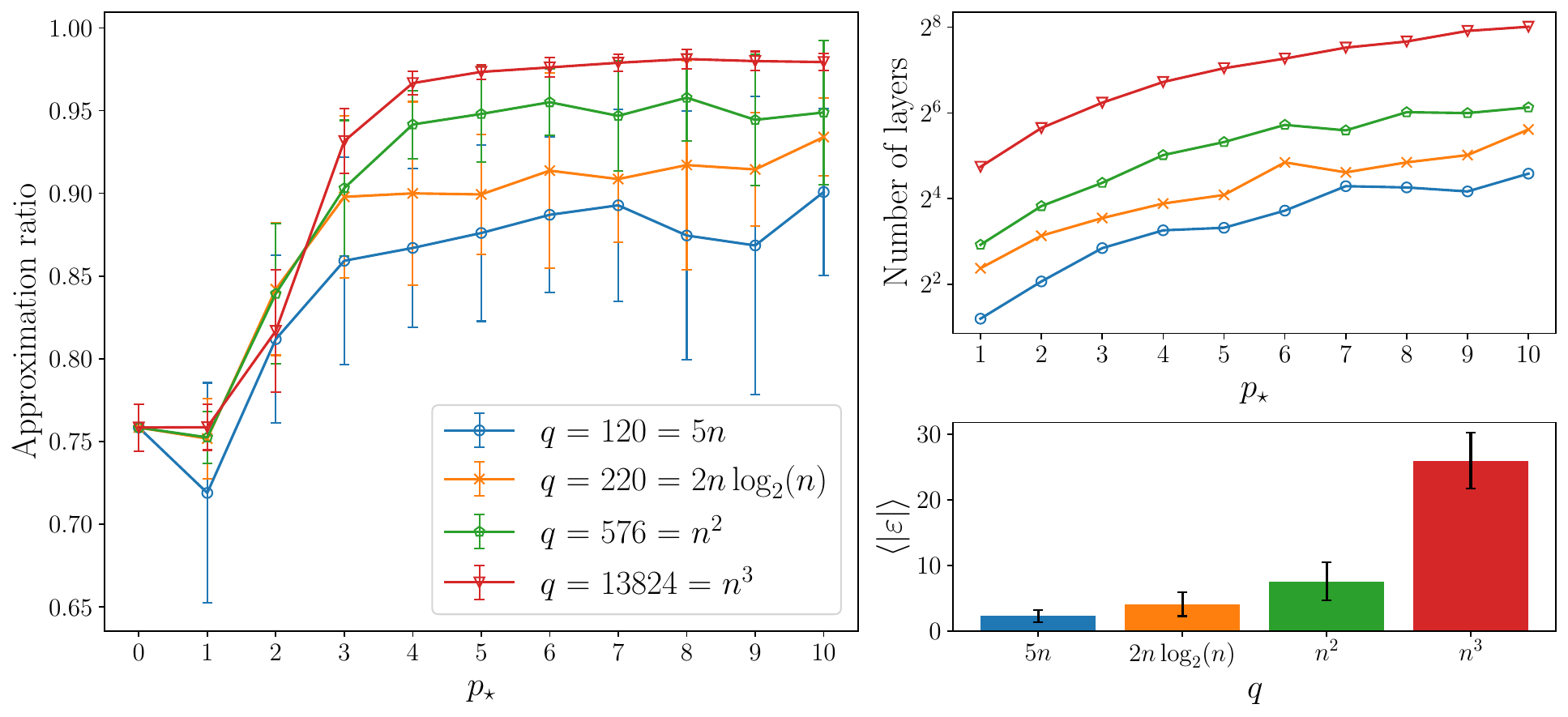}
        \includegraphics[width=1.\linewidth]{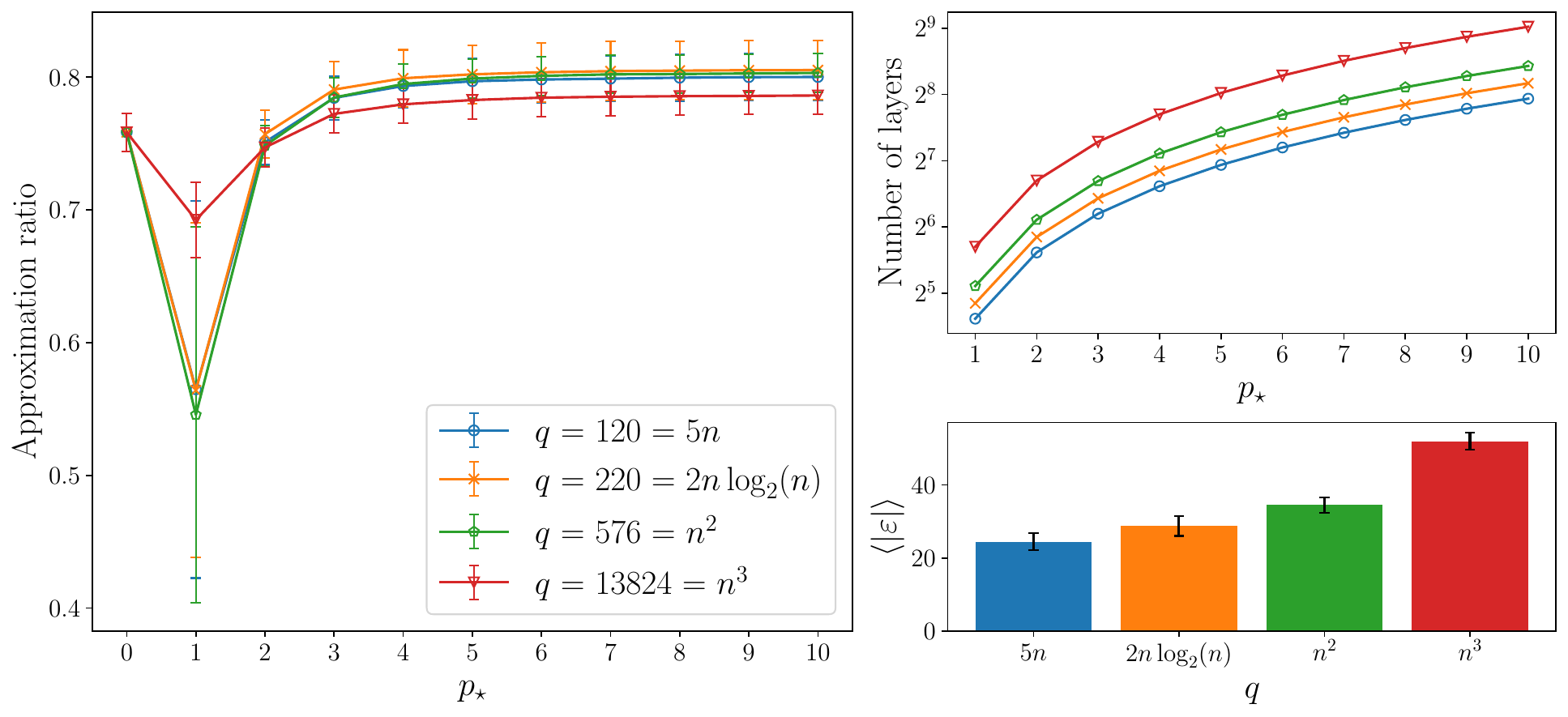}
        \caption{\textsc{SamBa--GQW} quantum circuit simulations for unweighted MaxCut on random Erdős–Rényi graphs for $n=24$ qubits with different sampling values $q\in\{5n,2n\log_2(n),n^2,n^3\}$. The discretization parameter $p_\star$ takes values in interval $[0,10]$. The upper (resp. lower) figure presents the results obtained when considering only positive gaps (resp. both positive and negative gaps).}
        \label{fig:appendix_unweighted_24}
\end{figure*}

\section{Absolute values of the energy gaps}\label{app:sampling_absolute}

In the sampling protocol presented in Algorithm~\ref{alg:sampling}, we consider only positive energy gaps. More precisely, at each iteration, a state is randomly selected and its energy gaps with all neighboring states are computed; only the positive gaps are retained, and the largest among them is stored. One may naturally ask why we do not instead consider both positive and negative gaps by taking their absolute values, since the sign of a gap depends solely on the randomly chosen reference state and does not carry intrinsic information. This choice was made to reduce the number of sampled gaps and, consequently, lower the computational cost of the classical simulations. Numerical results indicate that, for most of the considered optimization problems, this restricted sampling strategy remains sufficient to obtain high-quality solutions. However, as shown in Figs.~\ref{fig:samba_circuit_maxcut_samples_n} (unweighted MaxCut) and~\ref{fig:samba_circuit_labs_samples_n}, the number of sampled gaps decreases as $n$ increases. In the worst-case scenario, this may lead to situations where no energy gaps are sampled at all, implying that no information is extracted to construct the hopping rate and, consequently, no information is available to determine the circuit angles.

\begin{figure}
    \centering
        \includegraphics[width=1.\linewidth]{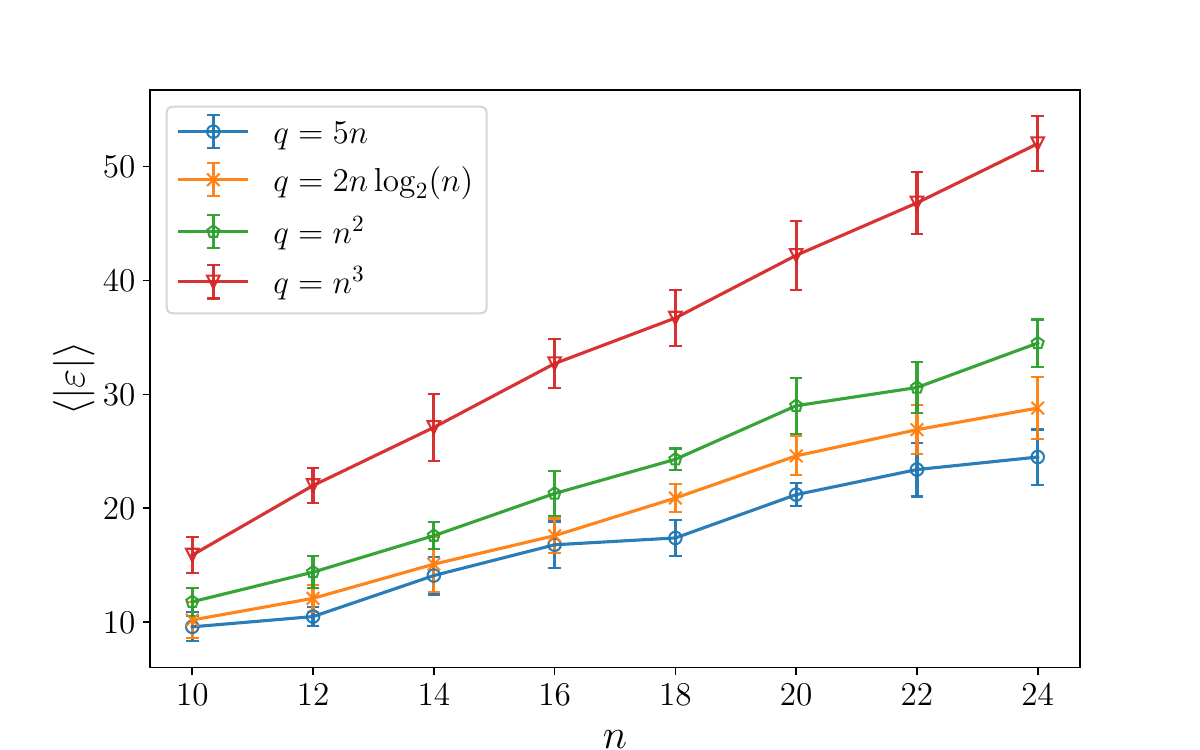}
        \caption{\textsc{SamBa--GQW} quantum circuit simulations for unweighted MaxCut on random Erdős–Rényi graphs for $n\in[10,24]$ qubits with different sampling values $q\in\{5n,2n\log_2(n),n^2,n^3\}$. The discretization parameter $p_\star$ takes values in interval $[0,10]$.}
        \label{fig:appendix_unweighted_real_samples}
\end{figure}

\begin{figure}
    \centering
        \includegraphics[width=1.\linewidth]{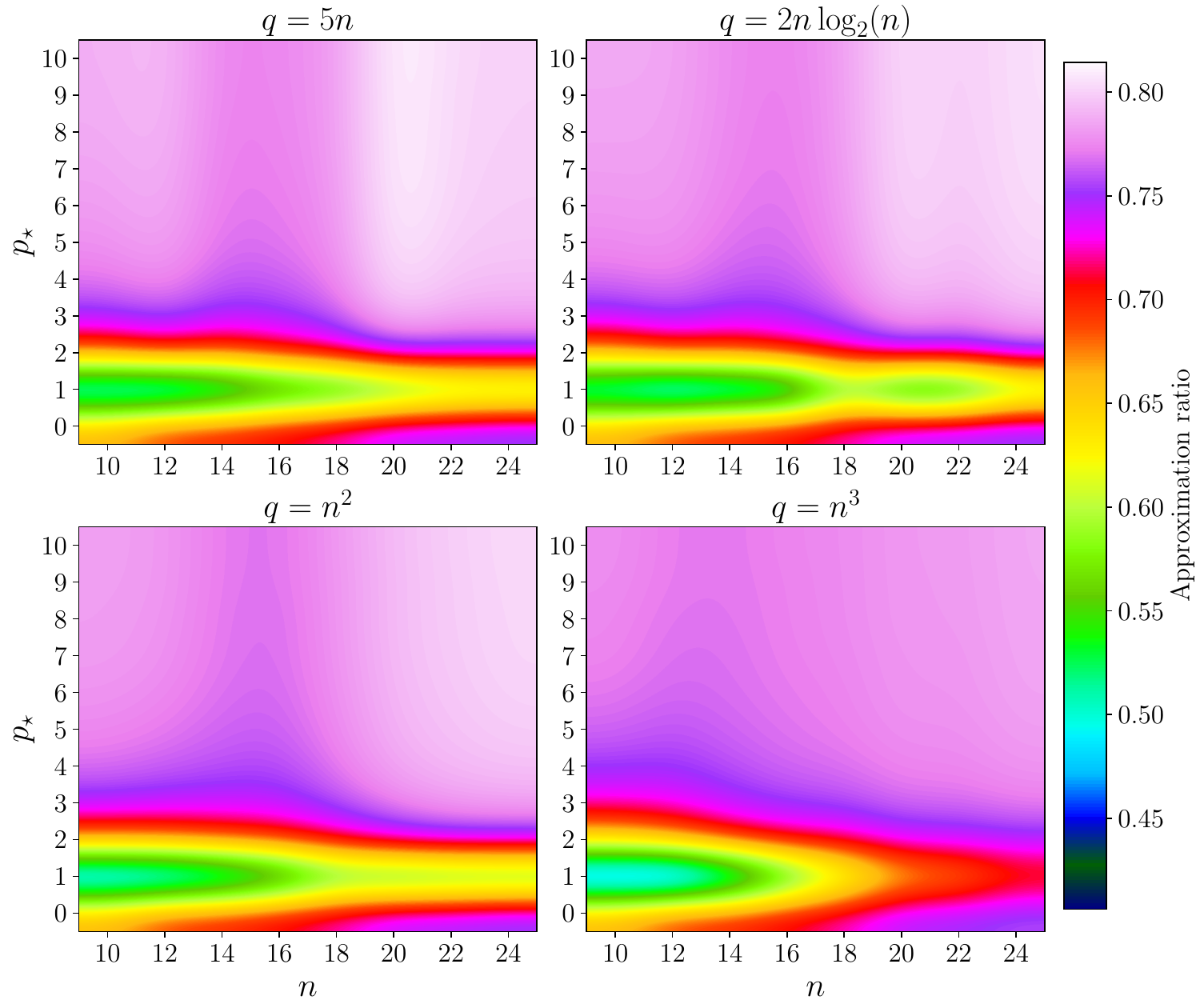}
        \includegraphics[width=1.\linewidth]{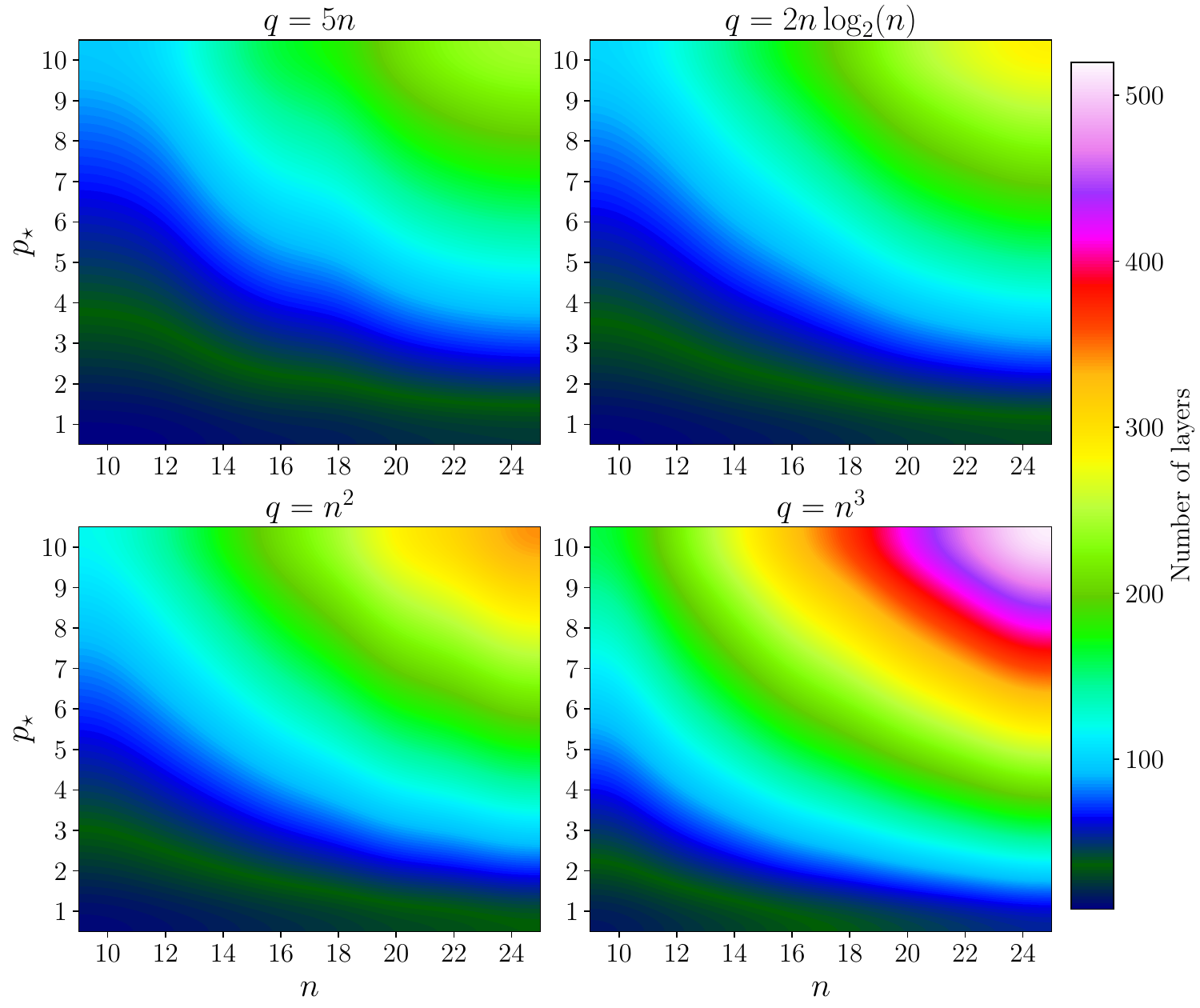}
        \caption{Approximation ratio (top) and number of layers in the circuit (bottom) as a function of the number of qubits $n\in[10,24]$ and the discretization parameter $p_\star\in[1,10]$ for unweighted MaxCut on random Erdős–Rényi graphs with different sampling values $q\in\{5n,2n\log_2(n),n^2,n^3\}$.}
        \label{fig:appendix_unweighted_ratios_layers}
\end{figure}

\begin{figure*}
    \centering
        \includegraphics[width=1.\linewidth]{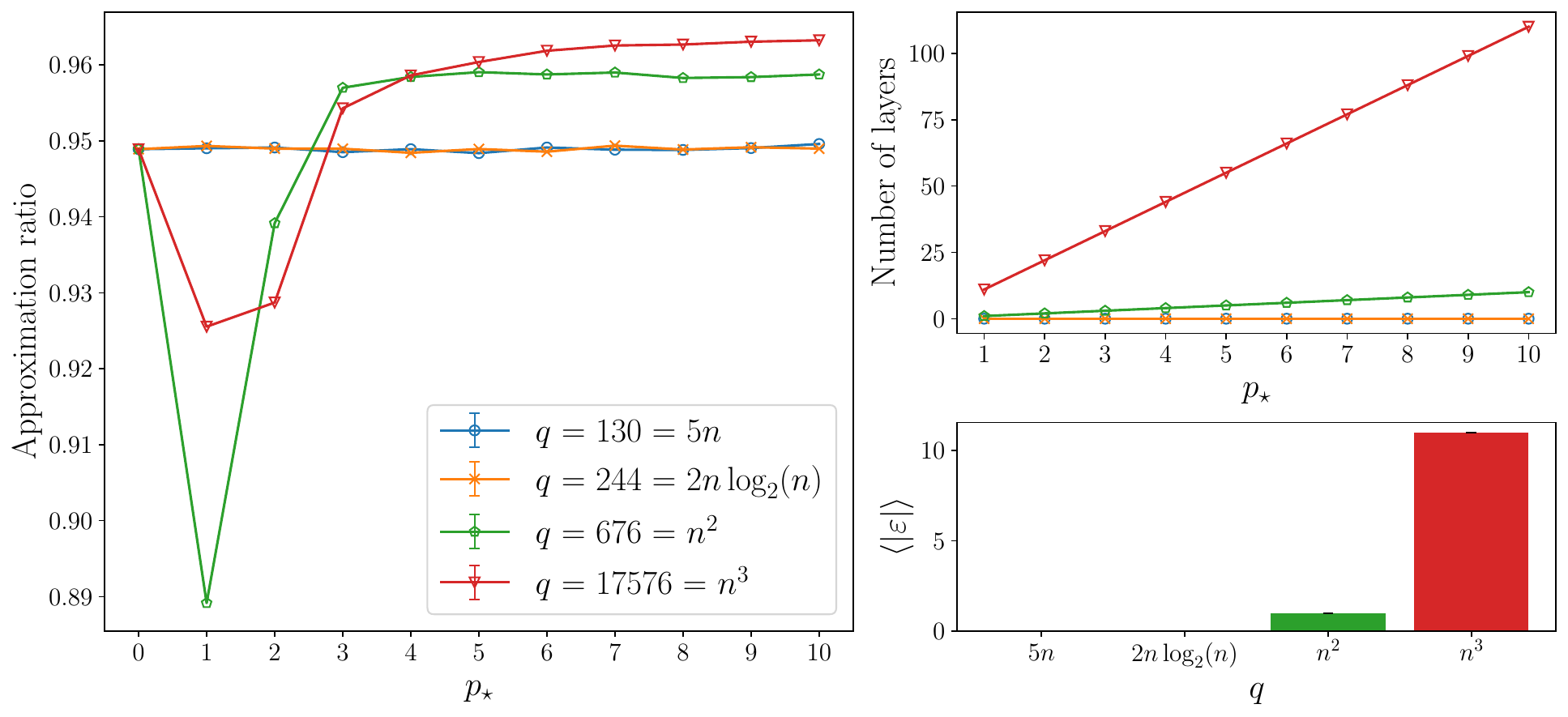}
        \includegraphics[width=1.\linewidth]{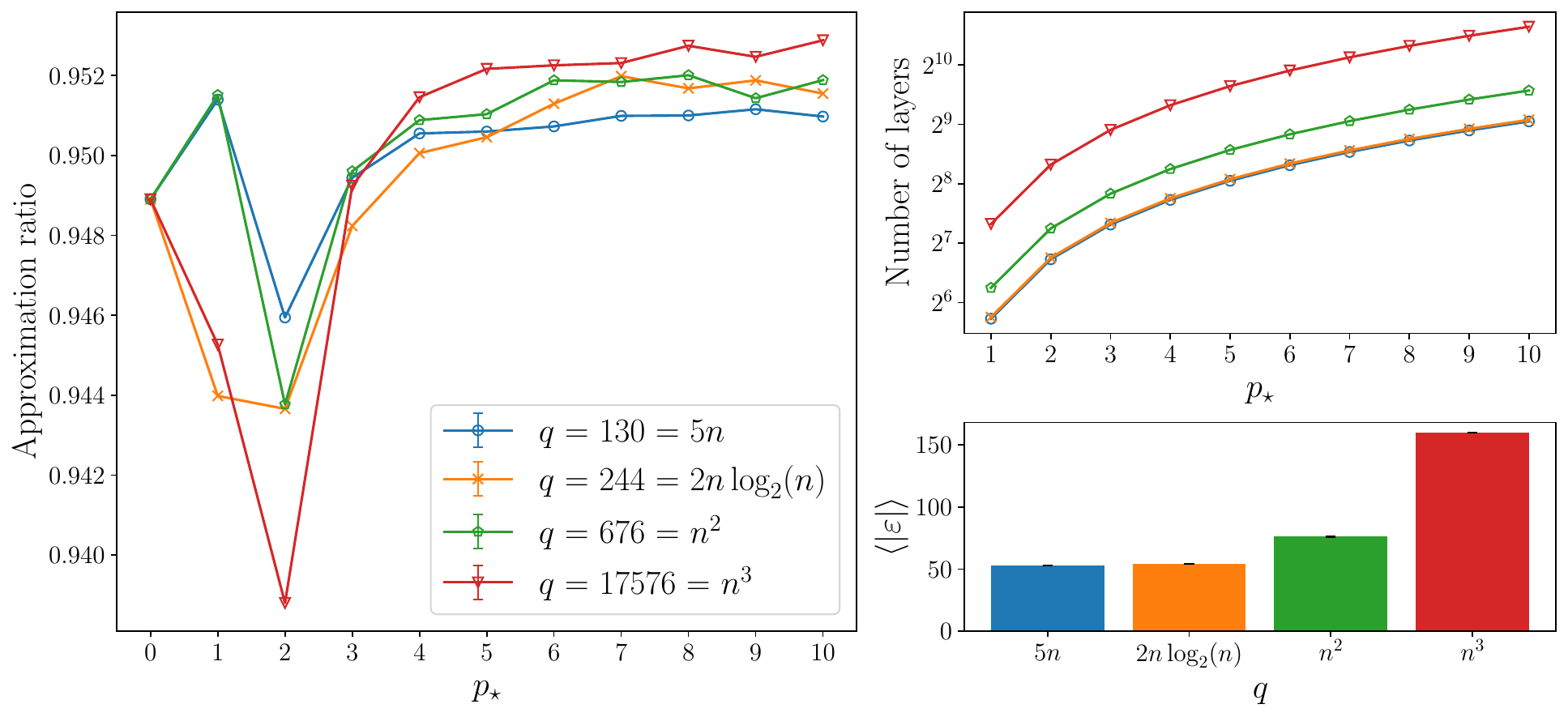}
        \caption{\textsc{SamBa--GQW} quantum circuit simulations for LABS for $n=26$ qubits with different sampling values $q\in\{5n,2n\log_2(n),n^2,n^3\}$. The discretization parameter $p_\star$ takes values in interval $[0,10]$. The upper (resp. lower) figure presents the results obtained when considering only positive gaps (resp. both positive and negative gaps).}
        \label{fig:appendix_labs_26}
\end{figure*}

\begin{figure}
    \centering
        \includegraphics[width=1.\linewidth]{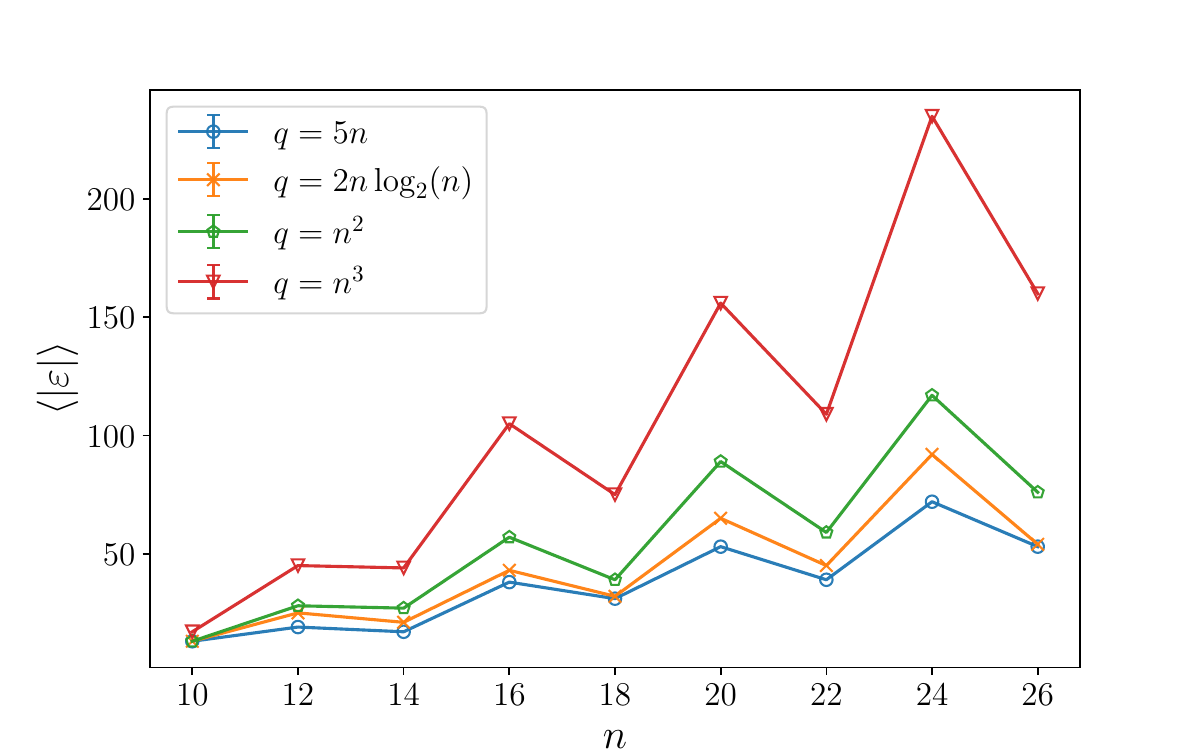}
        \caption{\textsc{SamBa--GQW} quantum circuit simulations for LABS for $n\in[10,26]$ qubits with different sampling values $q\in\{5n,2n\log_2(n),n^2,n^3\}$. The discretization parameter $p_\star$ takes values in interval $[0,10]$.}
        \label{fig:appendix_labs_real_samples}
\end{figure}

\begin{figure}
    \centering
        \includegraphics[width=1.\linewidth]{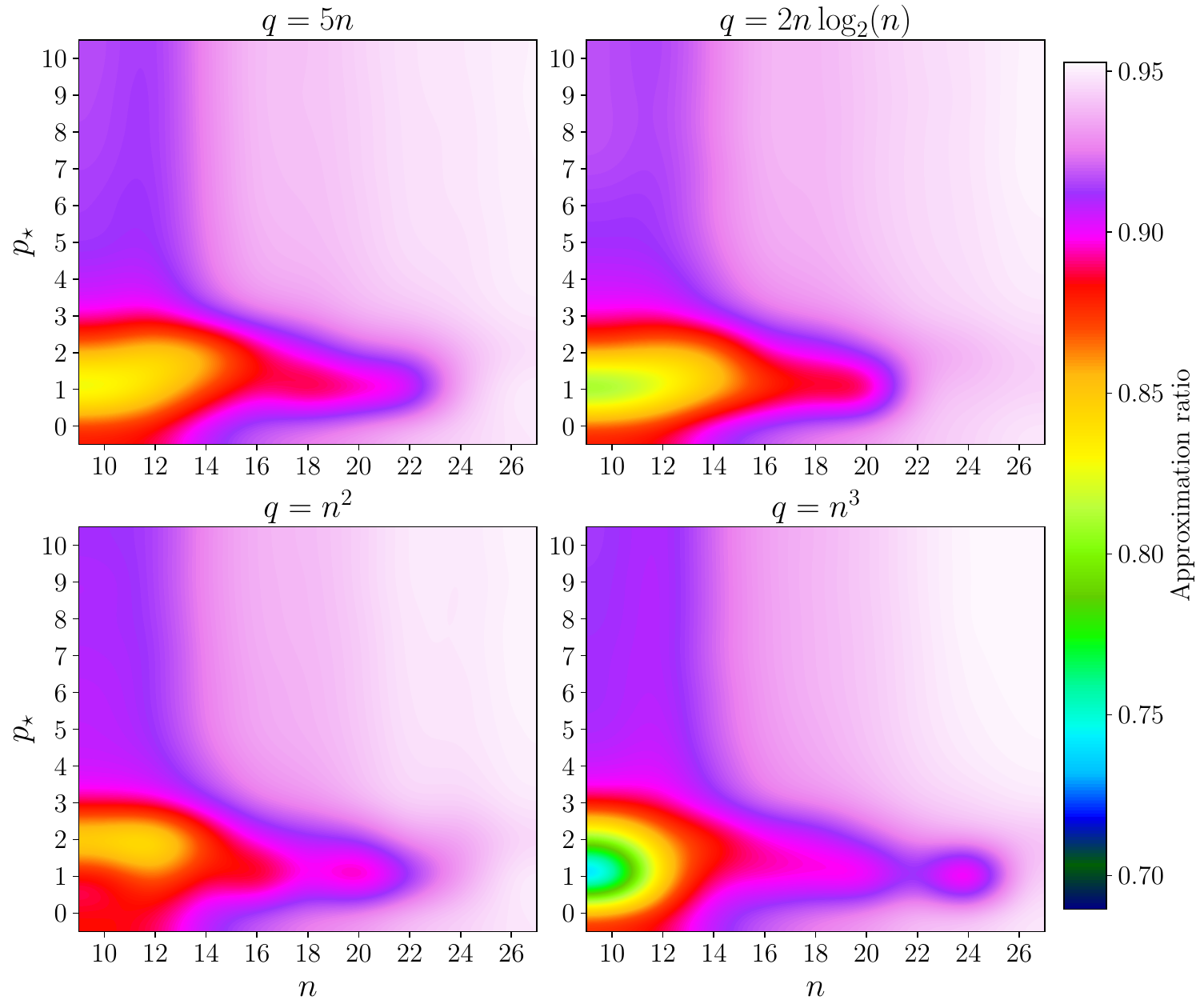}
        \includegraphics[width=1.\linewidth]{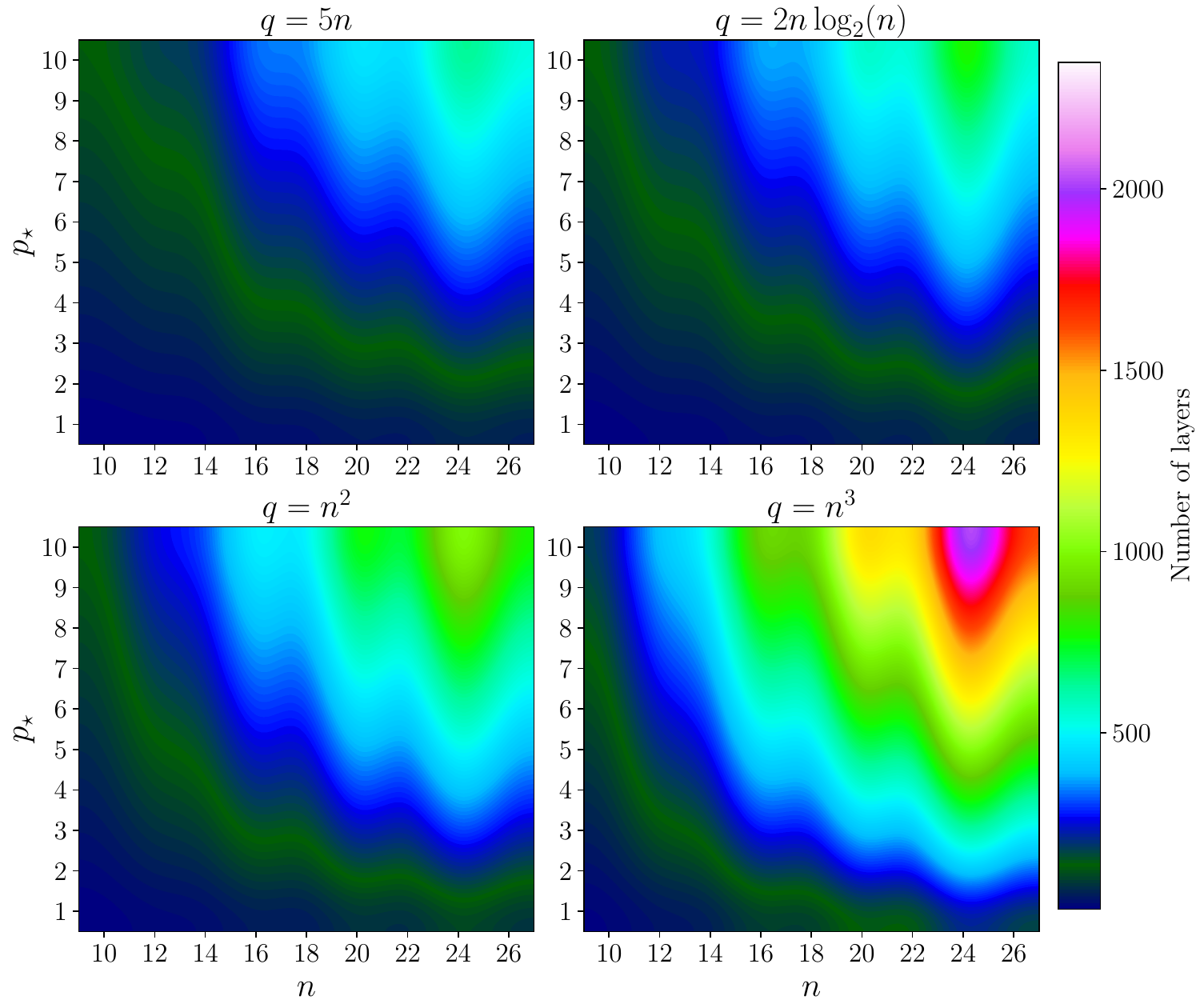}
        \caption{Approximation ratio (top) and number of layers in the circuit (bottom) as a function of the number of qubits $n\in[10,26]$ and the discretization parameter $p_\star\in[1,10]$ for LABS with different sampling values $q\in\{5n,2n\log_2(n),n^2,n^3\}$.}
        \label{fig:appendix_labs_ratios_layers}
\end{figure}

In this section, we modify the sampling protocol by considering the absolute values of the energy gaps rather than restricting to positive ones, and we study the impact of this modification on the results for unweighted MaxCut and LABS. These two problems were selected because they are the only cases in our simulations where the number of sampled gaps decreases significantly with increasing problem size. With this modification, line~7 of Algorithm~\ref{alg:sampling} becomes
\begin{equation}
	\Delta \gets \left\{\frac{\lvert\Delta^C_{x,y}\rvert}{\lvert\Delta^M_{x,y}\rvert} \;\middle|\; \forall y \in \mathrm{B}(x)\right\}.
\end{equation}
We present the results of the quantum circuit simulations for unweighted MaxCut for $n=24$ qubits on Fig.~\ref{fig:appendix_unweighted_24}. Unsurprisingly, relaxing the restriction to positive gaps results in a larger number of sampled gaps. This modification also impacts the approximation ratio, as sampling only positive gaps leads to systematically higher values, with a discrepancy of roughly $0.2$ between the two methods. In addition, Fig.~\ref{fig:appendix_unweighted_real_samples} indicates that, when gap magnitudes are considered irrespective of their sign, the number of sampled gaps continues to increase as the problem size grows. Furthermore, we illustrate the approximation ratio and number of layers in the quantum circuits as a function of $n$ in Fig.~\ref{fig:appendix_unweighted_ratios_layers}. We also present the results for LABS with $n=26$ qubits in Fig.~\ref{fig:appendix_labs_26}. As in the case of unweighted MaxCut (see Fig.~\ref{fig:appendix_unweighted_24}), including both positive and negative gaps leads to a substantially larger number of sampled gaps, increasing from zero to approximately $50$ for $q\in\{5n,2n\log_2(n)\}$. Consistent with the observations for unweighted MaxCut, considering all gaps together with their sign results in lower approximation ratios than those obtained when only positive gaps are considered. However, in contrast to unweighted MaxCut, the difference remains below $0.1$.
Furthermore, Fig.~\ref{fig:appendix_labs_real_samples} illustrates the evolution of the number of sampled gaps as a function of the problem size. Although this quantity exhibits oscillatory behavior, alternating between increases and decreases as $n$ grows, the magnitude of the increases appears to consistently exceed that of the decreases, resulting in an overall upward trend.
Finally, Fig.~\ref{fig:appendix_labs_ratios_layers} reports the approximation ratio and the number of layers in the quantum circuits as functions of the problem size $n$.

\end{document}